\documentclass[12pt,a4paper]{article}

\usepackage{a4wide}
\usepackage{graphicx}
\usepackage{amsmath}
\usepackage{amssymb}
\usepackage{hyperref}
\usepackage{url}
\usepackage{algorithm}
\usepackage{algorithmic}
\usepackage{rotating}
\usepackage{booktabs}
\usepackage[vflt]{floatflt}
\usepackage[small,bf]{caption}
\usepackage{xspace}

\newcommand{\ee}{\ensuremath{e^+e^-}}
\newcommand{\ie}{i.e.\ }
\newcommand{\cf}{cf.\ }
\newcommand{\eg}{e.g.\ }
\newcommand{\tw}{\textwidth}
\newcommand{\cut}{\mathrm{cut}}
\newcommand{\as}{\alpha_s}
\newcommand{\aew}{\alpha_\mathrm{EW}}
\newcommand{\order}[1]{{\cal O}\left(#1\right)}
\newcommand{\cG}{{\cal G}}
\newcommand{\alg}{\mathrm{alg}}
\newcommand{\quark}{\mathrm{quark}}

\newcommand{\jet}{\mathrm{jet}}
\newcommand{\SISCone}{\mathrm{SISCone}}

\newcommand{\CamAachen}{\text{C/A}}
\newcommand{\kt}{\ensuremath{k_t}}
\newcommand{\akt}{\ensuremath{\textrm{anti-}k_t}}
\newcommand{\NP}{\mathrm{NP}}
\newcommand{\UE}{\mathrm{UE}}
\newcommand{\cNg}{{\cal N}_g}
\newcommand{\JA}{\text{JA}}
\newcommand{\oPJ}{\ensuremath{\mathrm{1PJ}}}
\newcommand{\GJ}{\ensuremath{\mathrm{GJ}}}
\newcommand{\qq}{\ensuremath{q\bar q}\xspace}
\newcommand{\gluglu}{\ensuremath{gg}\xspace}

\newcommand{\Qa}[1]{\ensuremath{Q_{f=#1}^{w}}\xspace}
\newcommand{\Qb}[1]{\ensuremath{Q_{w=#1\sqrt{M}}^{1/f}}\xspace}
\newcommand{\rhoL}{\ensuremath{\rho_{L}}\xspace}

\newcommand{\cM}{{\cal M}}
\newcommand{\cA}{{\cal A}}

\newcommand{\Dzero}{D0\!\!\!/\xspace}

\def\CF{C_F}
\def\TR{T_R}
\def\CA{C_A}

\def\nf{n_{\!f}}

\newcommand{\second}{\;\mathrm{s}}
\newcommand{\GeV}{\;\mathrm{GeV}}
\newcommand{\TeV}{\;\mathrm{TeV}}
\newcommand{\mb}{\;\mathrm{mb}}

\newcommand{\cm}{\;\mathrm{cm}}
\newcommand{\s}{\;\mathrm{s}}
\newcommand{\ns}{\;\mathrm{ns}}

\newcommand{\filt}{\mathrm{filt}}

\newcommand{\toplevel}{section\xspace}
\newcommand{\Toplevel}{Section\xspace}

\title{\bf Towards Jetography}
\author{{\sc Gavin P. Salam}\\[1em]
  LPTHE,
  UPMC Univ. Paris 6,\\
  CNRS UMR 7589,
  75252 Paris 05, France
}
\date{}

\begin{document}

\maketitle

\begin{abstract}
As the LHC prepares to start taking data, this review is intended to
provide a QCD theorist's understanding and views on jet finding at
hadron colliders, including recent developments.
My hope is that it will serve both as a primer for the newcomer to
jets and as a quick reference for those with some experience of the
subject.
It is devoted to the questions of how one defines jets, how
jets relate to partons, and to the emerging subject of how best to
use jets at the LHC.
\end{abstract}
\newpage

\tableofcontents

\newpage



\section{Introduction}

It is common to discuss high-energy phenomena involving quantum
chromodynamics (QCD) in terms of quarks and gluons. Yet quarks and
gluons are never visible in their own right. Almost immediately after
being produced, a quark or gluon fragments and hadronises, leading to
a collimated spray of energetic hadrons --- a jet.
Jets are obvious structures when one looks at an event display, and
by measuring their energy and direction one can get close to the idea
of the original ``parton''.
The concept of a parton is, however, ambiguous: the fact that partons
have divergent branching probabilities in perturbative QCD means that
one must introduce a prescription for defining what exactly one means
by the term.
Similarly, jets also need to be defined --- this is generally done
through a jet definition, a set of rules for how to group particles
into jets and how to assign a momentum to the resulting jet.
A good jet definition can be applied to experimental measurements, to
the output of parton-showering Monte Carlos and to partonic
calculations, and
the resulting jets provide a common representation of all these
different kinds of events.

Jets are used for a wide range of physics analyses. One way of
classifying their uses is according to the different possible origins
for the partons that give rise to the jets.
At hadron colliders (and in photoproduction), one of the best studied
jet observables is the inclusive jet spectrum, related to the
high-momentum-transfer $2\to2$ scattering of partons inside the
colliding (anti)protons. In this kind of process the energy of the jet
(in the partonic centre-of-mass frame) is closely related to that of
the parton in the proton that underwent a hard scattering and the
inclusive jet spectrum therefore contains information on the
distributions of partons inside the proton (e.g.\
refs.~\cite{CDFkt,Abazov:2008hua,Adloff:2003nr,Breitweg:1998ur}), and
also on the strength of their interaction.

Another origin for the partons that lead to jets is that they come
from the hadronic decay of a heavy particle, for example a top quark,
a Higgs boson, or some other yet-to-be discovered resonance. If, at
tree-level, the heavy particle decays to many partons (e.g. through a
cascade of decays) then a high multiplicity of corresponding jets may
be a sign of the presence of that particle (as used for example in
SUSY searches, such as ref.~\cite{Wang:2007mqa}); and the sum of the momenta of the
jets (or of the jets, leptons and missing transverse energy
information) should have an invariant mass that is close to that of
the heavy particle, a feature used for example in measurements of the
top-quark mass~\cite{Abazov:2009eq,Aaltonen:2010pe}.
%

Jets may also originate radiatively, for example from the emission of
a gluon off some other parton in the event. The rate of production of
such jets provides information on the value of strong coupling
%
(for example refs.~\cite{Aaron:2009vs,Chekanov:2002be,Abdallah:2004uq,Abbiendi:2005vd,Heister:2002tq})
and is related also to the colour structure of events.
One use of this is, for
example, to help discriminate between Higgs-boson production through
electroweak vector-boson fusion (which radiates less) and through
gluon-fusion (which radiates more).
Radiative emission of partons is also one of the main backgrounds to
multi-jet signals of new physics; consistently predicting such
backgrounds involves matching tree-level matrix-element calculations
with Monte Carlo parton showers, for which jet-definitions provide a
powerful way of avoiding double counting, by prescribing which
emissions should be the responsibility of the matrix-element (those
that lead to extra jets) and which ones should come from the parton
showers (those that don't) \cite{Catani:2001cc,Alwall:2007fs}.

Though most uses of jets essentially identify a jet as coming from a
single parton, one should never forget how ambiguous this association
really is, and not just because partons are an ill-defined concept.
For example, when a highly boosted W or Z boson decays to two partons,
those partons may be so collimated by the boost that they will lead to
a single jet, albeit it one with substructure.
And QCD radiative corrections also inevitably give substructure to
jets.
Much as the number of jets and their kinematics can be used to learn
about the properties of the event, so can the structure within the
jets.

Given the variety of these and other related possible uses of jets,
it should not be surprising that there is no single optimal way of
defining jets and, over the 30 years that have passed since the first
detailed proposal for measuring jets~\cite{StermanWeinberg}, many jet
definitions have been developed and used.
The ideas behind jet definitions are rather varied. 
One of the aims of this review
(\toplevel~\ref{sec:2005-state-of-the-art}) is to provide an overview of
the different kinds of jet definition that exist.
Given that the main use of jets in the coming years will be at the
Large Hadron Collider at CERN (LHC), the emphasis here and throughout
the review will be on hadron-collider jets, though a number of the
ideas in jet finding actually have their origins in studies of \ee\
and $ep$ collisions.

One of the characteristics of the LHC is that its particle
multiplicity is expected to be much higher than in preceding
colliders.
Some part of the increase is due to the LHC's higher energy, but most
of it will be a consequence of the multiple minimum-bias interactions
(pileup) that will occur in each bunch crossing.
High multiplicities pose practical challenges for the computer codes
that carry out jet finding, because the computing time that is
required usually scales as some power of the multiplicity, $N$.
Until a few years ago, this was often a limiting factor in
experimental choices of jet finding methodology.
Recent years' work (described in \toplevel~\ref{sec:resolving-snowmass})
has shown how these practical issues can be resolved by exploiting
their relation to problems in computational geometry.
This makes it easier for LHC's jet-finding choices to be based on
physics considerations rather than practical ones.

Given a set of practical jet algorithms, the next question is to
establish their similarities and differences. 
Any jet algorithm will form a jet from a single hard isolated
particle.
However, different jet definitions may do different things when two
hard particles are close by, when a parton radiates a soft gluon, or
when the jet is immersed in noise from pileup.
\Toplevel~\ref{sec:understanding} examines standard and recent results
on these issues, for the most important of the current jet algorithms.

Once one has understood how jets behave, the final question that needs
to be addressed is that of determining the jet definitions and methods
that are optimal for specific physics analysis tasks. 
One might call this subject ``jetography'', in analogy with
photography, where an understanding of optics, of one's light sensor,
and of properties of the subject help guide the choice of focus,
aperture and length of exposure.
Ultimately, it is neither the photons in photography, nor the jets in
jetography that are of interest; rather it is the objects (new
particles, PDFs, etc.) that they help you visualise or discover.
In the context of the LHC, it is probably fair to say that jetography
is still in its infancy, hence the title of the review.
Nevertheless some first results have emerged in the past couple of
years, notably (as discussed in \toplevel~\ref{sec:using-jets}) with
respect to simple dijet mass reconstructions, hadronic decays of
boosted heavy particles, and the question of limiting the effect of
pileup.


One thing that this review does not do is examine the wide range of
uses of jets in LHC and other experiments' analyses, aside from the
brief discussion given above.
This is a vast subject, and to obtain a full overview probably
requires that one consult the main ATLAS and CMS physics analysis
programme documents~\cite{Aad:2009wy,Bayatian:2006zz} and the ``LHC
primer'' \cite{Campbell:2006wx}, as well as recent work by the
Tevatron and HERA, summarised for example in \cite{Hatakeyama:2009mn,Jung:2009eq}.
Other reviews of jets in recent years include
\cite{RunII-jet-physics,Ellis:2007ib}. 
Finally a topic that is barely touched upon here is the nascent field
of jet finding in heavy-ion collisions, for which the reader is
referred to~\cite{d'Enterria:2009vg,Salur:2008hs}.



\section{Jet algorithms}
\label{sec:2005-state-of-the-art}

Jet algorithms provide a set of rules for grouping particles into
jets. They usually involve one or more parameters that indicate how
close two particles must be for them to belong to the same
jet. Additionally they are always associated with a recombination
scheme, which indicates what momentum to assign to the combination of
two particles (the simplest is the 4-vector sum). Taken together, a
jet algorithm with its parameters and a recombination scheme form a
``jet definition''.

An accord as to some general properties of jet definitions, the
``Snowmass accord'', was set out in 1990~\cite{Huth:1990mi} by a group
of influential theorists and experimenters, and reads as follows
\begin{quotation}\small\sf
  \noindent Several important properties that should be met by a jet
  definition are [3]:
  \begin{enumerate}
  \item Simple to implement in an experimental analysis;
  \item Simple to implement in the theoretical calculation;
  \item Defined at any order of perturbation theory;
  \item Yields finite cross sections at any order of perturbation
    theory;
  \item Yields a cross section that is relatively insensitive to
    hadronisation.
  \end{enumerate}
\end{quotation}
where ref.~[3] is given below as \cite{Ellis:1989vm}.
It is revelatory that ref.~\cite{Huth:1990mi} is entitled ``{\it
  Toward} a standardization of jet definitions'' (my italics). If one
reads the rest of the article, one realises that it wasn't evident at
the time what the standard jet definition should actually be, nor was
there a clear path towards satisfying the Snowmass accords, at least
for hadron colliders.

When the next major community-wide discussion on jets took place, in
2000, in preparation for Run~II of the
Tevatron~\cite{RunII-jet-physics}, new jet algorithms had been
invented~\cite{Kt,KtHH,Kt-EllisSoper,Cam,Aachen}, old algorithms had
been patched~\cite{midpoint} and it is probably fair to say that the
community had \emph{almost} satisfied the Snowmass requirements.
%
Nevertheless, the recommendations
of the Run~II workshop were followed in only part of subsequent
Tevatron work and, until recently, had also been ignored in much of
the preparatory work towards LHC.

This means that
there are currently very many hadron-collider jet algorithms in use --- some
dating from the 80's, others from the 90's. 
The situation is further confused by the fact that different
algorithms share the same name (notably ``iterative cone''), and
that there is no single source of information on all the different
algorithms. 
Additionally, it has not always been clear how
any given algorithm fared on the Snowmass requirements.

The purpose of this \toplevel is to give an overview of all the main
different algorithms, including some of the most recently developed
ones, so as to provide the background for anyone reading current jet
work from both the theory and experimental communities.


The \toplevel's organisation reflects the split of jet algorithms into
two broad categories. Firstly those based in one form or another on
``cones''. They can be thought of as ``top-down'' algorithms, relying on
the idea that QCD branching and hadronisation leaves the bulk features
of an event's energy flow unchanged (specifically, energy flow into a
cone).
Secondly, sequential recombination algorithms, ``bottom-up''
algorithms that repeatedly recombine the closest pair of particles
according to some
distance measure, usually related to the divergent structure of QCD
matrix elements.

The nomenclature used to distinguish the types of jet algorithm
(notably cones) is currently not always uniform across the field.
That used here follows the lines set out
in~\cite{Cacciari:2008gp,Buttar:2008jx:All}.

Before continuing, a note is due concerning the completeness of this
section. Its aim is to communicate the essential ideas about many of
the main jet algorithms (a more concise overview is given in
\cite{Buttar:2008jx:All}). It will not describe every detail of every
single jet algorithm. Where possible, references will be supplied to
more complete descriptions. In some cases, no such reference exists,
and the interested reader is then advised to consult computer code for
the given jet algorithm.

\subsection{Cone algorithms}
\label{sec:cone-algs}

The first-ever jet algorithm was developed by Sterman and Weinberg in
the 1970's~\cite{StermanWeinberg}. It was intended for $\ee$
collisions and classified an event as having two jets if at least a
fraction $1-\epsilon$ of the event's energy was contained in two cones
of opening half-angle $\delta$ (and hence is known as a ``cone''
algorithm).
This definition made it possible to have a fully consistent perturbative QCD
calculation of the probability of having two jets in an event.

The two parameters $\delta$ and $\epsilon$ reflect the arbitrariness
in deciding whether an event has two or more jets. Typically one would
avoid taking extreme values ($\epsilon$ too close to $0$ or $1$,
$\delta$ too close to zero), but apart from that the optimal choice of
$\delta$ and $\epsilon$ would depend on the specific physics analysis
being carried out.
The presence of separate angular and energy parameters to dictate the
characteristics of the jet finding is typical of cone algorithms, as
we shall see below.


Cone algorithms have evolved substantially since
\cite{StermanWeinberg} and are today mostly used at hadron colliders.
The changes reflect the fact that in hadron collisions it doesn't make
sense to discuss the total energy (since most of it is not involved in
the hard reaction, and goes down the beam pipe), that it isn't always
obvious, physically or computationally, \emph{where} to place the
cones, and that issues arise when trying to define
events with more than two jets (with the associated problem of
``overlapping'' cones).

\subsubsection{Iteration}

Let us first examine the question of where to place the cones.
Most of today's widely used cone algorithms are ``iterative cones''
(IC). In such algorithms, a seed particle $i$ sets some initial
direction, and one sums the momenta of all particles $j$ within a
circle (``cone'')
of radius $R$ around $i$ in azimuthal angle $\phi$ and rapidity $y$
(or pseudorapidity $\eta$),%
\footnote{%
  These are standard hadron-collider variables. Given a beam along the
  $z$-direction, a particle with longitudinal momentum $p_z$, energy
  $E$ and angle $\theta$ with respect to the beam (longitudinal)
  direction has rapidity $y \equiv \frac12\ln \frac{E+p_z}{E-p_z}$ and
  pseudorapidity $\eta \equiv -\ln \tan \theta/2$. Massless particles
  have $y=\eta$.  Differences in rapidity are invariant under
  longitudinal boosts, whereas differences in pseudorapidity are
  invariant only for massless particles. 
  Where an analysis in $\ee$ will use particles' energies and the
  angles between the particles, an analysis in a $pp$ collider will
  often use $p_t$ (or $E_t$) and $\Delta R_{ij}^2$ (defined either
  with rapidities or pseudorapidities).
} %
i.e.\ taking all $j$ such that
\begin{equation}
  \label{eq:deltaij}
  \Delta R_{ij}^2 = (y_i - y_j)^2 + (\phi_i-\phi_j)^2 < R^2\,,
\end{equation}
where $y_i$ and $\phi_i$ are respectively the rapidity and azimuth of
particle $i$.
The direction of the resulting sum is then used as a new seed
direction, and one iterates the procedure until the direction of the
resulting cone is stable.
The dimensionless parameter $R$ here, known as the jet radius,
replaces the angular scale $\delta$ that was present in the original
Sterman-Weinberg proposal.
The Sterman-Weinberg $\epsilon$ parameter is less-directly mirrored in
hadron-collider cone algorithms. Rather, most physics analyses will
use a cone algorithm to obtain jets without any specific energy cut,
but then will consider only those jets that are above a certain
transverse-momentum threshold.

To be fully specified, seeded iterative jet algorithms must deal with
two issues:
\begin{itemize}
\item What should one take as the seeds?
\item What should one one do when the cones obtained by iterating two
  distinct seeds ``overlap'' (\ie share particles)?
\end{itemize}
Different approaches to these issues lead to two broad classes
of cone algorithm. 


\subsubsection{Overlapping cones: the progressive removal approach}

One approach is to take as one's first seed the particle (or
calorimeter tower) with the largest transverse momentum.
Once one has found the corresponding stable cone, one calls it a jet
and removes from the event all particles contained in that jet. 
One then takes as a new seed the hardest particle/tower among those
that remain, and uses that to find the next jet, repeating the
procedure until no particles are left (above some optional threshold).
This avoids any issue of overlapping cones.
A possible name for such algorithms is iterative cone with
progressive removal (IC-PR) of particles.

IC-PR algorithms' use of the hardest particle in an event gives them
the drawback that they are collinear unsafe:
the splitting of the hardest particle (say $p_1$) into a nearly
collinear pair ($p_{1a}$, $p_{1b}$) can have the consequence that
another, less hard particle, $p_2$, pointing in a different direction
and with $p_{t,1a},\,p_{t,1b}<p_{t,2}<p_{t,1}$, suddenly becomes the
hardest particle in the event, thus leading to a different final set
of jets. We will return to this in
section~\ref{sec:cone-algorithms-irc}.

\paragraph{Fixed cones.}
A widespread, simpler variant of IC-PR cone algorithms is one that
does not iterate the cone direction, but rather identifies a fixed
cone (FC)\footnote{``Fixed cone'' can be an ambiguous term. In
  particular, in some contexts it is used to refer to cones whose
  shape is fixed rather than cones whose position is fixed.} around
the seed direction and calls that a jet. It starts from the hardest
seed and progressively removes particles as the jets are identified
(thus FC-PR). It suffers from the same collinear unsafety issue as the
IC-PR algorithms.

IC-PR and FC-PR algorithms are often referred to as UA1-type cone
algorithms, even though the algorithm described in the original UA1
reference~\cite{Arnison:1983gw} is somewhat different.\footnote{ The
  UA1 algorithm~\cite{Arnison:1983gw} proceeds as follows:
  the particle (or cell) with highest $E_t$ starts a jet; working
  through the list of particles in decreasing $E_t$, each one is added
  to the jet to which it is closest, as long as it is within $\Delta
  R<R$ ($\Delta R^2 = \Delta \eta^2 + \Delta \phi^2$, $R$ taken to be
  $1$); otherwise, the particle initiates a new jet.
  Finally, once all remaining particles have $E_t < 2.5\GeV$, each
  particle is simply added to the jet nearest in $\eta,\phi$ if its
  transverse momentum relative to the jet axis is less than $1\GeV$ and
  it is no further than $45^\circ$ in direction from the jet axis.
  %
}
This may be due to different versions of the UA1 algorithm having been
presented at conferences prior to its final
publication~\cite{Arnison:1983gw}.\footnote{I am grateful to
  Torbj\"orn Sj\"ostrand for comments on this point.}


\subsubsection{Overlapping cones: the split--merge approach}

Another approach to the issue of the same particle appearing in many
cones applies if one chooses, as a first stage, to find all the stable
cones obtained by iterating from all particles or calorimeter towers (or those for
example above some seed threshold $\sim 1-2$GeV).\footnote{In one
  variant, CDF's JetClu~\cite{Abe:1991ui}, 
  ``ratcheting'' is included, which means that during iteration of 
  cone, all particles included in previous iterations are retained
  even if they are no longer within the geometrical cone, see also
  section~\ref{sec:cone-dark-towers}.}
One may then run a split--merge (SM) procedure, which merges a pair
of cones if more than a fraction $f$ of the softer cone's transverse
momentum is in particles shared with the harder cone; otherwise the shared
particles are assigned to the cone to which they are closer. 
A possible generic name for such algorithms is IC-SM.
The exact behaviour of SM procedures depends on the precise ordering
of split and merge steps and a fairly widespread procedure is
described in detail in~\cite{RunII-jet-physics}. It essentially works as
follows, acting on an initial list of ``protojets'', which is just the
full list of stable cones:
\begin{enumerate}
 \item Take the protojet with the largest $p_t$ (the `hardest'
   protojet), label it $a$.
 \item Find the next hardest protojet that shares particles with the
   $a$ (\ie overlaps), label it $b$. If no such protojet exists,
   then remove $a$ from the list of protojets and add it to the list of
   final jets.
 \item Determine the total $p_t$ of the particles shared between
   the two protojets, $p_{t,\mathrm{shared}}$.
   \begin{itemize}
   \item If $p_{t,\mathrm{shared}}/p_{t,b} > f$, where $f$ is a free
     parameter known as the overlap threshold, replace protojets $a$
     and $b$ with a single merged protojet.
   \item Otherwise ``split'' the protojets, for example assigning the shared
     particles just to the protojet whose axis is closer (in angle).
   \end{itemize}
 \item Then repeat from step 1 as long as there are protojets left.
\end{enumerate}
Generally the overlap threshold $f$ is chosen to be $0.5$ or $0.75$
(the latter is probably to be preferred~\cite{Cacciari:2008gn}).
%
An alternative to SM is to have a ``split-drop'' (SD) procedure, where
the non-shared particles that belong to the softer of two overlapping
cones are simply dropped, i.e.\ are left out of jets altogether.
The main example of an algorithm with a SD procedure is PxCone
(described for example in~\cite{Seymour:2006vv}).

The outcome of split--merge and split--drop procedures depends on the
initial set of stable cones. One of the main issues with IC-SM and
IC-SD algorithms is that the addition of a new soft seed particle can lead
to new stable cones being found, altering the final set of jets. 
This is infrared unsafety and we will discuss it in detail in the next
section.

\subsubsection{Infrared and collinear safety, midpoint cones}
\label{sec:cone-algorithms-irc}

Infrared and collinear (IRC) safety is the property that if one
modifies an event by a collinear splitting or the addition of a soft
emission, the set of hard jets that are found in the event should
remain unchanged.
IRC safety is an important property for a range of reasons: 
\begin{itemize}
\item A hard parton undergoes many collinear
  splittings as part of the fragmentation process; and the
  non-perturbative dynamics also lead to collinear splittings, for
  example in the decay of energetic hadrons.
  Additionally there is always some emission of soft particles in QCD
  events, both through perturbative and non-perturbative effects.
  Collinear splittings and soft emissions effectively occur randomly
  and even their average properties are hard to predict because of the
  way they involve non-perturbative effects.
  The motivation for constructing jets is precisely that one wants to
  establish a way of viewing events that is insensitive to all these
  effects (this is also connected with point 5 of the Snowmass
  conditions).

\item In fixed-order perturbative QCD calculations, one of the main
  tools involved in making accurate standard-model predictions at
  high-energy colliders, soft emissions and collinear splittings are
  associated with divergent tree-level matrix elements. There are also
  corresponding divergent loop matrix elements that enter with the
  opposite sign.
  Normally the two sources of divergence should cancel, but for IRC
  unsafe jet algorithms the tree-level splittings may lead to one set
  of jets, while the loop diagrams may lead to another, breaking the
  cancellation and leading to infinite cross sections in perturbation
  theory (point 4 of the Snowmass conditions). 
  Below, we shall illustrate this point in more detail.

\item Experimental detectors provide some regularisation of any
  collinear and infrared unsafety (because of their finite resolution
  and non-zero momentum thresholds), but the extent to which this
  happens depends on the particular combination of tracking,
  electromagnetic calorimetry and hadronic calorimetry that is used by
  the experiment. This can make it quite difficult to connect
  experimental results for IRC unsafe algorithms to the expectations
  at hadron-level.
\end{itemize}
Cone-type jet algorithms have, historically, been plagued by issues
related to IRC safety, and a significant amount of the work on them
has been directed towards understanding and eliminating these
problems.
Let us therefore examine the question for the two classes of algorithm
we have seen so far.

\begin{figure}
  \centering
  \includegraphics[width=0.8\textwidth]{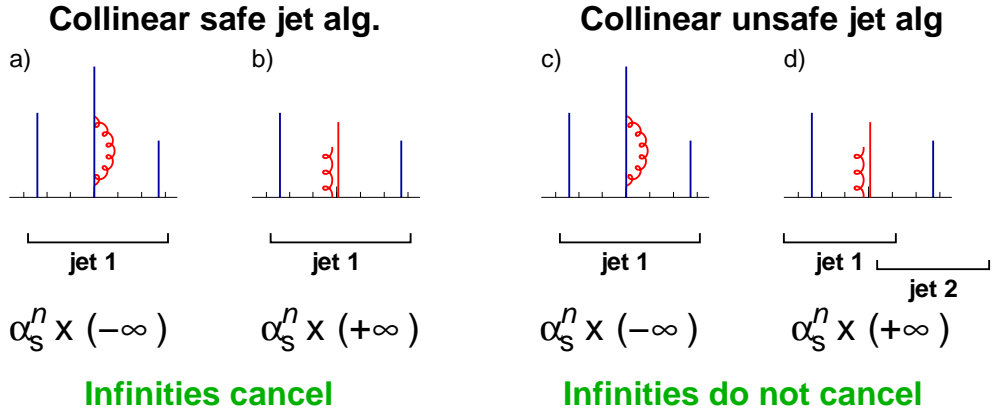}
  \caption{Illustration of collinear safety (left) and collinear
    unsafety in an IC-PR type algorithm (right) together with its implication for
    perturbative calculations (taken
    from the appendix of \cite{Cacciari:2008gp}).
    Partons are vertical lines, their height is proportional to their
    transverse momentum, and the horizontal axis indicates rapidity.
  }
  \label{fig:ICPR-coll}
\end{figure}

\paragraph{The IC-PR case.} IC-PR algorithms suffer from collinear
unsafety, as illustrated in fig.~\ref{fig:ICPR-coll}.
With a collinear safe jet algorithm, if configuration (a) (with an
optional virtual loop also drawn in) leads to one jet, then the same
configuration with one particle split collinearly, (b), also leads to
a single jet. In perturbative QCD, after integrating over loop
variables in (a) and the 
splitting angle in (b), both diagrams have infinite weights, but with
opposite signs, so that the total weight for the 1-jet
configuration is finite.

Diagrams (c) and (d) are similar, but for an IC-PR algorithm.
In configuration (c), the central particle is hardest and provides the
first seed. The stable cone obtained by iterating from this seed
contains all the particles, and one obtains a single jet.
In configuration (d), the fact that the central particle has split
collinearly means that it is now the leftmost particle that is
hardest and so provides the first seed.
Iteration from that seed leads to a jet (jet 1) that does not contain the
rightmost particle. 
That rightmost particle therefore remains, provides a new seed, and
goes on to form a jet in its own right (for full details, see the
appendix of~\cite{Cacciari:2008gp}).
As we have discussed above, it is problematic for the result of the
jet finding to depend on a collinear splitting.
The formal perturbative QCD consequence of this here is that the
infinities in diagrams (c) and (d) contribute separately to the 1-jet
and 2-jet cross sections. Thus both the 1-jet and 2-jet cross sections
are divergent.


\paragraph{The IC-SM case.} 
IC-SM (and IC-SD) type algorithms have the drawback that the addition
of an extra soft particle, acting as a new seed, can cause the
iterative process to find a new stable cone. Once passed through the
split--merge step this can lead to the modification of the final jets,
thus making the algorithm \emph{infrared} unsafe.
\begin{figure}
  \centering
  \includegraphics[width=0.7\tw]{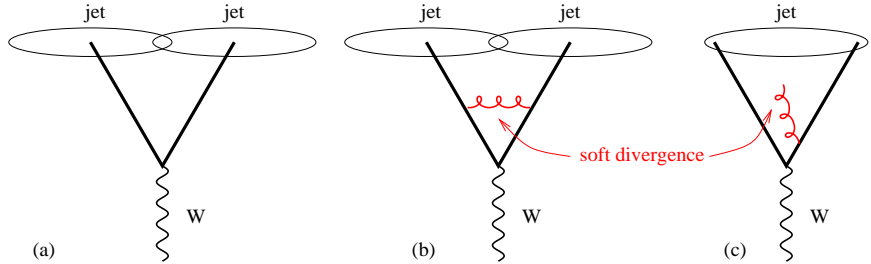}
  \caption{Configurations illustrating IR unsafety of IC-SM algorithms
    in events with a $W$ and two hard partons. The addition
    of a soft gluon converts the event from having two jets to just
    one jet. 
    In contrast to fig.~\ref{fig:ICPR-coll}, here the explicit
    angular structure is shown (rather than $p_t$ as a function of
    rapidity).}
  \label{fig:IR-unsafety}
\end{figure}
This is illustrated in fig.~\ref{fig:IR-unsafety}: in an event (a)
with just two hard partons (and a $W$, which balances momentum), both
partons act as seeds, there are two stable cones and two jets. 
The same occurs in the (negative) infinite loop diagram (b). 
However, in diagram (c) where an extra soft gluon has been emitted, the
gluon provides a new seed and causes a new stable cone to be found
containing both hard partons (as long as they have similar momenta and
are separated by less than $2R$). This stable cone overlaps with the
two original ones and the result of the split--merge procedure is that
only one jet is found.
So the number of jets depends on the presence or absence of a soft gluon
and after integration over the virtual/real soft-gluon momentum the
two-jet and one-jet cross sections each get non-cancelling infinite
contributions.
This is a serious problem, just like collinear unsafety. A good
discussion of it was given in \cite{SeymourJetShapes}.

%
\paragraph{The midpoint ``fix'' for IC-SM algorithms.} 
\begin{figure}
  \centering
  \includegraphics[width=0.53\textwidth]{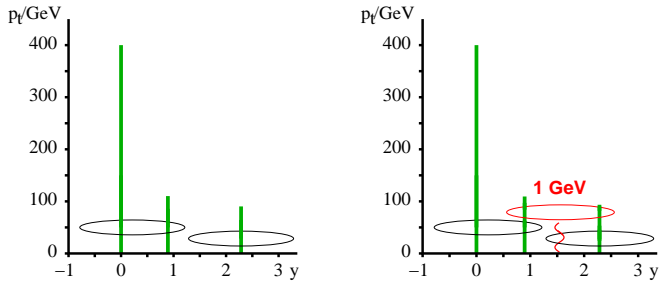}
  \caption{Configuration that is the source of IR unsafety in the
    midpoint (IC$_{mp}$-SM) algorithm, with the diagram on the right
    illustrating the extra stable cone that can appear with the addition
    of a new soft seed. Taken from~\cite{Salam:2007xv}.}
  \label{fig:midpoint-IR-problem-fig}
\end{figure}
A partial solution~\cite{midpoint} (described also in
\cite{SeymourJetShapes}), 
which was recommended
in~\cite{RunII-jet-physics}, is to additionally search for new stable
cones by iterating from midpoints between each pair of stable cones
found in the initial seeded iterations (IC$_{mp}$-SM).
This resolves the problem shown in fig.~\ref{fig:IR-unsafety} and the
resulting ``midpoint'' algorithm has often been presented as
a cone algorithm that was free of IR safety issues.
However, for configurations with three hard particles in a common
neighbourhood (rather than two for the IC-SM algorithms) the IR safety
reappears, as illustrated in fig.~\ref{fig:midpoint-IR-problem-fig}.

The ``midpoint algorithm'' has been widely used in Run~II of the
Tevatron within CDF (midpoint cone algorithm) and \Dzero (Run~II Cone
algorithm, or improved legacy cone algorithm). 
The two experiments have separate implementations, with slightly
different treatment of seeds (CDF imposes a threshold, \Dzero does
not), cone iteration (\Dzero eliminates cones below a $p_t$
threshold, CDF does not) and the split--merge stage.
In practice both algorithms incorporate a number of further technical
subtleties (for example an upper limit on the number of iterations, or
split--merge steps) and the best reference is probably the actual code
(available both within FastJet~\cite{FastJet} v2.4 and
SpartyJet~\cite{SpartyJet}).

\paragraph{Impact of IRC unsafety.} 
The impact of infrared and collinear (IRC) unsafety depends on the
observable in which one is interested.
For example for the IC-SM type algorithms, the configuration on the
right of fig.~\ref{fig:IR-unsafety} is a NNLO contribution to the
$W+\jet$ cross section, \ie a contribution $\as^3 \aew \times \infty$.
Physically, the infinity gets regularised by non-perturbative effects
and so is replaced by a factor of order $\ln p_t/\Lambda$, giving an
overall contribution $\as^3 \aew \ln p_t/\Lambda$. 
Since $\as \sim 1/\ln( p_t/\Lambda)$, this can be rewritten as $\sim
\as^2 \aew$, \ie the NNLO diagrams will give a contribution that is as
large as the NLO diagrams. Thus the perturbative series looks like:
\begin{multline}
  \label{eq:IR-bad-for-convergence}
  \underbrace{\as \alpha_{EW}}_{\mathrm{LO}} \;+\;
  \underbrace{\as^2 \alpha_{EW}}_{\mathrm{NLO}} \;+\;
  \underbrace{\as^3 \alpha_{EW} \ln \frac{p_t}{\Lambda}}_{\mathrm{NNLO}} \;+\;
  \underbrace{\as^4 \alpha_{EW} \ln^2 \frac{p_t}{\Lambda}}_{\mathrm{NNNLO}}\;+\; \cdots\,,
  \\\sim
  \underbrace{\as \alpha_{EW}}_{\mathrm{LO}} \;+\;
  \underbrace{\as^2 \alpha_{EW}}_{\mathrm{NLO}} \;+\;
  \underbrace{\as^2 \alpha_{EW}}_{\mathrm{NNLO}} \;+\;
  \underbrace{\as^2 \alpha_{EW}}_{\mathrm{NNNLO}}\;+\; \cdots\,,
\end{multline}
and it is meaningful to calculate the LO term, but no advantage is to
be had by calculating terms beyond, because the neglected pieces will
always be as large as the NLO term.
If one instead examines the $W+2$-jet cross section then the LO term
is $\as^2 \alpha_{EW}$. The NLO term, $\as^3\alpha_{EW}\ln
p_t/\Lambda \sim \as^2\alpha_{EW}$ is of the same size, so even the LO
prediction makes no sense.

The unsafety of the IC-SM algorithm can be labelled IR$_{2+1}$: its IR
unsafety is manifest for configurations with two hard particles in a
common neighbourhood plus one soft one. The midpoint algorithm is
IR$_{3+1}$, while the IC-PR and FC-PR algorithms are Coll$_{3+1}$ (the
collinear unsafety is manifest when there are 3 hard particles in a
common neighbourhood, of which one splits collinearly).

For an algorithm labelled as IR$_{n+1}$ or Coll$_{n+1}$, the last
meaningful order for the $W\!+\!\jet$ or the 2-jet cross section is
N$^{n-2}$LO.
The last meaningful order for the $W\!+\!2$-jet or the 3-jet cross
section is N$^{n-3}$LO.
The situation is summarised for various process in
table~\ref{tab:failure-cases}.

\begin{table}
  \centering
      \begin{tabular}{lcc}\toprule
        Observable        & IR$_{2+1}$ & IR$_{3+1}$, Coll$_{3+1}$\\ \midrule
        Inclusive jet cross section &  LO & NLO \\ 
        $W/Z/H$ + 1-jet cross section  &  LO & NLO \\
        $3$-jet       cross section &   none & LO  \\
        $W/Z/H$ + 2-jet cross section &   none & LO  \\
        jet masses in $3$-jet and $W/Z/H + 2$-jet events   &   none & none \\\bottomrule
      \end{tabular}
      \caption{Summary of the last meaningful order for various
        measurements with jet algorithms having different levels of IR and collinear
        unsafety.
        Adapted from~\cite{Salam:2007xv}.
      }
  \label{tab:failure-cases}
\end{table}

One way of visualising infrared and collinear unsafety (especially for
IR$_{2+1}$ algorithms) is that they
lead to an ambiguity in the effective jet radius $R$ --- a soft
emission or
collinear splitting affects how far the jet algorithm will reach for
particles.
For the IR$_{2+1}$ algorithms that ambiguity is of $\order{R}$ in the
reach (\ie the jet radius is devoid of meaning). 
For the IR$_{3+1}$ and Coll$_{3+1}$ algorithms this analogy is less
useful.

\paragraph{What to do with IRC unsafe measurements.}
Many IRC-unsafe jet measurements exist in the experimental
literature.\footnote{Strictly speaking, many algorithms incorporate a
  seed threshold, \eg $p_t > 1\GeV$. This means that they are not
  truly infrared unsafe, in that they don't lead to
  infrared infinities in perturbative calculations (though they are
  then collinear unsafe if applied to particles rather than to
  calorimeter towers).
  However a $1\GeV$ seed threshold fails to remove the large
  logarithms in eq.~(\ref{eq:IR-bad-for-convergence}) or to eliminate
  the non-perturbative uncertainties associated with IR unsafety.
  So the seed threshold does not make these algorithms any better than
  a formally IR unsafe one.  }
Some of these cases are like \cite{Aaltonen:2007ip}, where the
measurement for the $W$+$n$-jet cross section is carried out with
JetClu, an IR$_{2+1}$ unsafe algorithm for which no order of
perturbation theory is meaningful when $n>1$.%

The question then arises of how one can compare NLO theory predictions
like \cite{MCFM,NLOJet,Berger:2009zg,Ellis:2009zw,Ellis:2009bu} with the
experimental results.
One approach, specific to the IC-SM case, is to carry out the
NLO prediction with two somewhat different jet algorithms (for example
SISCone and anti-$k_t$, both discussed below), and use the difference
between the NLO calculations with the two algorithms as a measure of
the uncertainty in the prediction due to IR safety issues.
The logic behind this is that SISCone behaves as would an IC-SM
algorithm when there are soft particles everywhere (combining hard
partons into a common jet when they are as far as $2R$ apart in some
cases), while anti-$k_t$ behaves somewhat similarly to an IC-SM
algorithm when there are no soft particles present (hard partons
separated by more than $R$ usually do not end up in the same jet).
These differences are discussed in more detail in
section~\ref{sec:reach}. 

\begin{floatingfigure}{0.42\textwidth}
  \includegraphics[width=0.40\textwidth]{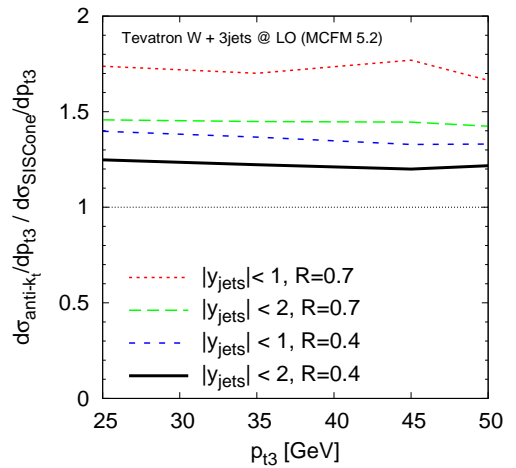}
  \caption{The ratio of the anti-$k_t$ and SISCone results for the
    $W+3$\,jet cross section, shown as a function of the transverse
    momentum of the third hardest jet, for two different $R$ values
    and rapidity acceptances for the jets, as calculated with
    MCFM~\cite{MCFM}. This ratio provides a measure of the ambiguity
    in perturbative predictions for an IR unsafe IC-SM jet algorithm
    such as JetClu. 
    \label{fig:W3j-akt-sis-review}
  }
\end{floatingfigure}
A comparison of SISCone and anti-$k_t$ was performed for example in
ref.~\cite{Ellis:2009bu}. It examined the $W+3$\,jets cross section at
the Tevatron (measured with JetClu, $R=0.4$ for jets with $|y|<2$
\cite{Aaltonen:2007ip}) and found that the SISCone prediction was
about $20\%$ smaller than the anti-$k_t$ prediction at LO (the
difference is reduced at NLO), because in the SISCone case there is a
higher likelihood that two of the three LO partons will be combined
into a single jet, giving $W+2$\,jets rather than $W+3$\,jets.
This may not seem like an enormous effect compared to typical
experimental systematic uncertainties, however one should remember
that the size of the difference depends also on the cuts and the
choice of $R$.
For example, with a larger $R$ value (e.g. $R=0.7$) or a smaller
rapidity range, the differences between the algorithms increase
noticeably, as illustrated in figure~\ref{fig:W3j-akt-sis-review}.

In the long-run, an alternative approach might be to
use tools like MC@NLO \cite{Frixione:2002ik} and
POWHEG \cite{Nason:2006hfa}, which may eventually include a range of
jet processes and thus provide both the NLO terms and an acceptable
estimate of the large higher-order logarithms and the non-perturbative
effects (with IRC jet safe algorithms another advantage of tools like
MC@NLO and POWHEG is that they provide a way of consistently including
both NLO corrections and non-perturbative hadronisation effects within a
single calculation).


\subsubsection{Exact seedless cones}

One full solution to the IRC safety issue avoids the use of seeds and
iterations, and instead
finds \emph{all} stable cones through some exact procedure. This type
of algorithm is often called a seedless cone (SC, thus SC-SM with a
split--merge procedure). 

In a seedless cone algorithm, the addition of a soft particle may lead
to the presence of new stable cones, however none of those new cones
will involve hard particles (a soft particle doesn't affect the
stability of a cone involving much larger momenta), and therefore the
set of hard stable cones is infrared safe.
As long as the presence of new soft stable cones (or of new soft
particles inside hard stable cones) doesn't change the outcome of the
split--merge procedure (a non-trivial requirement), then a seedless
cone will lead to an infrared safe collection of hard jets.

A computational strategy for identifying all cones was outlined in
ref.~\cite{RunII-jet-physics}: one takes all subsets of particles and
establishes for each one whether it corresponds to a stable cone ---
\ie one calculates its total momentum, draws a circle around the
resulting axis, and if the points contained in the circle are exactly
as those in the initial subset, then one has found a stable cone.
This is guaranteed to find all stable cones.

\begin{figure}[t]
  \centering
  \includegraphics[width=0.5\textwidth,angle=270]{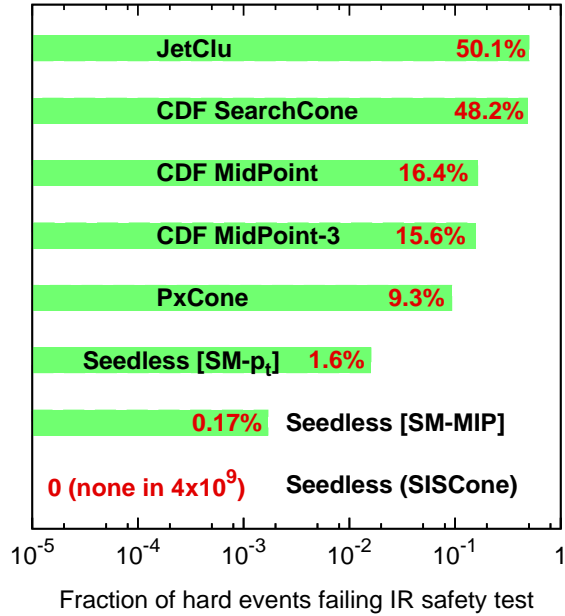}
  \caption{Failure rates for IR safety tests \cite{Salam:2007xv} with
    various algorithms, including a midpoint variant with $3$-way
    midpoints and some seedless algorithms with commonly used, but
    improper, split--merge procedures. 
    See
    table~\ref{jetalgs_conefeatures},
    p.~\pageref{jetalgs_conefeatures}, for the classification of the
    main different algorithms and \cite{Salam:2007xv} for a
    description of the different seedless variants. 
    CDF MidPoint-3 is like the standard MidPoint algorithm except that
    it also uses midpoints between triplets of stable cones.  
    Briefly, the IR safety test proceeds as follows: first one
    generates an event with between 
    2 and 10 hard particles, and applies a jet finder to the event; then
    one generates some number of random very soft particles ($p_t \sim
    10^{-100}\GeV$), and applies the jet finder to the event consisting of
    soft and hard particles. If the hard jets (those with $p_t \gg
    10^{-100}\GeV$) are the same in the two cases, then the jet finder
    passes the IR safety test for that event. One repeats the exercise for
    many events. SISCone passed the test for all $4\times10^{9}$
    events used. Other algorithms failed the test for some fraction of
    events, as given in the figure.
    \label{fig:IR_failures}
  }
\end{figure}

The above seedless procedure was intended for fixed-order
calculations, with a very limited number of particles. It becomes
impractical for larger numbers of particles because there are
$\order{2^N}$ possible subsets (think of an $N$-bit binary number where each
bit corresponds to a particle, and the subset consists of all
particles whose bit is turned on). Testing the stable-cone property
takes $\order{N}$ time for each subset and so the total time is
$\order{N2^N}$.
This exponential-time behaviour made seedless cones impractical for
use on events with realistic numbers of particles (the $N2^N$ approach
would take about $10^{17}$ years to cluster 100 particles). 
However in 2007 a polynomial-time geometrically-based solution was
found to the problem of identifying all stable
cones~\cite{Salam:2007xv}. The corresponding algorithm is known as
SISCone and it is described in section~\ref{sec:making-cone-ir}.
An explicit test of the IR safety of SISCone is shown in
fig.~\ref{fig:IR_failures}.

Seedless cone algorithms are also programmed into NLO codes like
NLOJET++ \cite{NLOJet} and MCFM \cite{MCFM}. Users should however be
aware that there is some degree of confusion in nomenclature --- for
example the cone algorithm in MCFM v.~5.2 is referred to as the midpoint
algorithm, but is actually a seedless implementation; in NLOJET++ v.~3
the algorithm is referred to as seedless, but has a midpoint option.
Users of NLO codes are therefore advised to make sure they know
\emph{exactly} what is implemented in the NLO code's native jet finder
(\ie they should carefully inspect the portion of code devoted to the
jet finder). Alternatively they may use appropriately documented 3rd
party libraries for their jet finding.

\subsubsection{Dark towers}
\label{sec:cone-dark-towers}

\begin{figure}
  \centering
  \includegraphics[width=0.24\textwidth]{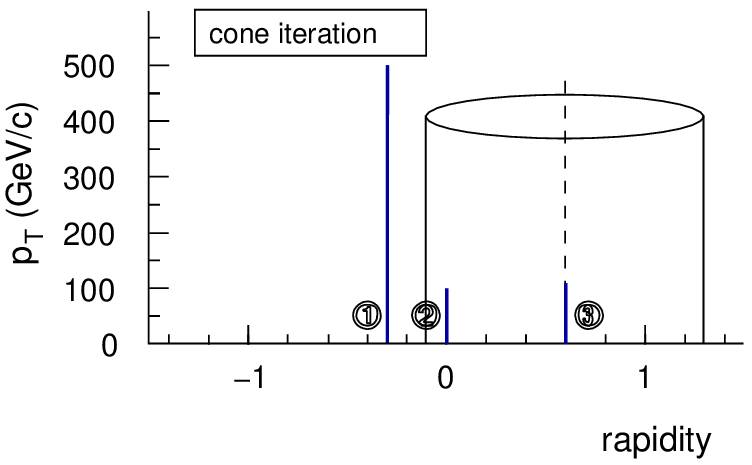}
  \includegraphics[width=0.24\textwidth]{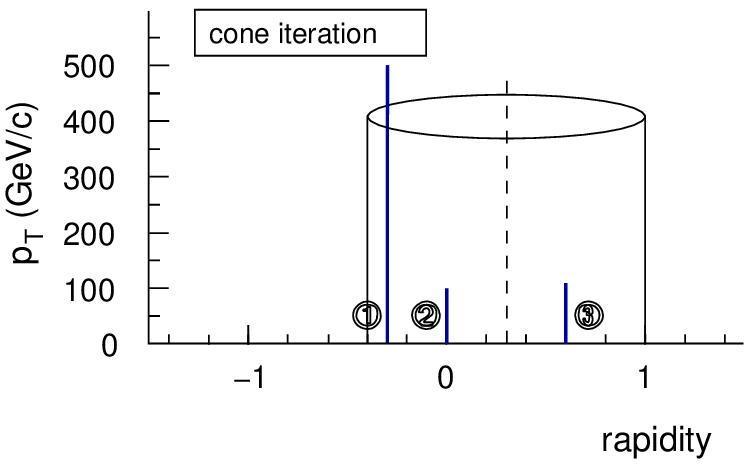}
  \includegraphics[width=0.24\textwidth]{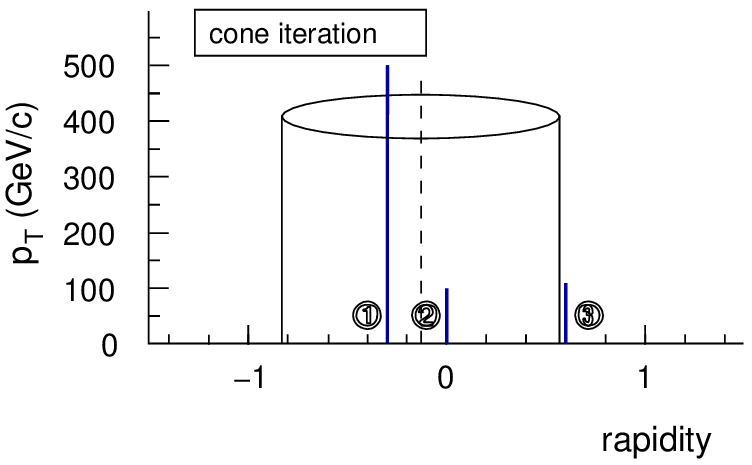}
  \includegraphics[width=0.24\textwidth]{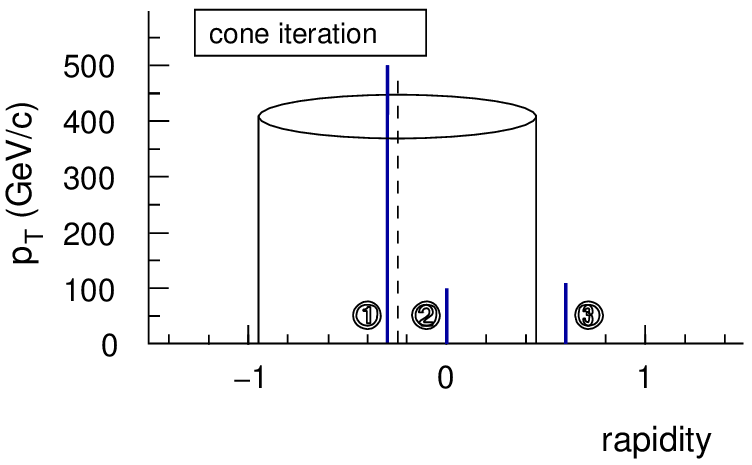}
  \caption{Some of the stages of a stable-cone iteration that leads to
    dark towers. 
    In the first panel, one starts the stable-cone iteration with the
    rightmost particle (3) as a seed. The cone contains particles 2
    and 3 and its momentum points roughly midway between them.  The
    direction of the momentum is then used as the centre of a cone for
    the next iteration  (second panel). 
    The cone in the 2nd panel contains all three particles. The
    resulting momentum direction provides the cone centre in the 3rd
    panel and now particle $3$ is no longer contained in the cone.
    One further iteration leads to the stable cone in panel 4, which
    does not contain particle $3$ even though it provided the initial
    seed direction.
    For this event, therefore, particle $3$ will never appear in any
    stable cone. 
    \label{fig:cone-iter-dark}
  }
\end{figure}

The xC-SM class of algorithms collectively suffers from a problem known
as dark towers~\cite{EHT}:
regions of hard energy flow that are not clustered into any jet.
Dark towers arise because there exist configurations in which some particles
will never end up in a stable cone. 
The stages of an iteration in which this is the case are shown in
fig.~\ref{fig:cone-iter-dark}, for which the rightmost particle cannot
be in a stable cone: even when one uses it as a starting point for
iteration, it is not contained in the final stable cone (nor is it
contained in any stable cone in a seedless algorithm).

One solution to this problem in iterative algorithms is ``ratcheting'':
a particle that was included at any stage of the iteration is manually
included in all subsequent stages of the iteration even if it is not
within the cone boundary.
This is used in CDF's JetClu algorithm (though it is not actually
described in the reference usually quoted by CDF for
JetClu~\cite{Abe:1991ui}).

Another fix to dark towers was proposed in~\cite{EHT}, and
referred to as the ``searchcone''. It eliminates a large fraction of
the dark towers by using a smaller radius to find stable cones and
then expands the cones to their full radius, without further iteration,
before passing them to the SM procedure.
Unfortunately, when applied together with the midpoint procedure
(IC$_{se,mp}$-SM) it worsens its IR unsafety status from IR$_{3+1}$
back to IR$_{2+1}$~\cite{Albrow:2006rt}.

Perhaps the simplest solution~\cite{Albrow:2006rt} to dark towers is
to identify any remainder energy flow that was not clustered by the
xC-SM algorithm and run an extra pass of the algorithm on that
remainder.
This is the approach used in SISCone (which by default runs multiple
passes until no energy is left).

\subsection{Sequential recombination jet algorithms}
\label{sec:jetalgs_seqrec}

Sequential recombination algorithms have their roots in $\ee$
experiments.
A detailed overview of their history in $\ee$ studies is given in the
introduction of~\cite{Moretti:1998qx}.
The intention here is not to repeat that history, but rather to walk
through the most widely used of the $\ee$ algorithms and then see how
they lead to corresponding hadron-collider algorithms.
It should be said that many of the ideas underlying today's sequential
recombination algorithms (including a momentum-related parameter to decide
jet resolution and the use of relative transverse momenta) actually
appeared first in the LUCLUS algorithm of
Sj\"ostrand~\cite{Sjostrand:1982am} (earlier work includes
\cite{Dorfan:1980gc,Daum:1980rp,Lanius:1980mz,Backer:1981cx}).
However computational constraints at the time led to the algorithm
including a preclustering phase, and it also involved a non-trivial
procedure of reassignment of particles between clusters at each
recombination.
These two characteristics made it somewhat more complicated than its
successors.

Today's sequential recombination algorithms are all rather simple to
state (far more so than the cone algorithms). Additionally they go
beyond just finding jets and implicitly assign a ``clustering sequence
to an event'', which is often closely connected with approximate
probabilistic pictures that one may have for parton branching.

\subsubsection{Jade algorithm}
\label{sec:jade}

The first simple sequential recombination algorithm was introduced by
the JADE collaboration in the middle of the
1980's~\cite{Bartel:1986ua,Bethke:1988zc}.
It is formulated as follows:
\begin{itemize}
\item[1.] For each pair of particles $i$, $j$ work out the 
  distance 
  \begin{equation}
    \label{eq:jade-mass}
    y_{ij} = \frac{2 E_i E_j (1-\cos\theta_{ij})}{Q^2}
  \end{equation}
  where $Q$ is the total energy in the event,\footnote{In experimental
  uses, it is often the total \emph{visible} energy in the event.} $E_i$ is
the energy of 
  particle $i$ and $\theta_{ij}$ the angle between particles $i$ and $j$.
  For massless particles, $y_{ij}$ is the just the (normalised)
  squared invariant mass of the pair.

\item[2.] Find the minimum $y_{\min}$ of all the $y_{ij}$. 
  
\item[3.]  If $y_{\min}$ is below some \emph{jet resolution threshold}
  $y_\cut$, then recombine $i$ and $j$ into a single new particle (or
  ``pseudojet'') and repeat from step 1.

\item[4.] Otherwise, declare all remaining particles to be jets and
  terminate the iteration.
\end{itemize}
The number of jets that one obtains depends on the value of $y_\cut$,
and as one reduces $y_\cut$, softer and/or more collinear emissions
get resolved into jets in their own right.
Thus here the number of jets is controlled by a single parameter
rather than the two parameters (energy and angle) of cone algorithms.

Quite often in $\ee$ analyses one examines the value of $y_\cut$ that
marks the transition between (say) an event being labelled as having
$n$ and $n+1$ jets, $y_{n (n+1)}$. Thus if $y_{23}$ is small, the event
is two-jet like, while if it large then the event clearly has 3 (or
more) jets.

The JADE algorithm is infrared and collinear safe, because any soft
particle will get recombined right at the start of the clustering, as
do collinear particles.
It was widely used up to the beginning of the 1990s (and still
somewhat beyond then),
however the presence of $E_{i} E_{j}$ in the distance measure means
that two very soft particles moving in opposite directions often get
recombined into a single particle in the early stages of the
clustering, which runs counter to the intuitive idea that one has of a
jet being restricted in its angular reach.
As well as being physically disturbing, this leads to very non-trivial
structure (non-exponentiated double logarithms) in higher-order
calculations of the distribution of
$y_{23}$~\cite{JadeDL,CataniJade,Leder} (later, this was also
discussed in terms of a violation of something called recursive
infrared and collinear safety \cite{Banfi:2004yd}).

\subsubsection{The $k_t$ algorithm in \ee}
\label{sec:eekt}

The $\ee$ $k_t$ algorithm~\cite{Kt} is identical to the JADE algorithm
except as concerns the distance measure, which is
\begin{equation}
  \label{eq:eekt-dist}
  y_{ij} = \frac{2 \min(E_i^2, E_j^2) (1-\cos\theta_{ij})}{Q^2}\,.
\end{equation}
In the collinear limit, $\theta_{ij} \ll 1$, the numerator just
reduces to $(\min(E_i,E_j) \theta_{ij})^2$ which is nothing but the
squared transverse momentum of $i$ relative to $j$ (if $i$ is the
softer particle) --- this is the origin of the name $k_t$-algorithm.%
\footnote{As mentioned above, the distance measured used in the
  earlier LUCLUS algorithm~\cite{Sjostrand:1982am}, $y_{ij} = 2
  \frac{|\vec p_i|^2|\vec p_j|^2}{(|\vec p_i|+|\vec p_j|)^2 Q^2}
  (1-\cos\theta_{ij})$ (in the version given
  in~\cite{Moretti:1998qx}), was also a relative transverse-momentum
  type variable.}
The use of the minimal energy  ensures that the distance between
two soft, back-to-back particles is larger than that between a soft
particle and a hard one that's nearby in angle.

Another way of thinking about eq.~(\ref{eq:eekt-dist}) is that the
distance measure is 
essentially proportional to the squared inverse of the splitting probability for one
parton $k$ to go into two, $i$ and $j$, in the limit where either
$i$ or $j$ is soft and they are collinear to each other,
\begin{equation}
  \label{eq:div-measure}
  \frac{dP_{k\to ij}}{dE_i d\theta_{ij}} \sim \frac{\as}{\min(E_i,E_j)
    \theta_{ij}}
\end{equation}
There is a certain arbitrariness in this statement, because of the
freedom to change variables in the measure on the left-hand side of
eq.~(\ref{eq:div-measure}). However the presence of a power of just
the minimum of the energy in the denominator (rather than some
function of both energies as in the JADE distance measure) is
robust.

The $k_t$ algorithm's closer relation to the structure of QCD
divergences made it possible to carry out all-order resummed
calculations of the distribution of $y_{n (n+1)}$
\cite{Kt,JetsDisSchmell,Banfi:2001bz} and of the mean number of jets
as a function of $y_\cut$ \cite{Catani:1992tm}.
This helped encourage its widespread use at LEP.
The relation to QCD divergences also means that the clustering
sequence retains useful approximate information about the sequence of
QCD splittings that occurred during the showering that led to the jet.
This is of interest both in certain theoretical studies (for example
CKKW matching of parton-showers and matrix
elements~\cite{Catani:2001cc}) and also for identifying the origin of a
given jet (for example quark versus gluon
discrimination~\cite{Chekanov:2004kz}).

\subsubsection{The $k_t$ algorithm with incoming hadrons}
\label{sec:ppkt}

In experiments with incoming hadrons two issues arise. Firstly (as
mentioned already for cone algorithms) the total energy is no longer
well defined. So instead of the dimensionless distance $y_{ij}$, one
might choose to use a dimensionful distance
\begin{equation}
  \label{eq:DISkt-dist}
  d_{ij} = 2 \min(E_i^2, E_j^2) (1-\cos\theta_{ij})\,,
\end{equation}
together with a dimensionful jet-resolution parameter $d_\cut$
(alternatively, one might maintain a dimensionless measure by choosing
some convention for the normalisation scale).
Secondly, the divergences in the QCD branching probability are not
just between pairs of outgoing particles, but also between an outgoing
particle and the incoming beam direction.

The first attempt at formulating a $k_t$ algorithm in such cases
was~\cite{Catani:1992zp}. It introduced the idea of an additional
\emph{particle-beam} distance.
\begin{equation}
  \label{eq:DISkt-beam-dist}
  d_{iB} = 2 E_i^2 (1-\cos\theta_{iB})\,,
\end{equation}
which, for small $\theta_{iB}$, is just the squared transverse momentum
of particle $i$ with respect to the beam.
The algorithm then remains the same as in $\ee$, except that if a
$d_{iB}$ is the smallest, then the particle is recombined with the
beam, to form part of the ``beam-jet''. If there are two beams, then
one just introduces a measure for each beam.

In $pp$ collisions it is standard to use variables that are invariant
under longitudinal boosts, however the $d_{ij}$ and $d_{iB}$ given
above only satisfy this property approximately. Thus ref.~\cite{KtHH}
introduced  versions of the distance measures
that were exactly longitudinally invariant
\begin{subequations}
  \label{eq:ppkt-dist}
  \begin{align}
    d_{ij} &= \min(p_{ti}^2, p_{tj}^2) \Delta R_{ij}^2\,,\qquad\quad
    \Delta R_{ij}^2 = (y_i - y_j)^2 + (\phi_i - \phi_j)^2
    \,,\\
    d_{iB} &= p_{ti}^2\,,
  \end{align}
\end{subequations}
(this variant does not distinguish between the two beam
jets).\footnote{Ref.~\cite{KtHH} also proposes a variant where $\Delta
  R_{ij}^2 \equiv 2 (\cosh (y_i-y_j) - \cos(\phi_i - \phi_j))$, more
  closely related to the precise structure of the QCD matrix elements;
  however, to the author's knowledge, it has not seen extensive use.}
It is straightforward to verify that in the relevant collinear limits,
these measures just reduce to relative transverse momenta, like those
in eqs.~(\ref{eq:DISkt-dist},\ref{eq:DISkt-beam-dist}). Furthermore,
since $(y_i - y_j)$, the $\phi_i$ and $p_{ti}$ are all invariant under
longitudinal boosts, the $d_{ij}$ and $d_{iB}$ are too.
Nowadays the procedure of section~\ref{sec:jade}, with the distance
measures of eqs.~(\ref{eq:ppkt-dist}), is referred to as the
\emph{exclusive} $k_t$ algorithm, in that every particle is assigned
either to a beam-jet or to a final-state jet.

\paragraph{Inclusive $\boldsymbol{k_t}$ algorithm.} 
At about the same
time that ref.~\cite{KtHH} appeared, a separate formulation was
proposed in~\cite{Kt-EllisSoper}, which has almost the same distance
measures as eq.~(\ref{eq:ppkt-dist}),
\begin{subequations}
  \label{eq:ppkt-dist-R}
  \begin{align}
    d_{ij} &= \min(p_{ti}^2, p_{tj}^2) \frac{\Delta R_{ij}^2}{R^2}\,,\qquad\quad
    \Delta R_{ij}^2 = (y_i - y_j)^2 + (\phi_i - \phi_j)^2
    \,,\\
    d_{iB} &= p_{ti}^2\,,
  \end{align}
\end{subequations}
where the difference lies in the presence of a new parameter $R$ (also
called $D$) in the $d_{ij}$, whose role is similar to $R$ in a cone
algorithm (see below). 
The other difference in this version of the algorithm is in
how the $d_{ij}$ get used:
\begin{itemize}
\item[1.] Work out all the $d_{ij}$ and $d_{iB}$ according to
  eq.~(\ref{eq:ppkt-dist}).

\item[2.] Find the minimum of the $d_{ij}$ and $d_{iB}$.
  
\item[3.]  If it is a $d_{ij}$, recombine $i$ and $j$ into a single
  new particle and return to step 1.

\item[4.] Otherwise, if it is a $d_{iB}$, declare $i$ to be a
  [final-state] jet, and remove it from the list of particles. Return
  to step 1.

\item[5.] Stop when no particles remain.
\end{itemize}
Here, all particles are included in final-state jets, there is no
concept of a beam jet, and there is no $d_\cut$ parameter --- the
question of what gets called a jet is determined by $R$: if a particle
$i$ has no other particles within a distance $R$ then the $d_{iB}$
will be smaller than the $d_{ij}$ for any $j$ and the particle will
then become a jet.
One consequence of this is that arbitrarily soft particles can becomes
jets in their own right and therefore (just as for cone algorithms),
one should additionally specify a minimum transverse momentum that a
jet should have for it to be of interest.

The above algorithm is most unambiguously referred to as the
\emph{inclusive} $k_t$ algorithm, though when people mention the
``$k_t$ algorithm'' in a collider context, it is nearly always the
inclusive variant that they have in mind.
It so happens that the exclusive and inclusive variants have identical
clustering sequences --- it is only the interpretation of those
clustering sequences that differs.

The $k_t$ algorithm has long been advocated by theorists because it
is free of any infrared and collinear safety issues. 
On the other hand it had been criticised by experimenters on the
grounds (a) that it was computationally slow, insofar as the two
public implementations that were available in 2005, \texttt{KtClus}
(Fortran) \cite{KtClus} and \texttt{KtJet} (C++)
\cite{KtImplementation}, both took times $\sim N^3$ to cluster $N$
particles; and (b) that it produces geometrically irregular jets,
which complicates certain detector and non-perturbative
corrections.\footnote{%
  For example, if jets are circular with radius $R$ in the $y-\phi$
  plane, then any jet whose momentum points at least a distance $R$
  away from the edge of the central part of the detector will always
  be fully contained in that central part.
  If jets can be irregular, with boundaries that sometimes extend
  beyond a distance $R$ from the jet momentum, then there is no such
  simple way of identifying the region of the detector in which all
  jets will be fully contained.  }
We will return to the speed issue in
section~\ref{sec:speeding-up-k_t}, while the irregularity is visible
as the jagged boundaries of the jets in fig.~\ref{fig:4algs},
p.~\pageref{fig:4algs} (related issues will be discussed in
section~\ref{sec:jet-areas}).

Given  the number of experimental objections that have been raised in the
past regarding the $k_t$ algorithm in a $pp$ environment, it is worth
commenting briefly on the two sets of hadron-collider measurements
that have been carried out with the $k_t$ algorithm.
One, from \Dzero \cite{D0kt,D0KtSubJets}, had to go to considerable
lengths (introducing preclustering) to get around the speed issue
(\Dzero's fine calorimeter meant that it had many input towers) and
found rather large non-perturbative corrections from the underlying
event (UE);
the latter issue perhaps discouraged further use of the $k_t$
algorithm until CDF performed a similar measurement in
2005~\cite{CDFkt,Abulencia:2007ez}.
CDF did not suffer particularly from the speed issue, largely because
their coarser calorimeter segmentation ensured modest input
multiplicities. Also, crucially, they showed that \Dzero's large UE
corrections were probably a consequence of taking the jet radius
parameter $R=1$. When CDF instead took $R=0.7$ (as is common for cone
algorithms), they found UE corrections that were commensurate with
those for cone algorithms.

It should also be added that the longitudinally invariant $k_t$
algorithm was the main jet algorithm used at HERA, both in
photoproduction (e.g.\ refs.~\cite{Adloff:2003nr,Chekanov:2007et}),
the first (published) experimental context in which it was
used~\cite{Breitweg:1997rx}, and deep inelastic scattering (e.g.\
refs.~\cite{Aktas:2007pb,Chekanov:2006yc}).
Compared to Tevatron this was probably facilitated by the lower
particle multiplicites in DIS and photoproduction and also by the
quieter underlying event.

\subsubsection{The Cambridge and Aachen algorithms}
\label{sec:pp-cam-aachen}

The Cambridge algorithm \cite{Cam} is a sequential recombination
algorithm for $\ee$ collisions that introduces \emph{two} distance
measures between pairs of particles. It has $v_{ij} = 2(1-\cos
\theta_{ij})$ (\ie the squared angle) as well as the $y_{ij}$ of
eq.~(\ref{eq:jade-mass}). It reads as follows
\begin{enumerate}
\item If only one particle is left, call it a jet and stop.
\item Otherwise find the pair of particles with smallest $v_{ij}$.
\item If the corresponding $y_{ij} < y_{cut}$, replace $i$ and $j$
  with the recombined one and go to step 1.
\item Otherwise: take the less energetic of $i$ and $j$, remove it
  from the list of particles, call it a jet, and go to step 1.
\end{enumerate}
The idea here was to combine the $y_{cut}$ jet resolution of the $k_t$
algorithm with a clustering sequence dictated by angular ordering, \ie
one that relates closely to the powerful concept of angular ordering
that arises when considering multiple gluon emission~\cite{Coherence}.

\paragraph{Cambridge/Aachen.} The most widely discussed extension (and
simplification) of the Cambridge algorithm to hadron colliders was
actually originally given in the context of DIS studies~\cite{Aachen} (another
one \cite{Pierce:1998sm} has seen less study).
It is like the inclusive $k_t$ algorithm in that it uses longitudinally
invariant variables, introduces an $R$ parameter, and does away with the
$y_{ij}$  cut on jets.
It procedes by recombining the pair of particles with the smallest
$\Delta R_{ij}$, and repeating the procedure until all objects are
separated by a $\Delta R_{ij} > R$. The final objects are then the
jets.\footnote{Alternatively, one can formulate it like the inclusive
  $k_t$ algorithm, but with $d_{ij} = \Delta R_{ij}^2/R^2$ and $d_{iB}
  = 1$.}

This algorithm was originally named the Aachen algorithm, though it is
often now called the Cambridge/Aachen (C/A) algorithm, reflecting its
angular-ordered Cambridge roots.

Like the $k_t$ algorithm, the C/A algorithm gives somewhat irregular jets, and
its original implementations took a time that scales as $N^3$.
The latter problem is now solved (as for the $k_t$ algorithm) and the
fact that the C/A has a clustering hierarchy in angle makes it
possible to consistently view a specific jet on many different angular
scales, a feature whose usefulness will become apparent in
section~\ref{sec:substructure} and is also relevant for a
``filtering'' method discussed below.

\subsubsection{The anti-$k_t$ algorithm}
\label{sec:anti-k_t-algorithm}

One can generalise the $k_t$ and Cambridge/Aachen distance measures
as~\cite{Cacciari:2008gp}:
\begin{subequations}
  \label{eq:genkt-dist-R}
  \begin{align}
    d_{ij} &= \min(p_{ti}^{2p}, p_{tj}^{2p}) \frac{\Delta R_{ij}^2}{R^2}\,,\qquad\quad
    \Delta R_{ij}^2 = (y_i - y_j)^2 + (\phi_i - \phi_j)^2
    \,,\\
    d_{iB} &= p_{ti}^{2p}\,,
  \end{align}
\end{subequations}
where $p$ is a parameter that is $1$ for the $k_t$ algorithm, and $0$
for C/A.
It was observed in \cite{Cacciari:2008gp} that if one takes $p=-1$,
dubbed the ``anti-$k_t$'' algorithm, then this favours clusterings that
involve hard particles rather than clusterings that involve soft
particles ($k_t$ algorithm) or energy-independent clusterings (C/A).
This ultimately means that the jets grow outwards around hard
``seeds''.
However since the algorithm still involves a combination of energy and
angle in its distance measure, this is a collinear-safe growth (a
collinear branching automatically gets clustered right at the
beginning of the sequence).\footnote{If one takes $p\to-\infty$ then
  energy is privileged at the expense of angle and the algorithm then
  becomes collinear unsafe, and somewhat like an IC-PR algorithm.}
The result is an IRC safe algorithm that gives circular hard jets,
making it an attractive replacement for certain cone-type algorithms
(notably IC-PR algorithms).

One should be aware that, unlike for the $k_t$ and C/A algorithms, the
substructure classification that derives from the clustering-sequence
inside an anti-$k_t$ jet cannot be usefully related to QCD branching
(essentially the anti-$k_t$ recombination sequence will gradually
expand through a soft subjet, rather than first constructing the soft
subjet and then recombining it with the hard subjet).

\subsubsection{Other sequential recombination ideas}
\label{sec:other-seq-rec}

The flexibility inherent in the sequential recombination procedure
means that a number of variants have been considered in both past and
recent work.
Some of the main ones are listed below.

\paragraph{Flavour-$k_t$ algorithms.}
If one is interested in maintaining a meaningful flavour for jets (for
example in purely partonic studies, or when discussing heavy-flavour
jets), then one may use a distance measure that takes into account the
different divergences for quark and gluon branching, as in
\cite{Banfi:2006hf,Banfi:2007gu}. The essential idea is to replace
eq.~(\ref{eq:eekt-dist}) with
\begin{equation}
  \label{eq:yij-flavour}
  y_{ij}^{(F)} = \frac{2(1-\cos\theta_{ij})}{Q^2} \times\left\{
    \begin{array}[c]{ll}
      \max(E_i^2, E_j^2)\,, & \quad\mbox{softer of $i,j$ is flavoured,}\\
      \min(E_i^2, E_j^2)\,, & \quad\mbox{softer of $i,j$ is flavourless,}
    \end{array}
  \right.
\end{equation}
where gluonic (or non-heavy-quark) objects are considered flavourless.
This reflects the fact that there is no divergence for producing a
lone soft quark, and correctly ensures that soft quarks are recombined
with soft antiquarks. In normal algorithms, in contrast, a soft quark
and anti-quark may end up in different jets, polluting the flavour of
each one.
Full details, and the hadron collider variants, are given in
\cite{Banfi:2006hf}, while an application to $b$-jets was given
\cite{Banfi:2007gu}, where it led to a much more accurate NLO
prediction for the inclusive $b$-jet spectrum.
Related ideas have also been used in a sequential-recombination
jet algorithm designed for combining QCD matrix elements and parton
showers~\cite{Hoeche:2009rj}.

\paragraph{Variable-$R$ algorithms.}
A recent proposal in \cite{VariableR} suggests a class of
hadron-collider distance measures of the following form
\begin{equation}
  \label{eq:genkt-dist-varR}
  d_{ij} = \min(p_{ti}^{2p}, p_{tj}^{2p}) \Delta R_{ij}^2\,,\qquad\qquad
  d_{iB} = p_{ti}^{2p} R_\mathrm{eff}(p_{ti}^2)\,,
\end{equation}
where the radius of the jet (now placed in the $d_{iB}$ term rather
than $d_{ij}$) becomes a function of the jet's transverse momentum
$R_\mathrm{eff}(p_{ti}^2)$.
This provides an original way of having a jet radius that depends on
the event structure, a feature which in general can be useful (\cf
section~\ref{sec:choosing-radius}).
In \cite{VariableR} it was applied specifically to the question of
dijet resonance reconstruction, with the aim of producing larger jets,
$R_\mathrm{eff} \sim 1/p_t$, (appropriate with $p\le 0$) for resonances
that decay along the beam direction, and it led to improved resolution
on the reconstructed mass peak.

\paragraph{Filtering, pruning and trimming.}
As we shall see in section~\ref{sec:using-jets}, contamination from
non-perturbative effects associated with beam-remnants (underlying
event) in hadron colliders is a major cause of degradation of
resolution on jets' energies.
One way of reducing this~\cite{Butterworth:2008iy} is to first find
the jets (with some given $R$) and then reconsider each jet on a
smaller angular scale, $R_{\filt} < R$ (either by reclustering, or by
making use of the hierarchical angular information in the C/A
algorithm).
On that smaller angular scale one then takes (say) the two hardest
subjets, corresponding physically to a hard parton and its hardest gluon
emission, while rejecting the junk that comes from the underlying
event. 
A variant of this, referred to as ``trimming'' in \cite{Krohn:2009th}, is to
retain all subjets above some threshold in transverse momentum.
Initial studies~\cite{Butterworth:2008iy,Cacciari:2008gd,Krohn:2009th} indicate
that these can provide non-negligible advantages in kinematic
reconstructions.

A related idea, ``pruning,'' was suggested in
ref.~\cite{Ellis:2009su,Ellis:2009me}. During the (re)clustering of the jet, if two
objects $i,j$ are separated by $\Delta R_{ij} > R_{\filt}$ and the
softer one has $z=\min(p_{ti},p_{tj}) < p_{t,i+j} < z_{\mathrm{cut}}$
(with $z_{\mathrm{cut}}=0.1$ say), then that softer one is simply
discarded.

One issue with filtering, pruning and the variable-$R$ approach
discussed above, is that they all introduce extra degrees of freedom
in the jet finding. Thus the gains that they may provide come at the
expense of having to tune those choices to the specific physics
analysis that is being carried out.

\paragraph{3 $\to$ 2 recombination.}
Most sequential recombination algorithms are related to the idea of
inverting successive $1\to2$ perturbative branchings (as used in many
parton-shower Monte Carlo programs).
When simulating QCD branching it can also be useful to consider
``dipole'' branchings, \ie $2\to 3$ splittings, as in
Ariadne~\cite{Lonnblad:1992tz}.
Correspondingly one can imagine a sequential-recombination jet
algorithm that inverts these branchings by carrying out $3\to2$
clusterings. This is the principle of the ARCLUS
algorithm~\cite{Lonnblad:1992qd} for $\ee$ collisions.  In practice
its performance is similar to that of other $\ee$ algorithms (as
discussed in~\cite{Moretti:1998qx}).

\subsection{Jet finding as a minimisation problem}

Several groups have considered jet finding as a minimisation
problem. Though not the main subject of this review, for completeness
it is worth devoting a few lines to describe these ideas, which fit
into the top-down approach to jet finding, and have been explored by
several groups over the past decade.

One approach \cite{Chekanov:2005cq} relates to a method known as
$k$-means in the more general computer science field of
clustering~\cite{K-means}.
It introduces a partition of particles $i$ into $n$ clusters $L_k$
($k=1\ldots n$, with $n$ chosen a priori). For a given partition, each
cluster has a centroid $C_k$ and one can evaluate a measure
\begin{equation}
  \label{eq:k-means-chekanov}
  S = \sum_k \sum_{i \in L_k} d(p_i,C_k)
\end{equation}
where $d(p_i,C_k)$ is some measure of the distance between particle
$i$ and the centroid $k$. One then chooses the assignment of
particles into clusters that minimises $S$.
Part of the motivation given for the approach of
\cite{Chekanov:2005cq} is that it allows one to also include a range
of physical constraints (such as the $W$-mass in top reconstruction)
when carrying out the minimisation. However there are open questions
as to how it may fare in analyses where one doesn't actually know what
the number of jets should be (for example because of background
contamination).

Two other approaches, ``deterministic annealing'' (DA)
\cite{DeterministicAnnealing} and the ``optimal jet finder'' (OJF)
\cite{Tkachov} do away with the idea that a particle belongs to any
single jet. Essentially (and in a language closer to \cite{Tkachov}),
they argue that each particle $i$ is associated with jet $k$ with a
weight $w_{ik}$ such that $\sum_k w_{ik} = 1$ (or alternatively also
allowing for particles to be associated with no jet~\cite{Tkachov}).
The momentum $P_k$ of jet $k$ is then given by $P_k = \sum_i w_{ik}
p_i$.
In the OJF approach, one makes an a-priori choice for the number of
jets, and then a 
minimisation is carried out over all the entries of the $w_{ik}$
matrix, so as to find the lowest value of some cost function,
corresponding for example to some combination of the jet masses;
one can then repeat the minimisation for a different number of jets
and introduce some criterion for one's choice of the number $n$ of
jets, based on the value of the cost function for each $n$.
%
%
%
%
%
%
%
%
%
%

In the DA approach, roughly speaking, given some initial weights, one
calculates the jet momenta $P_k$, and then one recalculates the
weights according to
\begin{equation}
  \label{eq:DA-weight-recalc}
  w_{ik} = \frac{P_{t,k} e^{-\beta d(p_i,P_k)}}{\sum_m P_{t,m} e^{-\beta d(p_i,P_m)}}\,,
\end{equation}
where $\beta$ is an inverse temperature and $d(p_i,P_k)$ is some
distance measure (for example $\Delta R_{ik}^2$). One iterates until
the weights converge. 
This is accompanied by the observation that for $\beta=0$, whatever
the starting conditions, the $w_{ik}$ will be independent of $i$,
which implies that whatever the initial conditions and value of $n$,
all jets will have identical directions (\ie there is only one jet);
as one increases $\beta$, the system will then tend to develop a larger
number of distinct jets. Thus $\beta$ plays the role of $1/d_\cut$ in
sequential recombination algorithms.

Finally, just as this review was about to be made public, it was
brought to the author's attention that a code FFTJet had just been
released~\cite{FFTJet} which is a further approach involving
minimisation (using fast fourier transforms) as well as weighted
assignment of individual particles to multiple jets (the method is
discussed in detail in ref.~\cite{Volobouev:2009rv}).

The ideas behind the OJF, DA and FFTJet algorithms are certainly interesting,
especially the concept that a particle may be associated with more
than one jet, though it is perhaps not obvious that the extra
conceptual complexity that stems from this is offset by any particular
benefits.
In the corresponding initial studies of OJF and DA
\cite{Tkachov,DeterministicAnnealing} physics performances were found
to be comparable to that of the $k_t$ algorithm, though a practical
advantage at the time that OJF and DA were proposed (no longer
relevant nowadays) was a better scaling of the computational speed
with particle multiplicity, $\sim N$ for OJF, $\sim N^2$ for
DA. 
The study performed with FFTJet in ref.~\cite{Volobouev:2009rv}
suggests that it might be more resilient than other algorithms with
respect to the effects of magnetic fields, however the study was
lacking a number of important physical effects such as the underlying
event, which might well affect the conclusions.
A further point is that FFTJet's timing scales not as the number of
particles in the event but as $k\ln k$ where $k$ is the number of
cells used in the fast-fourier transform procedure used for the
minimisation.
This is an advantage in very busy events, but can be a drawback for
parton-level calculations or other situations with low multiplicity,
because one still needs to use a large number of cells in order to
accurately carry out the minimisation.

The relation between minimisation and jet finding has also been
investigated in \cite{EHT}, where stable-cone finding (and cone
iteration) has been interpreted in terms of the search for the local
minima of the potential
\begin{equation}
  \label{eq:stable-cone-potential}
  F(\hat n) = \frac12 \sum_i E_{ti}\, \Delta R_{i\hat n}^2 \, \Theta(R^2 -
  \Delta R_{i\hat n}^2)\,,
\end{equation}
which is a function of the particle momenta and of a cone direction
$\hat n$ (a coordinate in $\eta$, $\phi$). Each stable cone corresponds to
a local minimum of the potential as a function of $n$.
Investigations have also been carried
out~\cite{Berger:2002jt,Lai:2008zp} into whether one can directly use
``potential'' approaches as a replacement for jet finding altogether.

For completeness, it should be stated that the above approaches are
infrared and collinear safe, as an almost direct consequence of the
way in which they are constructed.

\subsection{Recombination schemes}
\label{sec:recomb-schem}

The most widespread recombination scheme nowadays is the $E$-scheme,
or 4-vector recombination scheme. To merge two particles, it just adds
their 4-vectors (and it produces massive jets).
This is the current recommendation according to
\cite{RunII-jet-physics}.

A scheme that was widely used in the past at hadron-colliders was the
$E_t$ weighted recombination scheme, which had been put forward also
in the Snowmass accord. To recombine a set of particles into a jet, it
uses the following procedure:
\begin{subequations}
  \begin{align}
    E_{t,\jet} &= \sum_i E_{ti}\,,\\
    \eta_{\jet} &= \frac{1}{E_{t,\jet}} \sum_i E_{ti}\eta_i\,,\\
    \phi_{\jet} &= \frac{1}{E_{t,\jet}} \sum_i E_{ti}\phi_i\,,
  \end{align}
\end{subequations}
where the sum runs over the particles contained in the jet, and the
jet is taken to be massless. 
This procedure has the drawback that it is not invariant under
longitudinal boosts if the component particles are massive (though one
can formulate boost-invariant alternatives in terms of rapidity $y_i$
and and transverse momentum $p_{ti}$).

When other recombination schemes are used, this is usually stated
explicitly in the corresponding publication.
One should be aware that in some cases the recombination scheme used
during the clustering (\eg in the iteration of stable cones) differs
from the  recombination scheme that is used to obtain the final
jet momenta once the particle assignments to the jets are known.

\subsection{Summary}
\label{sec:algs-summary}

We have seen many different jet algorithms in this section.
A summary of the main ones in common use in hadron-collider studies is
given in table~\ref{jetalgs_conefeatures}.
Many of the algorithms (and all the IRC safe ones) are available from
the FastJet~\cite{FastJet} or SpartyJet~\cite{SpartyJet} packages (the
latter provides access to the IRC safe algorithms via FastJet).

\begin{figure}[t]
  \centering
  \includegraphics[width=0.49\textwidth]{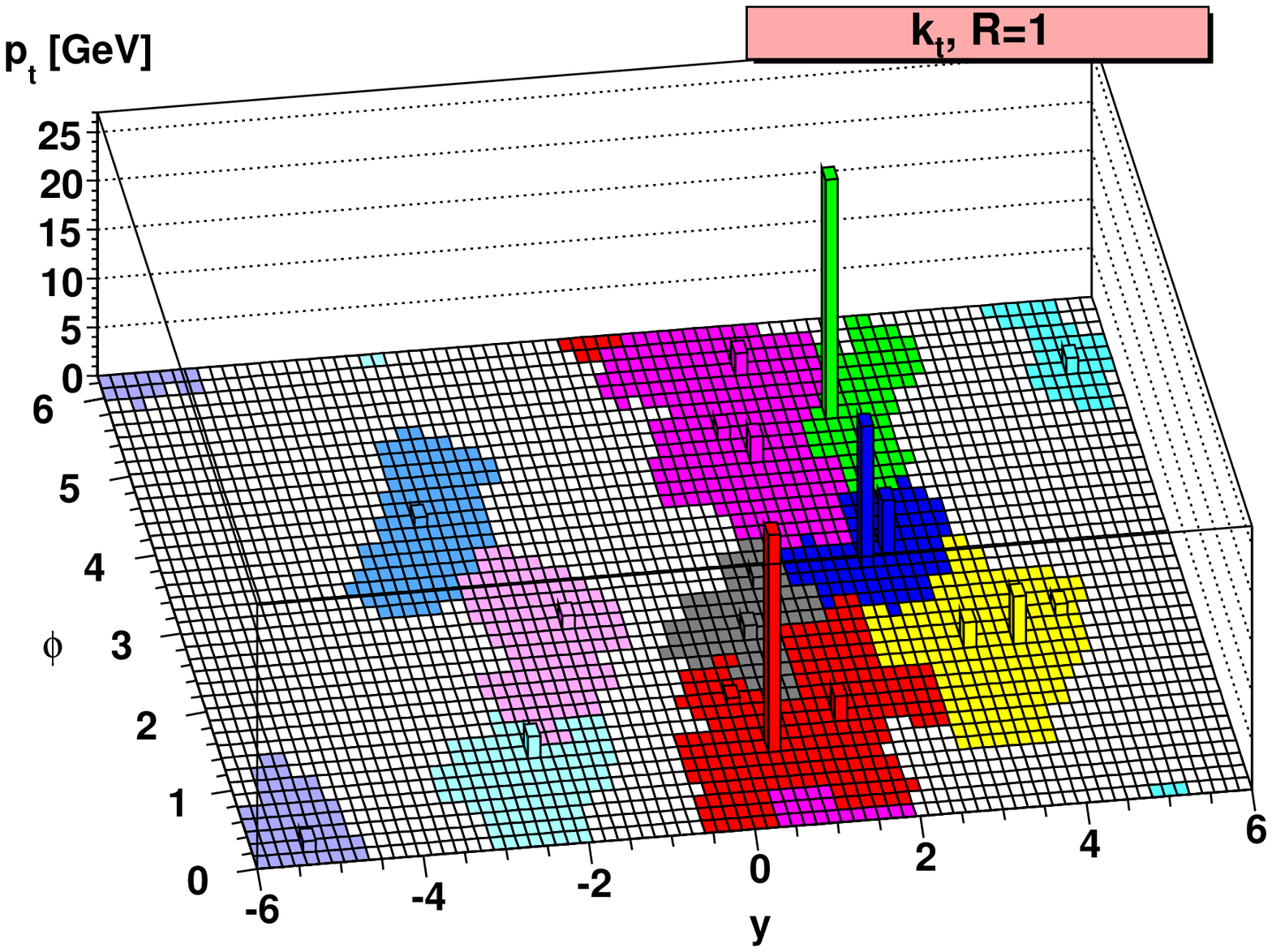}%
  \hfill                                 
  \includegraphics[width=0.49\textwidth]{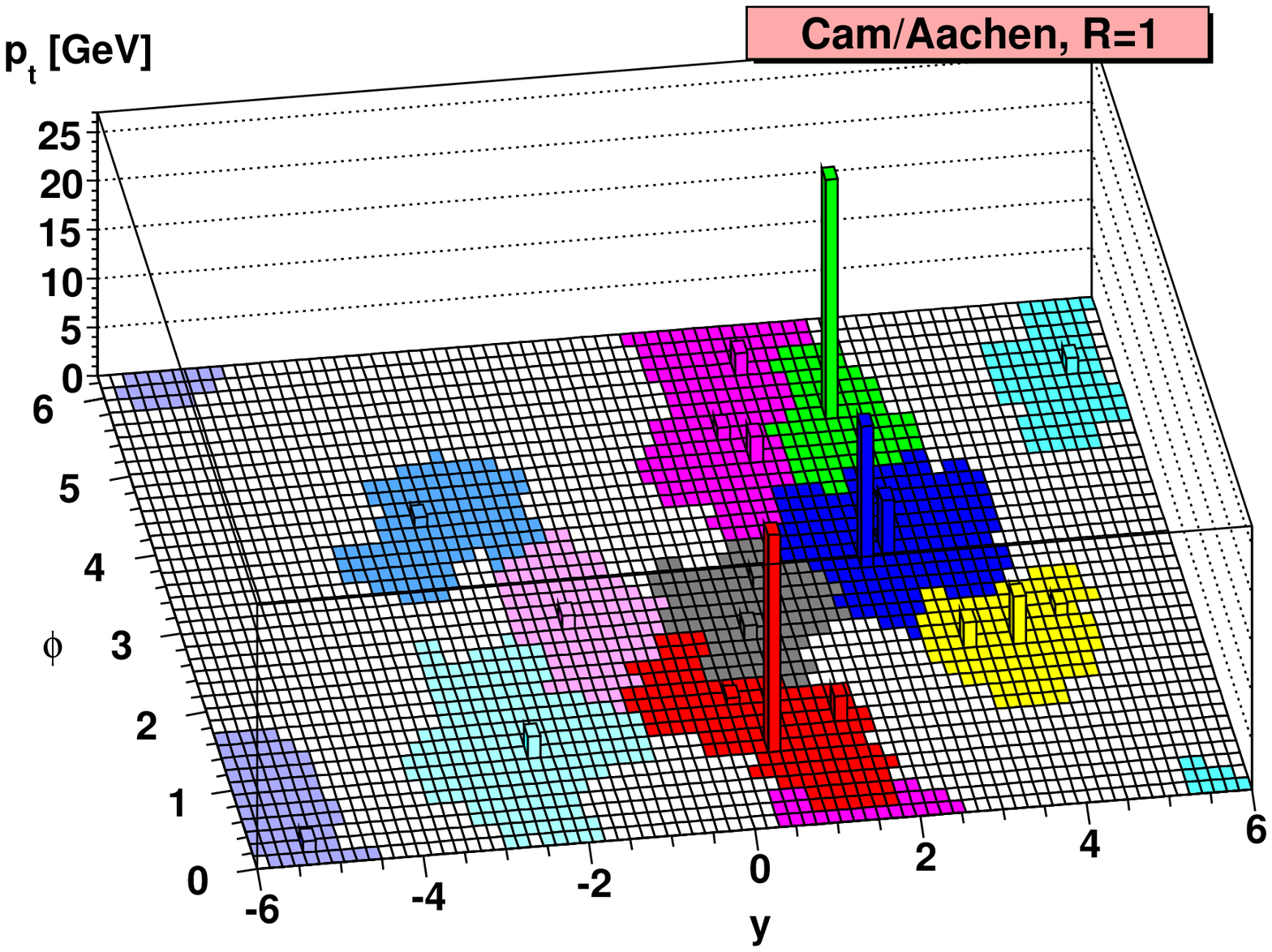}%
  \\[1em]                                     
  \includegraphics[width=0.49\textwidth]{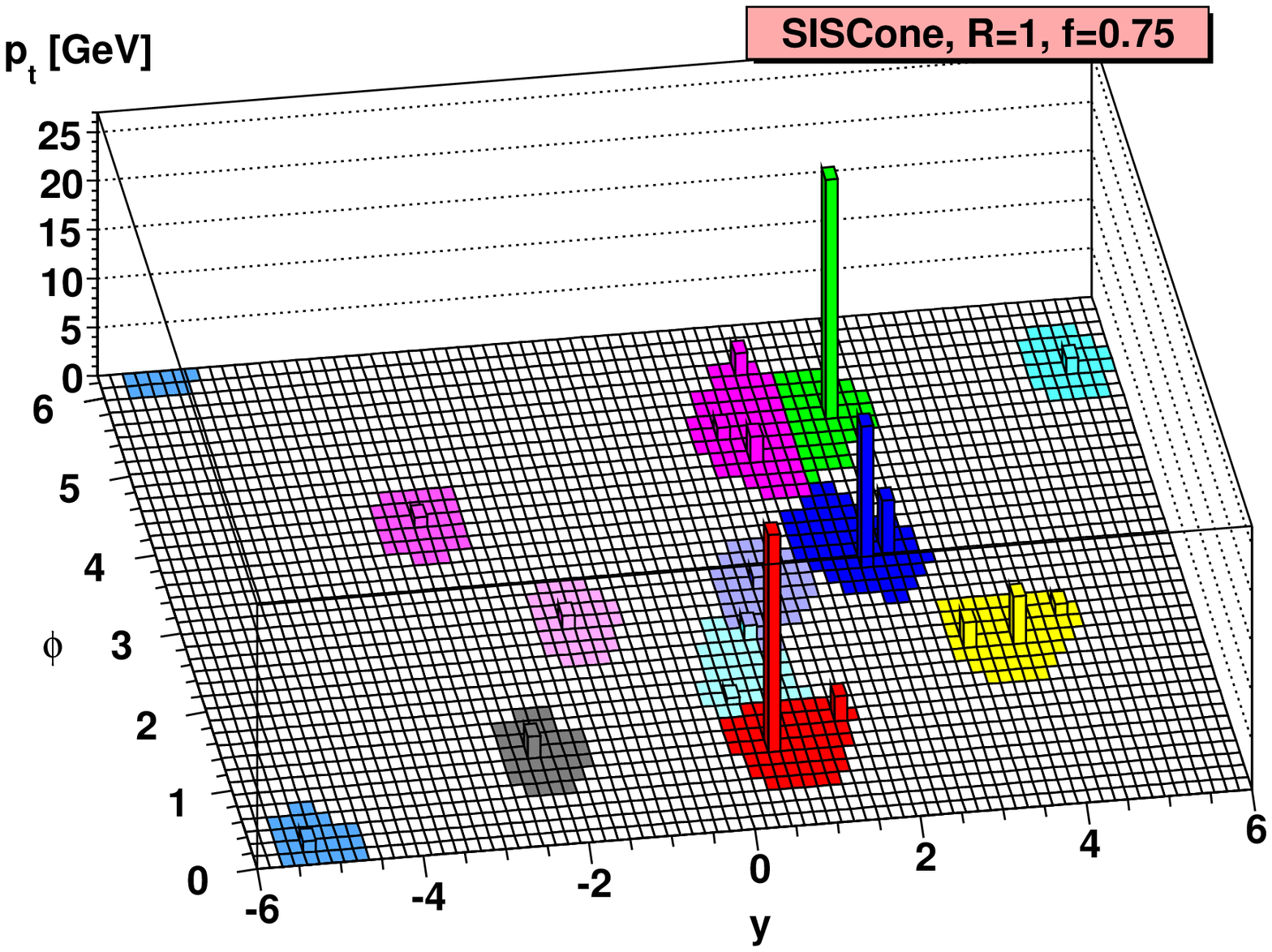}%
  \hfill                                 
  \includegraphics[width=0.49\textwidth]{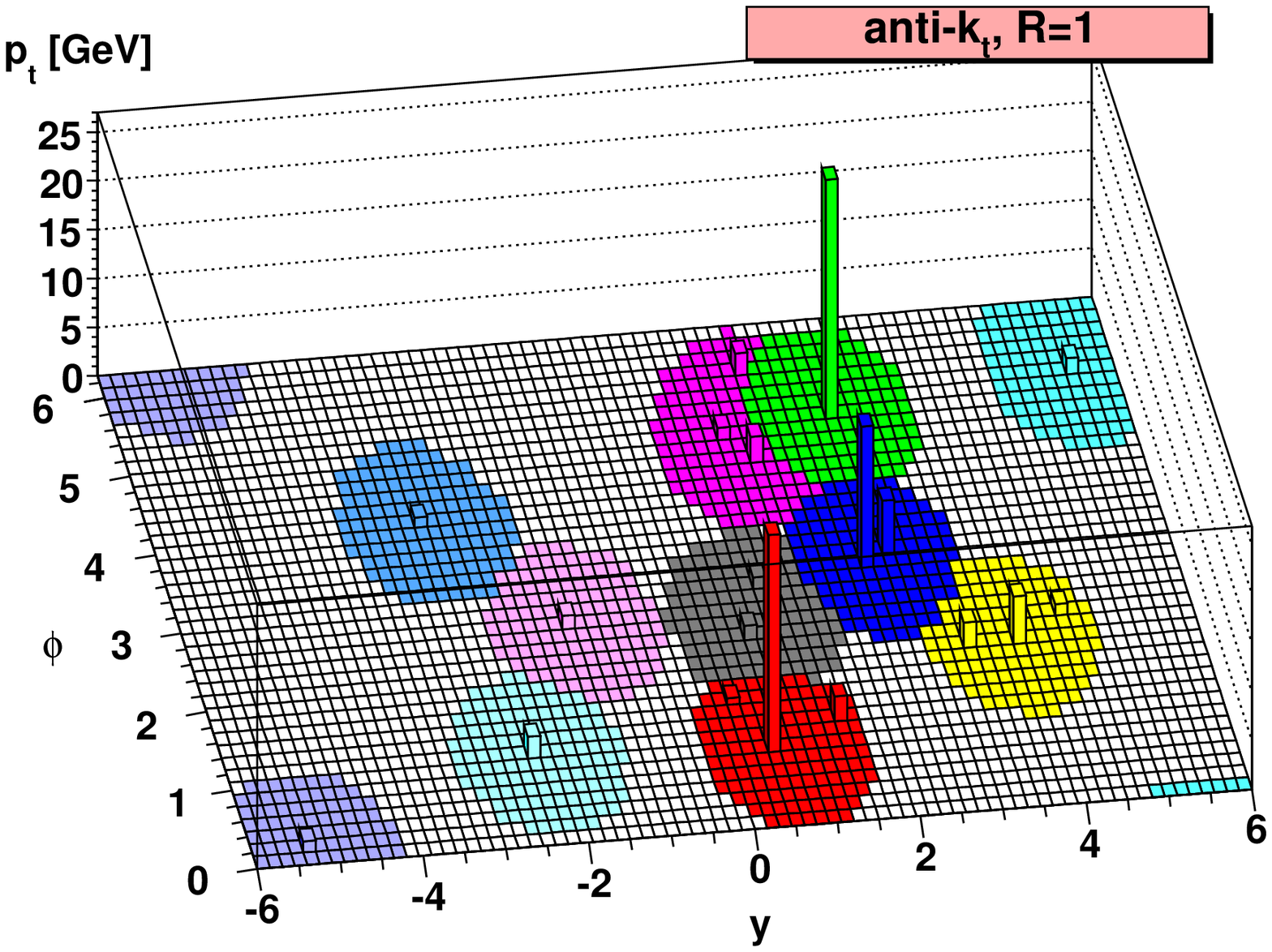}%
  \caption{A sample parton-level event (generated with
    Herwig~\cite{Herwig}), together with many random soft ``ghosts'',
    clustered with four different jet algorithms, illustrating the
    ``active'' catchment areas of the resulting hard jets (cf.\
    section~\ref{sec:jet-areas}). For $k_t$ and Cam/Aachen the
    detailed shapes are in part determined by the specific set of
    ghosts used, and change when the ghosts are modified.}
  \label{fig:4algs}
\end{figure}

A general recommendation is that hadron-collider algorithms that are
IR or collinear unsafe should in future work be replaced by IRC safe
ones, of which the inclusive $k_t$, C/A (possibly with ``filtering''),
anti-$k_t$ and SISCone are good choices.
Specifically the xC-PR class of algorithms is naturally replaced by
the anti-$k_t$ algorithm (which produces circular jets, as illustrated
in figure~\ref{fig:4algs}, and has similar low-order perturbative
properties),
while SISCone is very much like the IC-SM algorithms, but ensures that
the stable-cone finding is IRC safe.

Figure~\ref{fig:4algs} illustrates the jets that are produced with the
4 ``choice'' IRC-safe algorithms in a simple, parton-level event
(generated with Herwig), showing among other things, the degree of
regularity (or not) of the boundaries of the resulting jets and their
extents in the rapidity-azimuth place.

%

  \begin{table}
    \centering
    \begin{tabular}{|l|l|c|l|p{4cm}|}
      \hline
      Algorithm & Type  & IRC safe?  & Ref. & Notes \\
      \hline
      \hline
      inclusive $k_t$     & 
      SR$_{p=1}$ &
      OK        &
      \cite{Kt,KtHH,Kt-EllisSoper} &
      also has exclusive variant
      \\ \hline
      Cambridge/Aachen &
      SR$_{p=0}$ &
      OK        &
      \cite{Cam,Aachen} &
      \\ \hline
      anti-$k_t$ %
      &
      SR$_{p=-1}$ &
      OK         &
      \cite{Cacciari:2008gp} &
      \\ \hline
      SISCone     &
      SC-SM  &
      OK     &
      \cite{Salam:2007xv}&
      multipass, with optional cut on stable cone $p_t$
      \\ \hline
      CDF JetClu      &
      IC$_r$-SM   &
      IR$_{2+1}$  &
      \cite{Abe:1991ui}&
      \\ \hline
      CDF MidPoint cone &
      IC$_{mp}$-SM  &
      IR$_{3+1}$    &
      \cite{RunII-jet-physics} &
      \\ \hline
      CDF MidPoint searchcone &
      IC$_{se,mp}$-SM  &
      IR$_{2+1}$    &
      \cite{EHT} &
      \\ \hline
      \Dzero Run II cone     &
      IC$_{mp}$-SM  &
      IR$_{3+1}$    &
      \cite{RunII-jet-physics} &
      no seed threshold, but cut on cone $p_t$
      \\ \hline
      ATLAS Cone  &
      IC-SM       &
      IR$_{2+1}$  &
      \cite{Aad:2009wy} &
      \\ \hline
      PxCone     &
      IC$_{mp}$-SD  &
      IR$_{3+1}$    &
      \cite{Seymour:2006vv} &
      no seed threshold, but cut on cone $p_t$, 
      \\ \hline
      CMS Iterative Cone     &
      IC-PR  &
      Coll$_{3+1}$    &
      \cite{Bayatian:2006zz} &
      \\ \hline
      PyCell/CellJet (from Pythia)     &
      FC-PR  &
      Coll$_{3+1}$    &
      \cite{Sjostrand:2006za}
      &
      \\ \hline
      GetJet (from ISAJET) &
      FC-PR  &
      Coll$_{3+1}$    &
      &
      \\ \hline
    \end{tabular}
    \caption{\small Overview of some jet algorithms used in experimental
      or theoretical work in hadronic collisions in the past few
      years. SR$_{p=x}=$
      sequential recombination, with $p=-1,0,1$ characterising the exponent
      of the transverse momentum scale, 
      eq.~(\ref{eq:genkt-dist-R}); SC = seedless cone (finds all cones); IC = 
      iterative cone (with midpoints 
      $mp$, ratcheting $r$, searchcone $se$), using either
      split--merge (SM), split--drop (SD) or progressive removal (PR)
      in order to address issues with overlapping stable cones;
      FC = fixed-cone.
      In the characterisation of infrared and collinear (IRC) safety
      properties (for the algorithm as applied to particles), 
      IR$_{n+1}$ indicates that given $n$ hard particles in a common
      neighbourhood, the addition of 1 extra soft particle can modify
      the number of final hard jets; Coll$_{n+1}$ indicates that given
      $n$ hard 
      particles in a common 
      neighbourhood, the collinear splitting of one of the particles
      can modify the number of final hard jets.
      Where an algorithm is labelled with the name of an experiment,
      this does not imply that it is the only or favoured one of the
      above algorithms used within that experiment.
      Note that some of the corresponding computer codes for
      jet finding first project particles onto modelled calorimeters.
      \label{jetalgs_conefeatures}
    }
  \end{table}

\section{Computational geometry and jet finding}
\label{sec:resolving-snowmass}


It takes the human eye and brain a fraction of a second to identify
the main regions of energy flow in a calorimetric event such as
fig.~\ref{fig:4algs}. A good few seconds might be needed to
quantify that energy flow, and to come to a conclusion as to how many
jets it contains.
Those are timescales that usefully serve as a reference when
considering the speed of jet finders --- if a jet finder takes a few
seconds to classify an event it will seem somewhat tedious, whereas a
few milliseconds will seem fast.
One can reach similar conclusions by comparing to the time for a Monte
Carlo event generator to produce an event (from tens of milliseconds
to a fraction of a second), or for a fast detector simulation to
process it.
Or by considering the number of CPU hours needed to process a typical
event sample, which might consist of $\order{10^{7}}$ events.

\begin{table}[t]
  \centering
  \begin{tabular}{l|r}
    Type of event                                    & $N$\\ \hline
    $\ee\to$\,hadrons event on the $Z$ peak          & $40$ \\
    HERA direct photoproduction (dijet) or DIS       & $40$ \\
    HERA resolved photoproduction (dijet)            & $60$ \\
    Tevatron ($\sqrt{s} = 1.96\TeV$) dijet event     & $200$     \\
    LHC ($\sqrt{s} = 14\TeV$) dijet event            & $400$     \\
    LHC low-luminosity event (5  pileup collisions)  & $1000$     \\
    RHIC  Au\,Au event ($\sqrt{s} = 200\GeV$/nucleon)& $3000$     \\
    LHC high-luminosity event (20 pileup collisions) & $4000$     \\
    LHC Pb$\,$Pb event ($\sqrt{s} = 5.5\TeV$/nucleon)& $30000$     \\
  \end{tabular}
  \caption{Orders of magnitude of the event multiplicities $N$
    (charged + neutral) for 
    various kinds of event. The $\ee$, photoproduction, DIS and $pp$ results have been
    estimated with Pythia~6.4\cite{Pythia63,Sjostrand:2006za}, LHC PbPb
    with Pythia + Hydjet~\cite{hydjet} 
    and RHIC has been deduced from~\cite{Adler:2001yq}.
    Note that experimentally, algorithms may run on calorimeter towers
    or cells, which may be more or less numerous than the
    particle multiplicity.
  }
  \label{tab:event-multiplicites}
\end{table}

The time taken for jet finding by computer codes depends strongly on
the number of input particles (or towers, etc.), $N$. We don't yet
know the exact average multiplicities of LHC events, but rough
estimates are given in table~\ref{tab:event-multiplicites}.
With the $k_t$ algorithm's ``standard'' $N^3$ timing, assuming about
$10^9$ computer operations per second, one expects a time for
clustering a low-luminosity LHC event of $1\second$ (this is also what
one finds in practice).
So this is close to being ``tedious,'' and becomes dissuasive for
high-luminosity LHC and heavy-ion collisions, or if one wishes to try
out many distinct jet definitions (\eg several different $R$ values to
see which is best).
A more extreme example is the exact seedless cone algorithm following
the method in~\cite{RunII-jet-physics}, which has a timing of $N 2^N$.
In practice (NLOJET++ implementation~\cite{NLOJet}), an event with
$\sim20$ particles takes about a second, so one can extrapolate that
even just $100$ particles will take $10^{17}$ years. This is beyond
prohibitive.

To speed up jet finders
one may consider the general class of computational algorithm that the
jet finder belongs to.
For instance, all the SR jet finders are examples of ``hierarchical
clustering'', with a range of different distance measures. General
solutions to the problem were discussed long ago in the
computer science literature by Anderberg~\cite{Anderberg}, with a set
of rather 
good solutions proposed by Cardinal and Eppstein more recently
in~\cite{EppsteinHierarchical,LazyClosestPairs}, which scale roughly
as $N^2$.

Generic hierarchical clustering is, however, a broad problem. For
example, given three ``points'', $A$, $B$ \& $C$, generic distances
are not `transitive': if $A$ is close to $B$ and $B$ is close to $C$,
this does \emph{not} imply that $A$ is close to $C$ (the reader is
encouraged to think up a concrete example for the $k_t$ distance
measure).
On the other hand, jet finding often has a geometrical component
(since, at hadron colliders, the rapidity and azimuth coordinates
represent the surface of a cylinder). In geometry, if $A$ is close to
$B$ and $B$ to $C$, then $A$ and $C$ are necessarily also close. This
is of significant help, and a whole research field exists for such
geometric proximity problems, computational geometry.
Section~\ref{sec:speeding-up-k_t} will show how we can make use of
this to obtain $N\ln N$ scalings for the $k_t$ algorithm rather than
the $N^2$ of generic hierarchical clustering, or $N^3$ of the older
$k_t$-clustering codes.
Then in section~\ref{sec:making-cone-ir} we will examine how to apply
computational geometry to cone algorithms.

\subsection{Sequential recombination algorithms}
\label{sec:speeding-up-k_t}

The original implementations of the $k_t$ algorithm 
\cite{KtClus,KtImplementation} set up a two-dimensional array of the
$d_{ij}$, and at each stage of the clustering run through all entries
of it in order to find the minimum, and then update the array with the
entries for the newly created particle. Since the $d_{ij}$ array is of
size $\order{N^2}$ and the minimum is searched for $\order{N}$ times
in total (\ie $\order{N}$ clusterings), these implementations take a
time $\sim N^3$.

We have seen briefly above that there exist generic methods for
hierarchical clustering, \ie repeated recombination of the closest
pair of objects, that take $N^2$ time.
In general $N^2$ time is a lower bound because, at the very least,
one has to consider all entries of the $d_{ij}$ distance matrix in
order to find the smallest.
One may then be clever in keeping track of distance information as
points are recombined, so as to side-step the $N^3$ growth of some
$k_t$ algorithm implementations,\footnote{Essentially, observing that
  at most $\order{N}$ distances change at every pair recombination.}
but the initial search for the minimum among all pairs of points
seems unavoidable.

\subsubsection{$k_t$ algorithm}
\label{sec:kt-alg-fast}

To see whether we can evade the $N^2$ bound, let us examine the $k_t$
algorithm's distance measure in more detail
\begin{equation}
  \label{eq:dij-again}
  d_{ij} = \min(p_{ti}^2, p_{tj}^2) \Delta R_{ij}^2,\quad\qquad\Delta
  R_{ij}^2 = (y_i - y_j)^2 + (\phi_i - \phi_j)^2\,.
\end{equation}
We could equally well have considered a distance measure
\begin{equation}
  \label{eq:Dij}
  D_{ij} = p_{ti}^2 \Delta R_{ij}^2,
\end{equation}
The smallest of the $D_{ij}$ across all $i,j$ coincides with the
smallest of all the $d_{ij}$, since $\min(D_{ij},D_{ji}) = d_{ij}$. So
it is irrelevant whether we use eq.~(\ref{eq:dij-again}) or
(\ref{eq:Dij}) in the $k_t$ algorithm.

Eq.~(\ref{eq:Dij}) has the important property that the
transverse-momentum part depends on just one of the two particles, so
we can write
\begin{equation}
  \label{eq:1}
  \min_{j} \{D_{ij}\} = p_{ti}^2 \min_{j} \{\Delta R_{ij}^2\}\,,
\end{equation}
\ie fixing $i$, the smallest of the $D_{ij}$ involves $i$'s geometrical
nearest neighbour (let's refer to it as $\cG_i$). So if we can find some
efficient way of establishing and tracking that geometrical
information, then rather than finding the minimum of $N^2$ $D_{ij}$
values, the
sequential recombination problem involves only finding the minimum of
$N$ $D_{i \cG_i}$ values. This was the key observation
of~\cite{Cacciari:2005hq}.

One is then left with the question of how to find the minimum of the
$\Delta R_{ij}^2$ for each $i$, since this still seems to involve a
total of $N^2$ points.
Technically, the problem is that of establishing and maintaining a
nearest-neighbour graph on the 2-dimensional surface of a cylinder.
A rule of thumb when faced with such problems is to first ask how one
might deal with them in 1 dimension, say rapidity $y$.
That is easy: one sorts the points according to their $y$ coordinate,
and the nearest neighbour of a point is the one that immediately
precedes or follows it.

Let us do the bookkeeping for this case with just a rapidity
coordinate:\medskip\\
\underline{Initialisation:}
\begin{itemize}
\item Sort points according to $y$ coordinate (with a balanced binary
  tree), find nearest neighbours, and find all $d_{i\cG_i}$. %
  \mbox{ } \hfill{[$\sf N \ln N$]}
\item Place the $d_{i\cG_i}$ in a ``priority queue'' (a structure for
  efficient minimum-finding and updates; often simply a balanced
  binary tree)
  \mbox{ } \hfill{[$\sf N \ln N$]}
\end{itemize}
\underline{Iteration:}
\begin{itemize}
\item Recombine the pair with smallest $d_{i\cG_i}$,
  remove the corresponding two points from the
  rapidity-sorted tree, add the new one, establish the new point's
  nearest neighbours and establish if it has become the nearest
  neighbour of any of the existing points.\\
  \mbox{ } \hfill{[\sf $\sf \ln N$ \sf per recomb.]}
\item Update the priority queue of $d_{i\cG_i}$
  values and find the new minimum (only a finite number of
  $d_{i\cG_i}$ will change per round).
  \hfill{[\sf $\sf \ln N$ \sf per recomb.]}
\end{itemize}
This gives a total time of $\order{N \ln N}$.
To understand the origin of the $\ln N$ factor, observe, for example,
that if you organise $N$ objects into a binary tree structure, then
the depth of the tree will be $\ln_2 N$ (equivalently, given $k$
levels to the tree, it can contain up to $\sum_{\ell=0}^{k-1} 2^{\ell}
= 2^k-1$ objects).
Any operation such as adding or removing an entry in the tree involves
working through the depth of the tree, and so paying a price
$\order{\ln N}$. Since building the tree can be seen as adding $N$
objects this costs $\order{N \ln N}$ time.

\begin{figure}[t!]
\begin{center}
\includegraphics[width=0.3\tw]{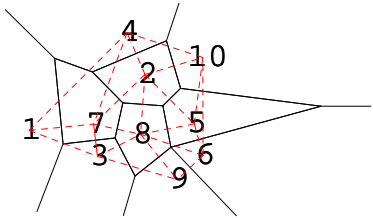}
\caption{\label{fig:voronoi} \small The Voronoi diagram for ten random
  points. The Delaunay triangulation (dashed, red) connecting the ten points
  is also shown. In this example the points 1, 4, 2, 8 and 3 are the
  `Voronoi' neighbours of 7, and 3 is its nearest neighbour. Adapted
  from~\cite{Cacciari:2005hq}.}
\end{center}
\end{figure}

With two geometrical dimensions, nearest-neighbour finding is more complex,
however it has been the subject of research by the computational
geometry community.
One structure that can help is the Voronoi diagram~\cite{Voronoi}, or
its dual, the Delaunay triangulation.
A Voronoi diagram divides the plane into cells (one per vertex),
such that every point in the cell surrounding a vertex $i$ has $i$
as its nearest vertex.
The structure is useful for nearest-neighbour location because the
vertex $\cG_i$ nearest to vertex to $i$ is always in one of the (few,
\emph{expected}%
\footnote{``Expected'' means that there can be special cases where the
  number is parametrically larger.} %
$\order{1}$)
cells that share an edge with the cell of vertex $i$.  An example is
shown in figure~\ref{fig:voronoi}.

Voronoi diagrams for $N$ points can be constructed with $\order{N \ln
  N}$ operations \cite{Fortune}. Maintaining dynamic point sets is
more complicated, however there exists an
approach~\cite{DelaunayDeletion} that takes $\order{N \ln N}$ for the
initial construction and expected $\ln N$ per insertion/deletion and
it is available as a public code, CGAL~\cite{CGAL,CGALTriang}.

A complete expected $N \ln N$ implementation that makes use of CGAL is
available in the FastJet program~\cite{FastJet}.
In practice, $\ln N$ terms in computational geometry come with a large
coefficient --- typically a $\ln N$ term might be smaller than 
a term linear in $N$ only for $N \gtrsim 10^3$.
Therefore for moderate $N$ it is useful to include alternative
computational strategies. One particularly successful one (optimal in
the range $50 \lesssim N \lesssim 10^4$) makes use of the fact that
only for $\Delta R_{ij} < R$ will a $d_{ij}$ distance be smaller than
$d_{iB}, d_{jB}$ in eq.~(\ref{eq:DISkt-dist})--- therefore one can
restrict one's search for $i$'s geometric nearest neighbours to the
region within $R$ of $i$. Denoting by $n$ the typical number of points
in such a region, one then has a $\order{N n}$ algorithm.

Further details are available in the FastJet documentation (from the
web site~\cite{FastJet}) and from an unpublished
preprint~\cite{CGTA-not}.



\subsubsection{Special cases}
\label{sec:SR-special-cases}

The above approaches can be used for the whole class of generalised
(longitudinally invariant) $k_t$ algorithms, however some special
cases deserve comment.

\paragraph{Cambridge/Aachen.} For the C/A jet finder, there is no
momentum scale, so rather than having a dynamic planar
nearest-neighbour problem, one has a dynamic planar closest pair
problem, 
$d_{ij} \equiv \Delta R_{ij}$. 
This is simpler --- essentially one can maintain a nearest neighbour
candidate for each point, but it only need be correct for the closest
pair.
One remarkable solution to this problem\footnote{Which can be stated
  in a paragraph, though this does not mean that it is simple to
  understand!} was given by Chan in~\cite{Chan}. It is included
natively in FastJet, and is slightly faster than the CGAL based
solution (as well as avoiding the need for a separate package).

Note: Chan's solution relies on the use of integer arithmetic (part of
its cleverness lies in its implicit use of the binary representation of
integers). However since rapidities and azimuths do not extend to
large values, one can safely rescale them by some large constant and
represent them as integers.

\paragraph{Anti-$k_t$.} The generalised $k_t$ algorithm with $p<0$ (and
specifically anti-$k_t$, with $p=-1$) has the property that it
effectively produces jets that grow outwards in a circular pattern
around a high-$p_t$ seed.
This leads to configurations with one particle at the centre of a
circle and many on the edge (the first layer of points on the edge
contains $\order{\sqrt{n}}$ particles).

This is precisely the configuration in which the `expected' $N \ln N$
behaviour of the clustering breaks down: the central point has many
Voronoi neighbours, and, furthermore, is involved in each clustering
(so it is removed, reinserted, and then one must go around all the
points on the edge to see which now is its nearest neighbour).
This means that for very large $N, n$, the timing for anti-$k_t$ type
algorithms is closer to $N \sqrt{n}$ than to $N \ln N$.

According to \cite{TOPP} there exist approaches to the planar
nearest-neighbour problem that have worst-case behaviour
$N^{\epsilon}$ with arbitrarily small $\epsilon$, however these have
not been investigated in the context of jet finding.

\subsection{A polynomial-time seedless cone}
\label{sec:making-cone-ir}

We saw in section~\ref{sec:cone-algorithms-irc} that the use of
particles as seeds, \ie starting points for cone iterations, gets us
into trouble with IRC safety: if one finds jets based on the ordering
of the seeds in $p_t$, then one is sensitive to collinear splittings;
if one uses the stable cones obtained from iterating all seeds, then
one becomes sensitive to the addition of new soft seeds.

We also saw an exact seedless approach that takes all subsets of
particles and establishes for each one whether it corresponds to a
stable cone --- \ie for each subset one calculates its total momentum, draws a circle
around the resulting axis, and if the points contained in the circle
are exactly those in the initial subset, then one has found a
stable cone.
This is guaranteed to find all stable cones.

For large multiplicities, this is inherently wasteful insofar as most
of the $2^N$ subsets of particles don't fit into a circle of radius
$R$ on the rapidity-azimuth plane, so there is no way for them to
form a stable cone.

The obvious corollary of that observation is that one should only
consider subsets of points in which the members of the subset are
contained within a circle of radius $R$, and any point not in the
subset is outside the circle.
It is only these subsets that can ever form a stable cone.
Therefore rather than considering all subsets of points, one can restrict
one's attention to all distinct ways of separating points on the
surface of a cylinder (or plane) into two subsets, those points inside
a circle of radius $R$, all others outside --- a ``planar all
distinct circular enclosures'' problem.

\begin{figure}
  \centering
  \includegraphics[width=0.5\textwidth]{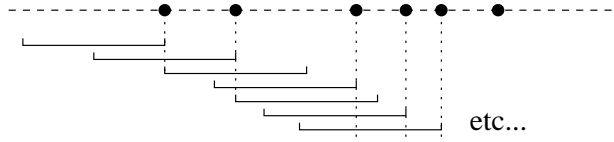}
  \caption{Representation of points on a line and the places where a
    sliding segment has a change in its set of enclosed points.}
  \label{fig:1dSegment}
\end{figure}

This problem is a clear example of a computational geometry problem.
Let's first see how we would deal with it in one dimension, for which
a ``circular enclosure'' just reduces to a line segment
enclosure. Given points on a line and a segment of length $2R$, we can
order the points, place the segment to the left of the leftmost point,
and then slide it sideways. Each time the left or right edge of the
segment touches a point, the contents of the enclosure change.
The cost of finding all enclosures is just that of ordering the points
($N \ln N$).

\begin{figure}
  \centering
  \includegraphics[width=\textwidth]{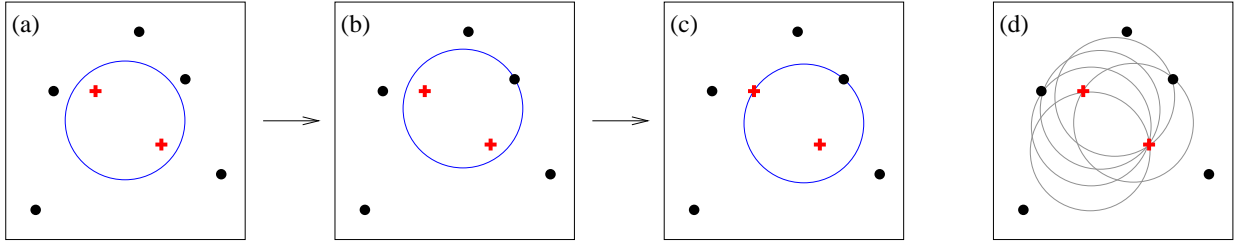}
  \caption{(a) Some initial circular enclosure; (b) moving the circle
    in a random direction until some enclosed or external point
    touches the edge of the circle; (c) pivoting the circle around the
    edge point until a second point touches the edge; (d) all circles
    defined by pairs of edge points leading to the same circular
    enclosure.}
  \label{fig:2dcircle}
\end{figure}

How do we extend this to two dimensions? The central idea is that the
enclosed point set changes when a point touches the enclosure. In 1d
we can always shift the enclosure, without changing its contents,
until its edge touches a point (either in or out of the enclosure).
In 2d we can first shift the circular enclosure until one point
touches the edge, then pivot the circle around that point until its
circumference touches a second point (fig.~\ref{fig:2dcircle}).
Conversely if we consider all pairs of points (within $2R$ of each
other) and draw all possible circles that go through those pairs, then
we will have found all possible enclosures (one should remember that
edge points can be either in or out of the enclosure; special
treatment is also needed for points that are alone, \ie further than
$2R$ from the nearest other point).

There are $\order{Nn}$ relevant pairs of points (recall, cf.\ the end
of section \ref{sec:kt-alg-fast}, that $n$ is the number of objects in
a region of area $\sim R^2$).%
\footnote{There is a correspondingly large number of distinct cones,
  and this has implications for
  proposal~\cite{Volobouev,Volobouev:2009rv} to use a Fast Fourier
  Transform for the stable-cone search (cf.\ also the FFTJet package
  \cite{FFTJet}, which was released just as this review was being
  finalised), essentially because it implies the need for a Fast
  Fourier Transform grid of size $\order{N n}$.}
One could directly check the stability of the cones defined by each
pair as follows: one sums the $\order{n}$ momenta contained within a
given cone and then checks to see whether a new cone centred on the
direction of the resulting momentum contains the same set of points.
This would give an $\order{N n^2}$ algorithm.
Alternatively one can establish a
traversal order in which the circle contents change by one point at a
time, avoiding (with the help of a few other tricks) the need to pay a
price of $\order{n}$ for the stable-cone check for each distinct
enclosure. This is the basis of the (expected) $N n \ln n$ algorithm
that is known as SISCone (Seedless Infrared Safe Cone)
\cite{Salam:2007xv}.\footnote{It has been pointed out by
  Sj\"ostrand~\cite{SjostrandThrust} that SISCone's use of all pairs
  of points to provide the full list of distinct circular enclosures
  bears a close relation to a technique used in the $N^3$ computation
  of the Thrust~\cite{Brandt:1978zm}. There, all pairs of particles
  are used to generate all relevant separations of the surface of a
  sphere into two hemispheres. 
  A corollary of this observation is that SISCone's idea of a
  traversal order could also be used in the context of the thrust, to
  reduce its computation time to $N^2 \ln N$.  }

Some comments are due concerning SISCone's timing. There are usually only
$\order{N}$ stable cones, parametrically fewer than the number of
distinct enclosures. Might there be a way of somehow skipping all the
unstable enclosures? 
It is not clear, because the upper bound on the number of stable cones
is actually $\order{N n}$ (the much lower expected value holds for
random point sets \cite{Salam:2007xv}).
This worst case can actually occur (for example with regular sets of
``ghosts,'' \cf section~\ref{sec:jet-areas}), with implications then
for the split--merge procedure.
Normally the split--merge
procedure is significantly faster than the stable-cone search, in that
it takes time $\order{N^2}$ ($\ll N n \ln n$ in
practice%
\footnote{In QCD events and with typical values of the jet finding
  radius $R=0.4-1.0$, $N/n$ is usually between $10$ and $100$; in
  2-dimensional problems, for the multiplicities that are of relevance
  here, a rule of thumb seems to be that $\ln n$ is very roughly
  equivalent to a factor of order $10^3$.}%
) in SISCone's
fairly straightforward implementation. However if the number of stable
cones is $\order{Nn}$, then the split--merge step becomes $\order{N^2
  n}$ unless one applies additional dedicated techniques (such as
quad-trees or $k$-d trees, as discussed in \cite{Salam:2007xv} and
also suggested in \cite{Volobouev}).

A further comment is due on memory usage: SISCone maintains a hash of
circular enclosures that it has already seen (and whether they are
candidates for stable cones or not). That hash has as many entries as
distinct enclosures, $\order{N n}$, and this can become problematic
for very large multiplicities.\footnote{If we suppose $n \sim 0.1 N$,
  and that each entry needs 12 bytes for the hash, two
  double-precision numbers (8 bytes each) to describe the center of
  the cone, and a pointer to the next hash element, plus various
  overheads, then we get a memory usage of about $4 N^2$ bytes, \ie
  nearly $4\,$GB for $N \sim 30\,000$, which is a typical expected LHC
  heavy-ion multiplicity.}
In such cases one could in principle reduce the memory use to
$\order{N}$, at the expense of a slower run-time $\order{N n^{3/2}}$,
but this has not been implemented.

\subsection{Speed summaries}
\label{sec:speed-summary}

Statements of timings in terms of their scaling with $N$ can hide
large coefficients and significant preasymptotic corrections. Another
issue is that as $N$ increases so does memory usage, requiring
(slower) access to the main memory rather than the CPU cache, and this
too can affect practical timing results.

\begin{figure}
  \centering
  \includegraphics[width=0.8\tw]{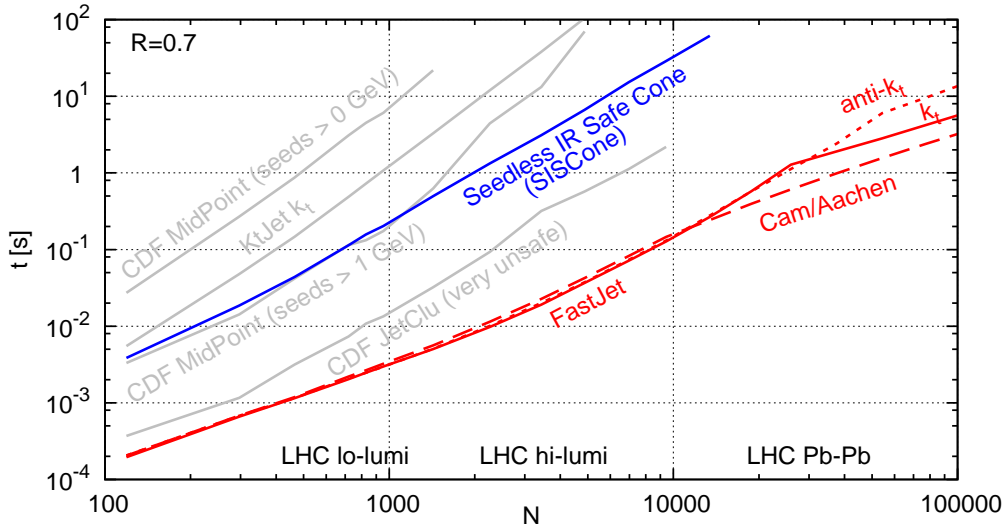}
  \caption{Timings for the clustering of a simulated $\sim 50\GeV$
    dijet event, to which increasing numbers of simulated minimum-bias
    events have been added (both simulated with Pythia).
    In dark colours one sees SISCone and the FastJet $k_t$, anti-$k_t$
    and Cambridge/Aachen implementations. For $k_t$ (anti-$k_t$), 
    the kink at $N\simeq 25000$ ($N\simeq 50000$)
    signals the point where FastJet switches between $N n$ and $N \ln
    N$ ($N n^{1/2}$) strategies. In grey one sees results for the KtJet
    implementation~\cite{KtImplementation} of the the $k_t$ algorithm,
    the Midpoint cone (IC$_{mp}$-SM) in CDF's implementation (with and
    without a $1\GeV$ cutoff on seeds) and the JetClu iterative cone
    (IC$_r$-SM, with a $1\GeV$ seed threshold). All non-FastJet
    algorithms (except KtJet) have been accessed through FastJet
    plugins.}
  \label{fig:timings}
\end{figure}

Timings for a subset of commonly used algorithms are shown in
fig.~\ref{fig:timings}. One conclusion from that figure is that
SISCone, the slowest of the IRC safe algorithms, is still competitive
in speed with the main public Midpoint-cone code and is acceptably
fast unless one goes to $N$ larger than several thousand.
FastJet's implementation of the $k_t$ algorithm (and C/A and
anti-$k_t$) is much faster, with clearly healthier scaling at large
$N$, and it beats even the fast IRC unsafe cone codes, like CDF's
JetClu.

\begin{figure}
  \centering
  \includegraphics[width=0.6\tw]{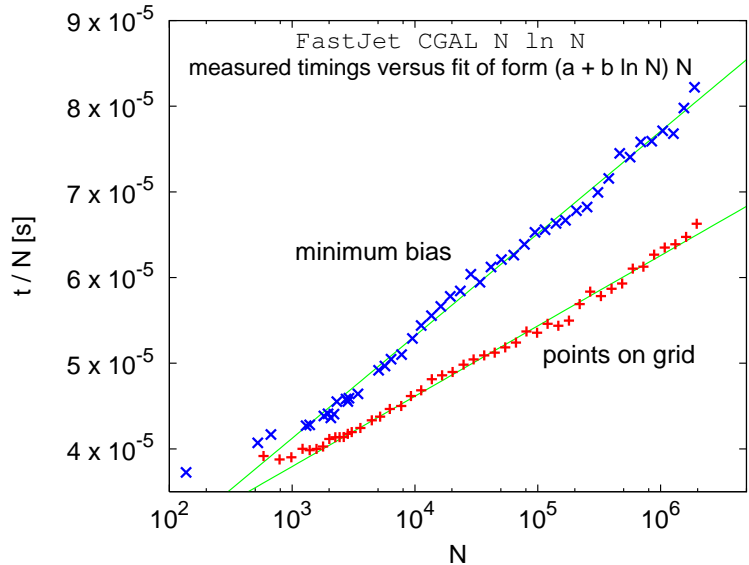}
  \caption{Verification of the $N\ln N$ timing behaviour of the
    CGAL-based implementation of the $k_t$ algorithm in FastJet. The
    timings are divided by $N$ so as to highlight the remaining $\ln
    N$ dependence. Taken from~\cite{CGTA-not}.}
  \label{fig:NlnN-check}
\end{figure}

The FastJet curve has a clear kink at $N \sim 15000$. This is the
point where FastJet switches from an $N n$ (tiled) algorithm to the $N
\ln N$ CGAL-based one.
One can explicitly verify that the CGAL-based algorithm does have an
$N\ln N$ behaviour, by dividing the run times by $N$ and plotting the
result. This is shown in fig.~\ref{fig:NlnN-check}.

To conclude this part, we have seen how the computational geometry
aspect of jet-related problems can be exploited to help resolve many of the practical
computational issues that arise if one is to carry out infrared-safe
hadron-collider jet finding.
It is probably fair to say that this is playing a crucial role in
encouraging the LHC experiments to switch to QCD-compatible
jet finders.
In particular, all the LHC experiments now incorporate FastJet
and its SISCone plugin in their software frameworks (and ATLAS also
has its own $\sim N^2$ implementation of the
sequential recombination algorithms)~\cite{Aad:2009wy,CMS-Jet-PAS}.



\section{Understanding jets}
\label{sec:understanding}

Ideally, one would really like to be able to measure partons in
experiments. Jets are the closest, physically, that we get to
partons. How close are
they exactly? And what about the fact that a ``parton'' isn't actually
a well-defined concept in the first place?
An understanding of these questions is part of the key to knowing how
best to use jet algorithms at colliders, both in terms of choosing
which algorithm to use and setting its parameters.

The precise issues that one might investigate fall into various
categories.
For example, how broadly will a jet reach for its constituents
(section~\ref{sec:reach})? This information is important in terms of
one's ability to disentangle different partons in heavy-particle
decays (for example hadronic $t\bar t$ events, which decay to 6 hard
partons).
Often, one will use jets to reconstruct the kinematics of some `parent
object' (again, a heavy particle that decayed; or, in inclusive-jet
measurements, a scattered parton from the incoming proton). How does the jet's
energy relate to that of the `parent object'? This is affected both by
perturbative (section \ref{sec:pert-p_t-loss}) and non-perturbative
(section~\ref{sec:hadronisation}) radiation.
Finally LHC is special in that it will have significant
underlying-event activity (maybe $15\GeV$ per unit rapidity) and even
larger pileup (easily $100\GeV$ per unit rapidity). How do jets react
to this (section~\ref{sec:jet-areas})?

It is probably fair to say that our understanding of all these
questions is still incomplete. But the material below outlines some of
what we do know.


\subsection{Reach}
\label{sec:reach}

\subsubsection{Two-particle case}
\label{sec:two-particle-reach}

Given just two massless particles, separated by a distance $\Delta R$
on the $y-\phi$ cylinder, will they be recombined into a single jet?
This is the simplest of the questions one might ask about a
jet definition's reach.
It was discussed in~\cite{Kt-EllisSoper} for the inclusive
$k_t$ algorithm and for a partially specified cone, which behaves
somewhat like SISCone.

It is convenient to take the transverse momenta of the two particles,
$p_{t1}$, $p_{t2}$ (defined as the softer one) to be related by
\begin{equation}
  \label{eq:pt2-pt1}
  p_{t2} = x p_{t1}\,, \qquad\quad x < 1\,.
\end{equation}
Writing the sum of the two particles' momentum as $p_J$, with $p_{tJ}
= p_{t1} + p_{t2}$,\footnote{In the widespread 4-vector ($E$-scheme)
  recombination scheme, this is exact only if the particles are very
  close in angle. However it remains a good approximation even for $\Delta
  R_{12} \sim 1$ and so it is an approximation that will recur in this
  \toplevel.} %
and imagining the two partons as coming from a common ancestor, we can
also write
\begin{equation}
  \label{eq:pt12-z}
  p_{t1} = (1-z) p_{tJ}\,,\qquad\quad p_{t2} = z p_{tJ}\,,\qquad
  \left(z  = \frac{x}{1+x} <
   \frac12\right)\,.
\end{equation}
According to the context, results are more simply expressed either in
terms of $x$ or $z$, which is why it is useful to introduce both.

In the $k_t$ algorithm, the two particles will form a single jet if
$d_{12} < d_{1B}, d_{2B}$ (as defined in eqs.~(\ref{eq:ppkt-dist-R})),
or equivalently if 
\begin{equation}
  \label{eq:kt-2particle-condition}
  \Delta R_{12} < R\,.
\end{equation}
In the case of an algorithm like SISCone, based on stable cones, the
question is whether particles $1$ and $2$ can both belong to a single
cone. This happens if both are within $R$ of $p_{J}$,
\begin{equation}
  \label{eq:stable-cone}
  \left.
    \begin{array}{lr}
      \Delta R_{1J} =&     z \Delta R_{12}  \\
      \Delta R_{2J} =& (1-z) \Delta R_{12} 
    \end{array}
  \right\}
  < R\,,
\end{equation}
where the relations between $\Delta R_{iJ}$ and $\Delta R_{12}$ follow
from $y_{J} = (1-z)y_1+zy_2$, $\phi_{J} = (1-z)\phi_1+z\phi_2$.
Since we have defined $z<1/2$, it is the lower condition of
eq.~(\ref{eq:stable-cone}) that is more constraining and it leads to
\begin{equation}
  \label{eq:cone-2particle-condition}
  \Delta R_{12} < (1 + x) R\,,
\end{equation}
\ie the same as the $k_t$ algorithm condition,
eq.~(\ref{eq:kt-2particle-condition}), when $p_2$ is soft, $x\ll 1$,
but reaching twice as far when $p_{t2} \simeq p_{t1}$.

The conditions
eq.~(\ref{eq:kt-2particle-condition},\ref{eq:cone-2particle-condition})
basically account for the behaviours of nearly all jet algorithms:
\begin{itemize}
\item eq.~(\ref{eq:kt-2particle-condition}) holds for the $k_t$
  algorithm, as well as Cambridge/Aachen, anti-$k_t$, and the xC-PR
  cone algorithms;
\item eq.~(\ref{eq:cone-2particle-condition}) holds for IC$_{mp}$-SM
  (``midpoint''), IC$_{mp}$-SD (PxCone) and SC-SM (SISCone) algorithms;
\item IC-SM algorithms without midpoint seeds (JetClu, Atlas Cones)
  have ill-defined behaviour. For just two particles, they lead to a
  single jet based on eq.~(\ref{eq:kt-2particle-condition}), but if
  additional soft seeds are present then this transforms into
  eq.~(\ref{eq:cone-2particle-condition}). This is a manifestation
  of their infrared unsafety.
\end{itemize}

\subsubsection{General case}
\label{sec:reach-general}

The complexity of tracing the behaviour of a jet algorithm precludes
general results about the reach of different jet algorithms for
multi-particle configurations.
One question that has however seen some attention is that of how the
results of section~\ref{sec:two-particle-reach} get modified in the
presence of parton showering and hadronisation.

This is a delicate question because to answer it one has to know
something about the \emph{environment} that created the two partons:
did they come from the branching of a single parent quark, or a parent
gluon, or even the decay of a colour-singlet particle? And what was
the parent parton colour-connected to? All of these issues relate to
the fact that there is no rigorous way of defining partons in the
first place. Furthermore, even in a probabilistic Monte-Carlo type
approximation, the way partons shower and hadronise depends on the
environment.

One approach~\cite{Abe:1991ui,Abbott:1997fc} to the question involved
superposing pairs of events and establishing under what conditions
jets that had been identified in the individual events became a single
jet if one applied the jet finder to the two events combined together.
This study was performed for IC-SM type algorithms and came to the
conclusion that the individual jets were merged if there were within
$1.3 R$ of each other.%
\footnote{The value $1.3$ has also inspired a practice in NLO
  calculations, still current within the CDF
  collaboration~(e.g.~\cite{Aaltonen:2008eq}), that involves placing
  an artificial cut on the separation between partons within a jet at
  $\Delta R = R_{sep} \times R$, with $R_{sep} = 1.3$.
  Such ad-hoc modifications of the jet algorithm used in a theory
  prediction defeat the purpose of a NLO calculation. It is probably
  fair to say, however, that in most contexts where $R_{sep}$ has been
  used, its impact is smaller than the dominant theory and
  experimental uncertainties.} %
This corresponds roughly to expectations based on
eq.~(\ref{eq:cone-2particle-condition}), if one performs some
reasonable averaging over the jet momenta, \ie $x$ (indeed we will see
the value $1.3$ appear again below in
section~\ref{sec:jet-pt-Rdep}). However it fails to provide a direct
link with the $x$-dependence of
eq.~(\ref{eq:cone-2particle-condition}).

Another approach to testing
eqs.~(\ref{eq:kt-2particle-condition},\ref{eq:cone-2particle-condition})
was taken in \cite{Salam:2007xv}. There, jets were initially found
with a ``reference,'' $R=1$ hierarchical-clustering algorithm. The
hierarchy was used to identify the two main subjets, $S_1$ and $S_2$,
within the jet.
Each event was also clustered with a test algorithm $T$, with
$R_T=0.4$.
The test algorithm is the one whose clustering behaviour one wishes to
probe and need not be the same as the reference algorithm (and in our
tests will often not be).
One then looked to see if there was a jet with algorithm $T$ that
contained at least half of the $p_t$ of each of $S_1$ and $S_2$. If
there was, the conclusion was that the two subjets had ended up
(dominantly) in a single jet from the test algorithm.
The procedure was repeated for many events and one could then plot the
fraction of $k_t$-algorithm jets for which this occurs,
$P_{2\to1}(\Delta R, x)$, as a function of the distance $\Delta
R\equiv \Delta R_{S_1 S_2}$ and the momentum ratio $x \equiv
p_{t,S_2}/p_{t,S_2}$.

In ref.~\cite{Salam:2007xv} it was the $k_t$ algorithm that was used
as a reference. 
Here we shall instead use the C/A algorithm as a reference and the
subjets are the two objects whose merging in the reference jet's
clustering sequence involves the largest $k_t$ distance (\ie the
hardest merging).\footnote{This is inspired by the use of a $k_t$ and
  an angular distance measure in the original Cambridge algorithm, and
  gives clearer results than the $k_t$ algorithm's subjets.} %
The results will be based on dijet events simulated with Herwig
6.5~\cite{Herwig}, both at parton and hadron levels (the latter
including Herwig's default soft UE). Reference jets were required to
have $p_t > 50\GeV$.


\begin{figure}
  \centering
  \includegraphics[width=0.48\tw]{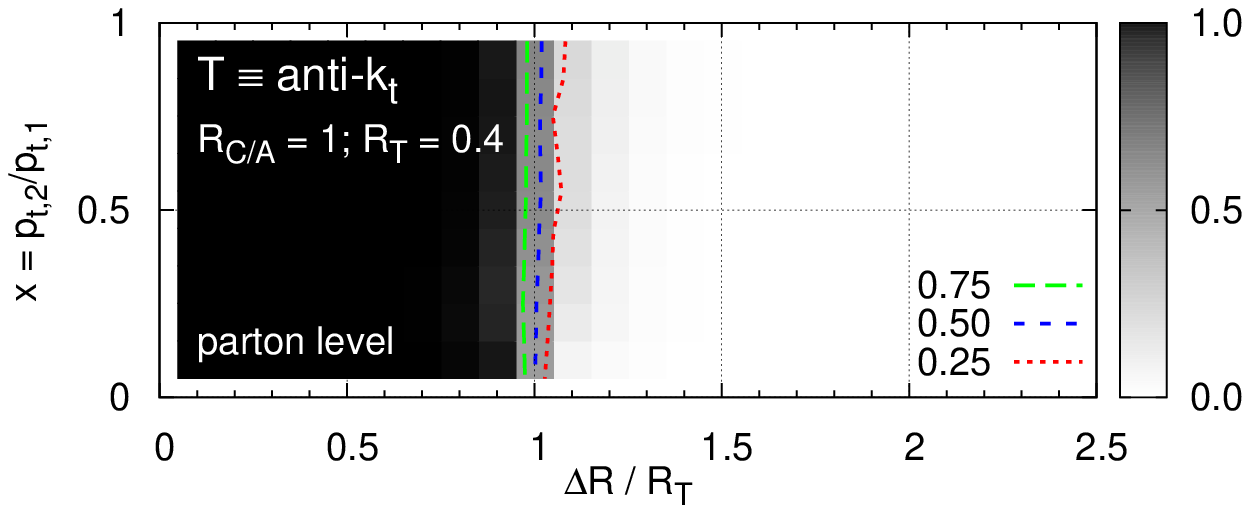}\hfill
  \includegraphics[width=0.48\tw]{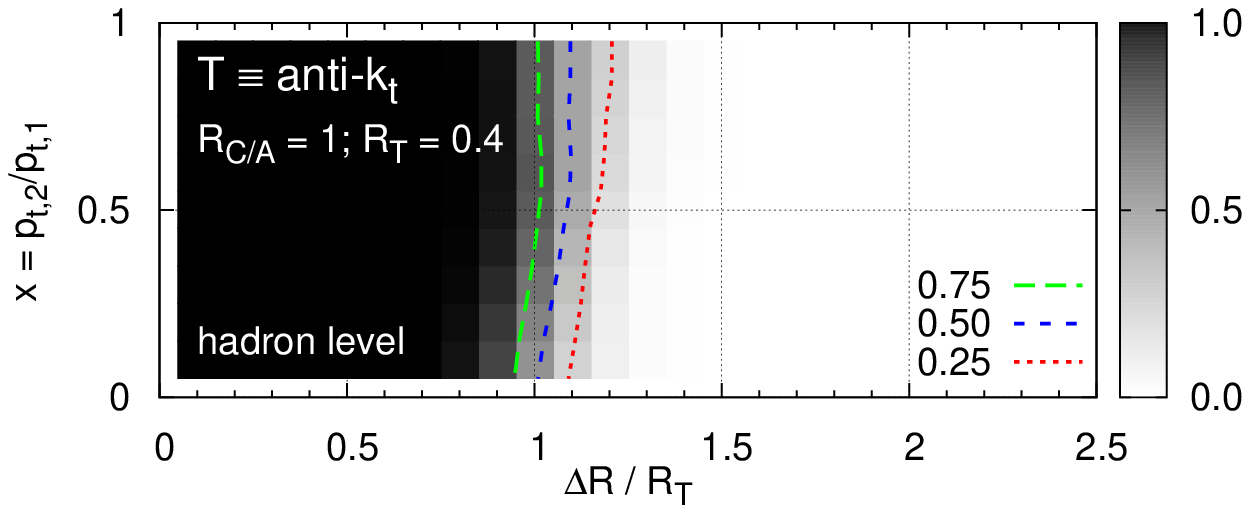}\medskip\\
  \includegraphics[width=0.48\tw]{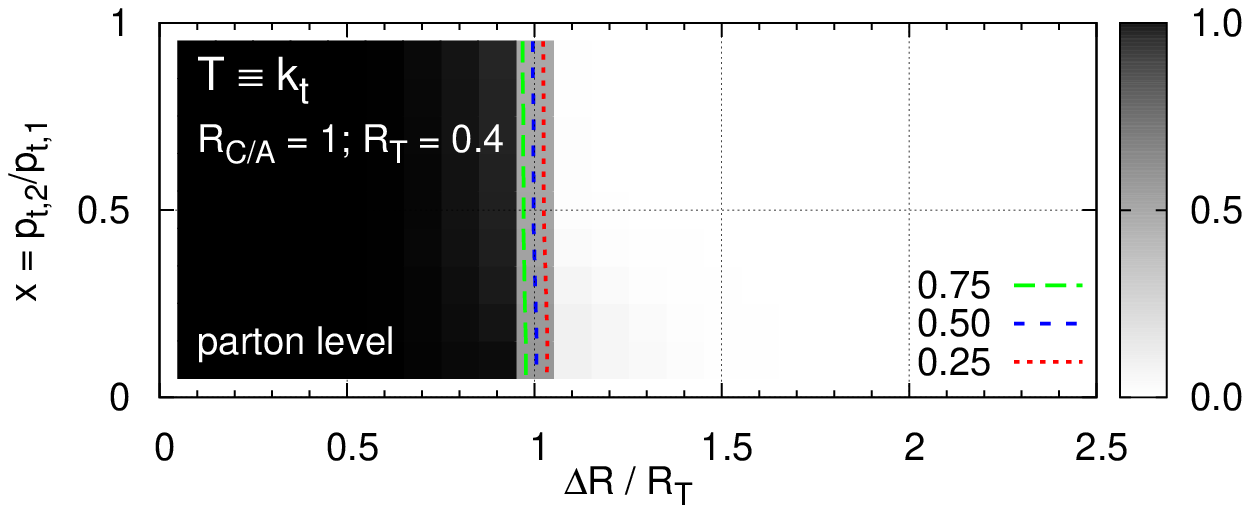}\hfill
  \includegraphics[width=0.48\tw]{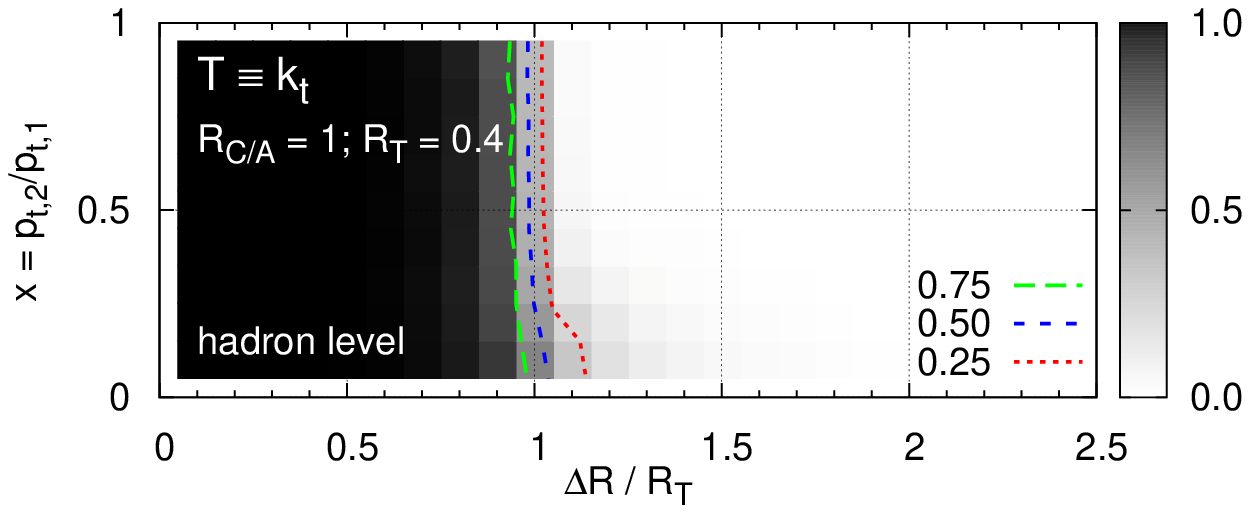}\medskip\\
  \includegraphics[width=0.48\tw]{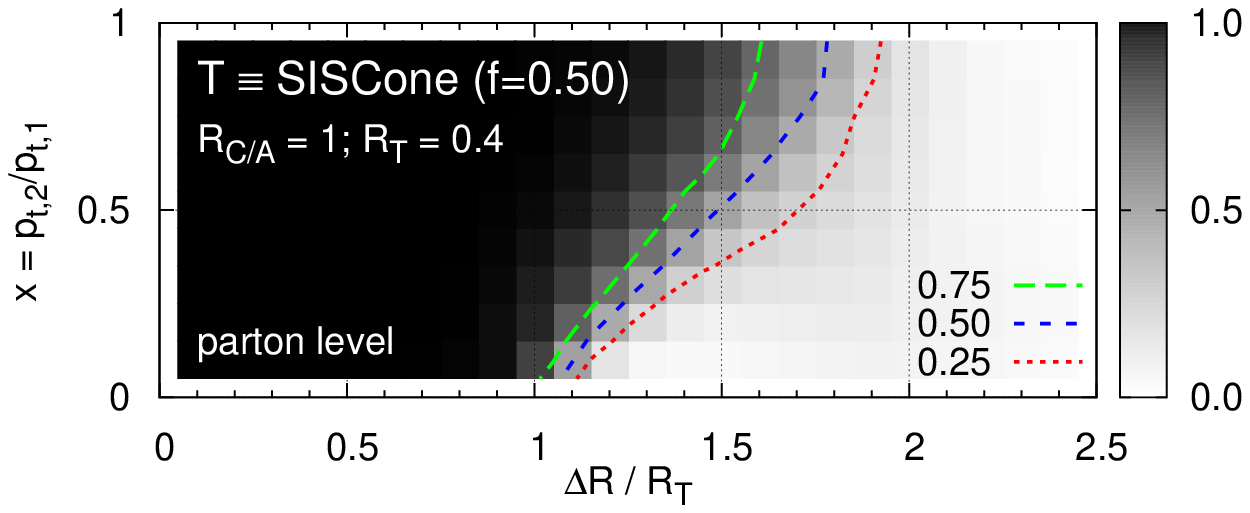}\hfill
  \includegraphics[width=0.48\tw]{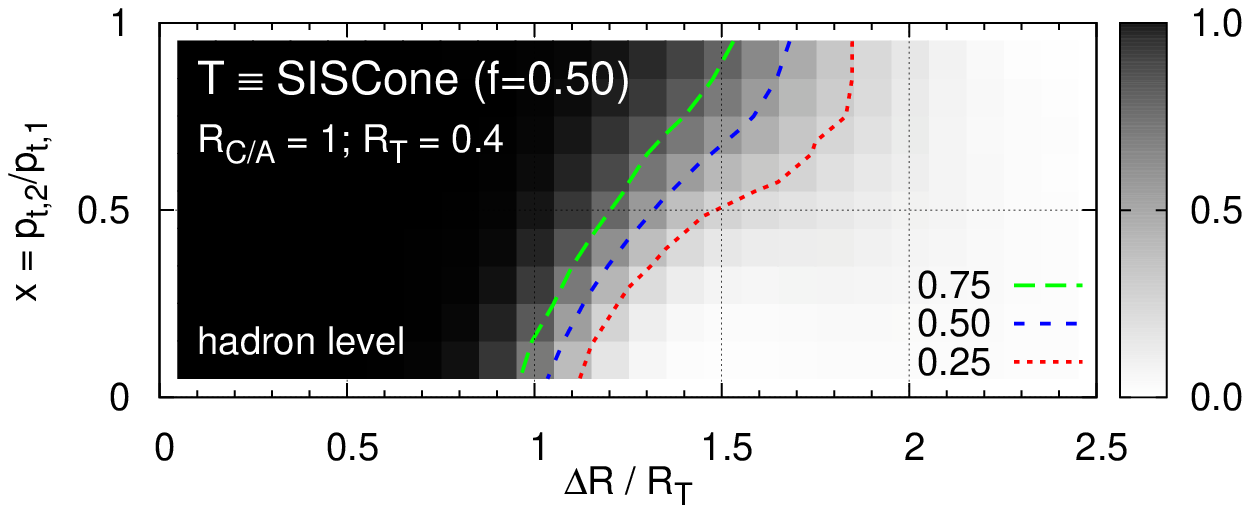}\medskip\\
  \includegraphics[width=0.48\tw]{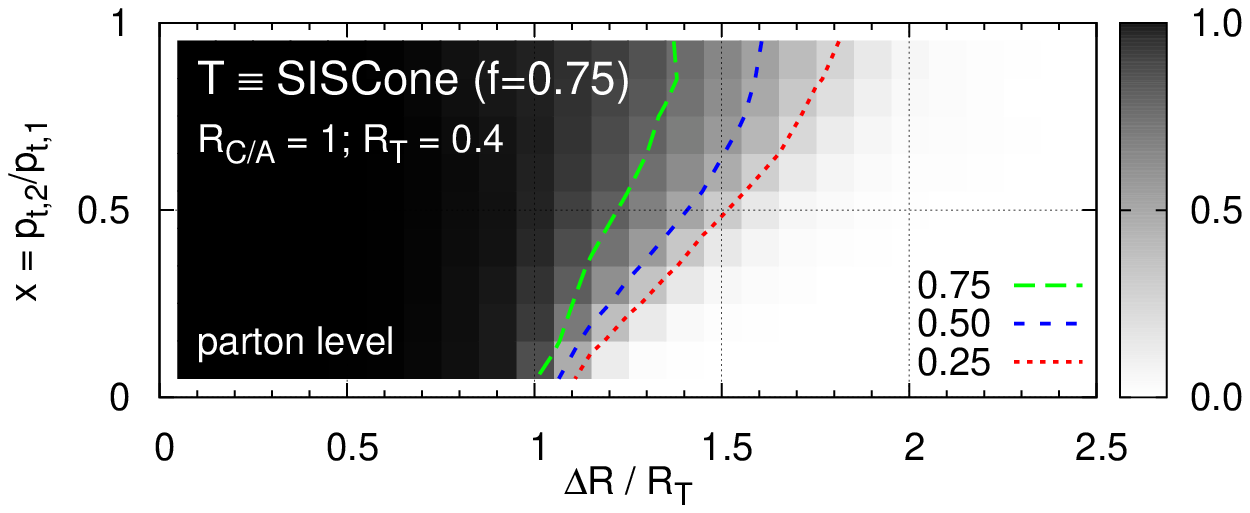}\hfill
  \includegraphics[width=0.48\tw]{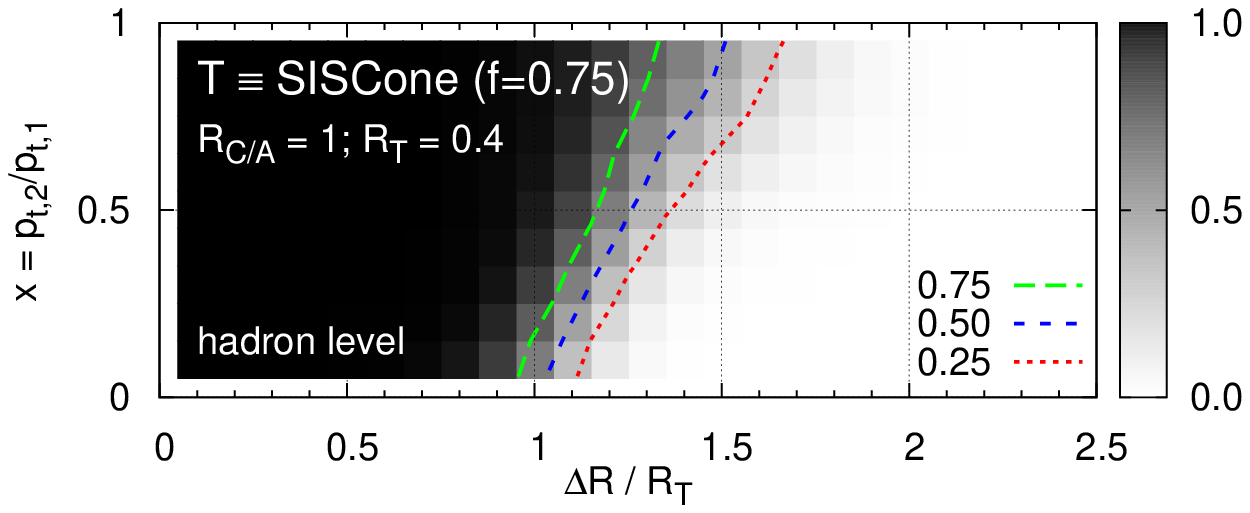}\medskip\\
  \includegraphics[width=0.48\tw]{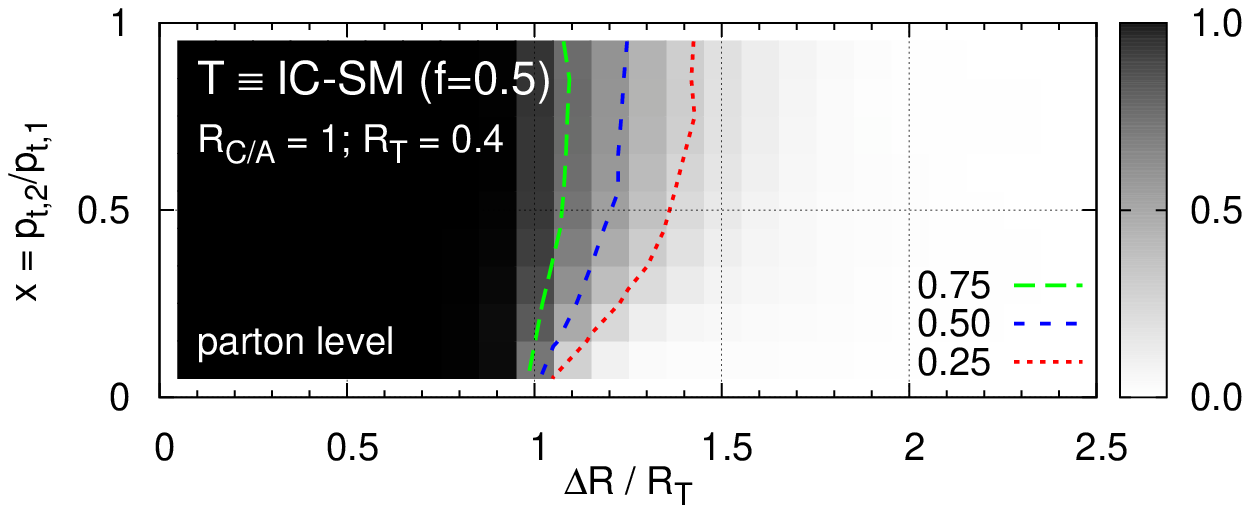}\hfill
  \includegraphics[width=0.48\tw]{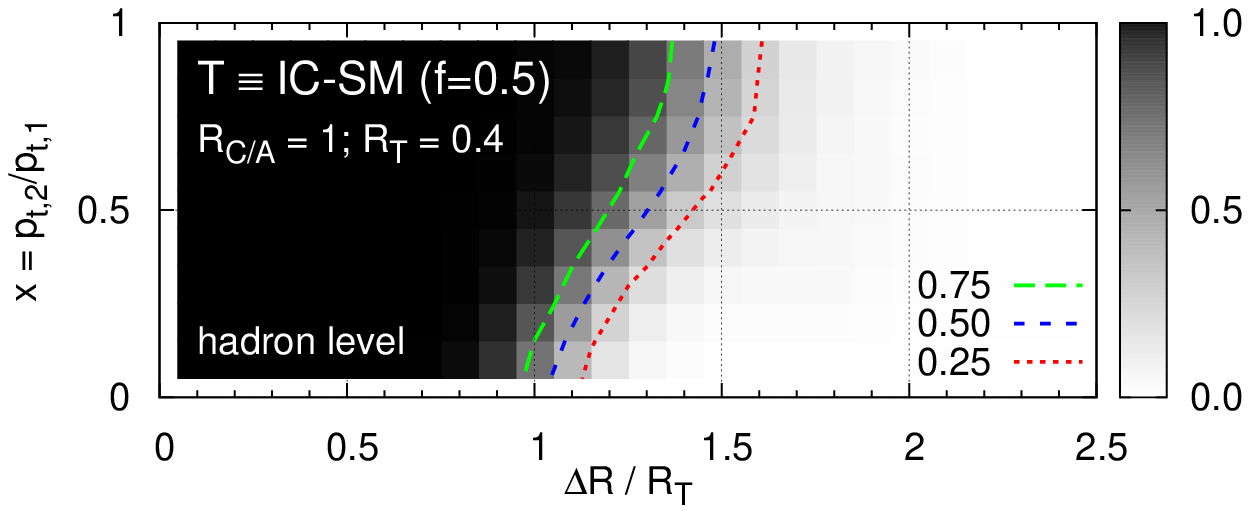}\medskip\\
  \includegraphics[width=0.48\tw]{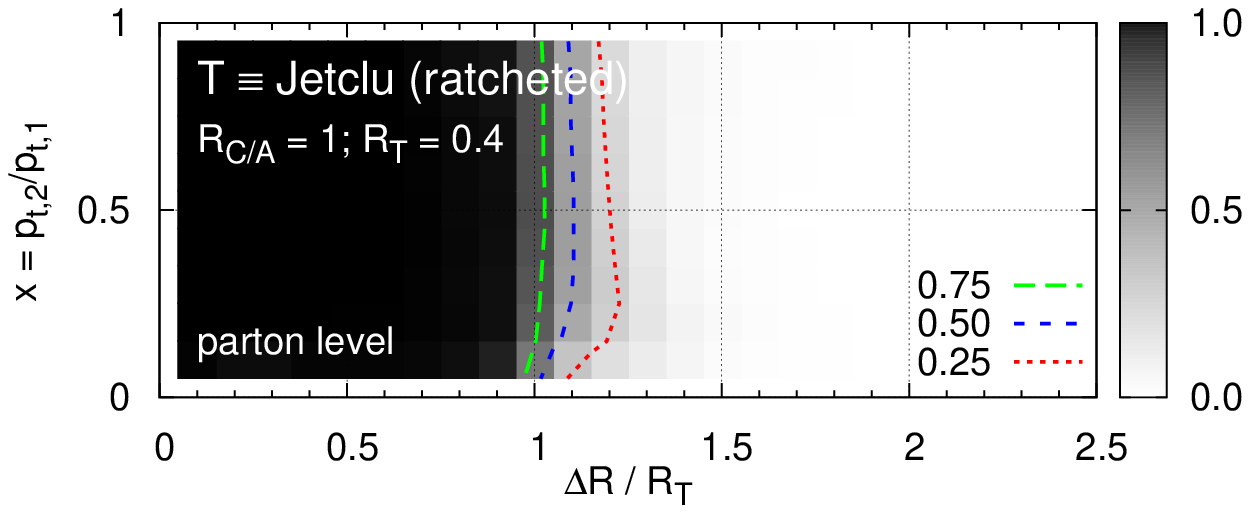}\hfill
  \includegraphics[width=0.48\tw]{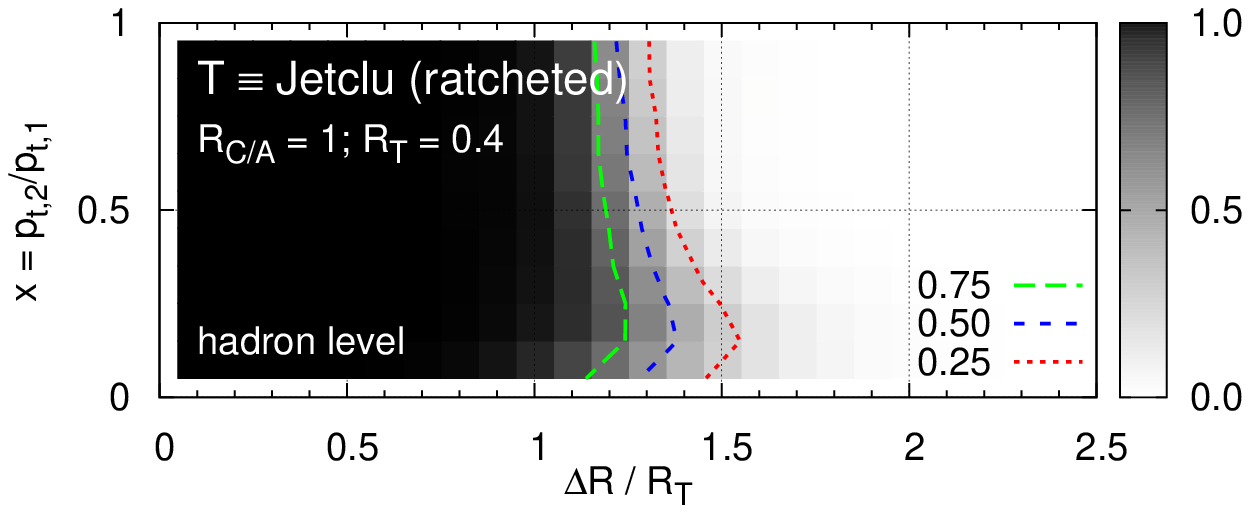}\medskip\\
  \caption{Shade/contour plot of the probability for two C/A subjets
    to each have at least 50\% (blue, short-dashed contour; 25\% and
    75\%: red dotted,
    green long-dashed contours) of their transverse momentum within a
    single test-algorithm (T) jet. Shown based on a sample of dijet
    events simulated with Herwig 6.5 at parton-level and at
    hadron-level (with Herwig's default soft UE), using all C/A jets
    with $p_t > 50\GeV$. 
    Carried out for Tevatron Run~II conditions, $p\bar p$ at $\sqrt{s}
    = 1.96\TeV$.  }
  \label{fig:merging-contours}
\end{figure}

Results are shown in fig.~\ref{fig:merging-contours}, for various
jet algorithms $T$, with  parton-level results on the left and hadron level
on the right. In the regions in black, the two C/A subjets always end
up in a common $T$-jet, while in the region in white this does not
occur.
For the IR-safe algorithms one sees rough agreement with the
expectations from
eqs.~(\ref{eq:kt-2particle-condition},\ref{eq:cone-2particle-condition}),
though for SISCone the boundary is quite broad and shifted to the left
of $\Delta R/R_T = 2$ at $x=1$. This is probably partly a consequence of the
showering and hadronisation, which limit the stability of
configurations in which the two subjets are near opposing edges of a
cone, as has been extensively discussed in \cite{EHT}. 
The dependence of the effect also on the split--merge overlap
threshold $f$ suggests that the split--merge dynamics have a
non-trivial impact as well.
Fig.~\ref{fig:merging-contours} also shows results two very IR unsafe
algorithms, a plain IC-SM variant (CDF's Midpoint algorithm with the
midpoint option turned off) and JetClu.
Among the relevant features, one notes the somewhat different shape
for the IC-SM algorithm at parton and hadron level, most visible if
one examines the contours (dashed lines), especially at higher x
values. This is a consequence of the IR unsafety. JetClu bears little
resemblance to the plain IC-SM algorithm, even though it is IC-SM
based. More detailed study reveals that this is only partially due to
its use of ratcheting.



\subsection{Perturbative properties, $p_t$ and mass}
\label{sec:pert-p_t-loss}

Gluon radiation is inevitable from fast-moving partons. How does it
affect the properties of a jet? Basically the gluon may be radiated
beyond the reach of the jet definition (``splash-out'') and thus
reduce the jet's energy compared to that of the parton. 
Alternatively it may be radiated within the reach of the
jet definition and then generate a mass for the jet (assuming a
4-vector-addition recombination scheme).
The aim of this section is to give some simple analytical
understanding of the effect of perturbative radiation on a jet's
transverse momentum and mass --- rules of thumb --- as well as
references to the literature for more detailed analyses.

For the reader who is interested principally in the results, the two
main ones can be summarised as follows. For small jet radii, $R\ll 1$,
the average fractional difference between a jet's transverse momentum
and that of the original parton is
\begin{equation}
  \frac{\langle p_{t,\jet} - p_{t,\text{parton}}  \rangle_{\rm pert}}{p_t} \simeq
  \left.
  \begin{array}{lc}
    \text{quarks:} & -0.43  \\
    \text{gluons:} & -1.02
  \end{array}\right\}
  \times \as \ln \frac1R + \order{\as}\,.
\end{equation}
where the $\order{\as}$ term depends both on the jet algorithm and the
global environment in which the parton is to be found (e.g.\ colour
connections to other partons) and is often ill-defined because of the
ambiguities in talking about partons in the first place.
Ignoring these important caveats, the above result implies that an
$R=0.4$ quark (gluon) jet has about $5\%$ ($11\%$) less momentum on
average that the original parton (for $\as=0.12$).

The second result is that the average squared jet mass for all
non-cone algorithms is
\begin{equation}
  \label{eq:jet-pert-mass}
  \langle M^2\rangle \simeq 
  \left.
  \begin{array}{lc}
    \text{quarks:} & 0.16  \\
    \text{gluons:} & 0.37
  \end{array}\right\}
  \times \as  p_t^2 R^2\,.
\end{equation}
For both the $p_t$ loss and the squared jet mass, SISCone results are
similar to $k_t$, anti-$k_t$ and C/A results when $R_\mathrm{SSICone}
\simeq 0.75 R_{k_t}$.

\subsubsection{Jet $p_t$}
\label{sec:jet-pt-Rdep}

In many uses of jets, one needs to know how a jet's energy (or $p_t$)
relates to the underlying hard scale of the process --- for example to
the mass of a decaying heavy particle (top quark, Higgs boson, new
particle), or to the momentum fraction carried by a scattered parton
in an inclusive jet cross section.

One approach to this is to take a Monte Carlo event generator, let it
shower a parton from some source and then compare the jet's $p_t$ to
that of the parton. This often gives a reasonable estimate of what's
happened, even if the Monte Carlo basically acts as a black box, and
brings a somewhat arbitrary definition of what is meant by the initial
``parton'' (or of the mass of the top quark).

Another approach is to take a program for carrying out NLO
predictions, like MCFM~\cite{MCFM} or NLOJET++~\cite{NLOJet}, and for
example determine the relation between the jet $p_t$-spectrum and the
parton distribution functions.
NLO calculations are perhaps even blacker boxes than Monte Carlo
generators, on the other hand they do have the advantage of giving predictions of
well-defined precision; however, one loses all relation to the
intermediate (ill-defined) ``parton'' (this holds also for tools like
MC@NLO~\cite{Frixione:2002ik} and POWHEG~\cite{Nason:2006hfa}).

Some insight can be obtained from analytical NLO calculations of jet
cross sections, such as
\cite{Furman:1981kf,Aversa:1988vb,Guillet:1990ez,Jager:2004jh,deFlorian:2007fv}. A
feature that is common to them is that at the first non-trivial order,
cross sections acquire a $\ln R$ dependence in the small-$R$ limit.
The small-$R$ limit is one case where one \emph{can} say something
meaningful the relation between a jet's $p_t$ and that of the original
parton (another is the threshold limit, for example
\cite{Kidonakis:1997gm,Kidonakis:1998bk,Kidonakis:1998nf,Kidonakis:2000gi,deFlorian:2007fv}),
because the emitting parton decouples from its environment, a
consequence of angular ordering.
Working in a collinear approximation and considering an initial quark,
with a gluon emission matrix element proportional to the real
$P_{qq}(z)$ splitting function ($P_{qq}(z) = \CF(1+z^2)/(1-z)$), one
can simply write the average difference $\delta p_t = p_{t,\jet} -
p_{t,\quark}$ as
\begin{equation}
  \label{eq:deltapt_pert_smallR}
  \langle \delta p_t \rangle_{\mathrm{pert}} = \int \frac{d \theta^2} 
  {\theta^2} \int dz\, \underbrace{p_t \big( \max[z, 1 - z] - 1
    \big)}_{\delta p_t} \,
  \frac{\as \big(\theta \, (1 - z) \, p_t \big)}{2 \pi} \, P_{qq} (z) \,
  \Theta \big( \theta - f_\alg(z) R \big) \, ,
\end{equation}
where one integrates over the angle $\theta$ between the quark and an
emitted gluon and over the momentum fraction $z$ that is kept by the
quark, weighting the matrix element with the loss of momentum from the
leading jet, $p_t(\max[z, 1 - z] - 1)$, when the gluon and quark form two
separate jets, $\theta > f_\alg(z) R$ (throughout this section, $\theta$
is to be understood as a boost-invariant angle, $\theta \equiv \Delta R_{qg}$).
The quantity $f_\alg(z)$ reflects the algorithm's reach, cf.\
eqs.~(\ref{eq:kt-2particle-condition},\ref{eq:cone-2particle-condition})
and is given by
\begin{equation}
  \label{eq:fz}
  f(z) =
  \begin{cases}
    1 & \mbox{$k_t$, C/A, anti-$k_t$}\\
    1 + \min(\frac{z}{1-z},\frac{1-z}{z}) & \mbox{SISCone}
  \end{cases}
\end{equation}
Carrying out the integration in a fixed-coupling approximation gives
\begin{equation}
  \label{eq:deltapt_pert_smallR_res}
  \frac{\langle \delta p_t \rangle_{\rm pert}}{p_t} = 
  \frac{\as}{\pi} \, L_i \,  \ln R + \order{\as},\qquad\quad
  R \ll 1\,,
\end{equation}
with $L_i$ a coefficient that depends on whether it is a quark or a
gluon that is the initiating parton (cf.\ \cite{Dasgupta:2007wa}):
\begin{subequations}
  \label{eq:Lqg-res}
  \begin{align}
    L_q &= 
    \CF \left(2 \ln 2 - \frac{3}{8} \right) 
    \simeq 1.01 \CF
    \, ,\\
    L_g &= 
    \CA \left( 2 \ln 2 - \frac{43}{96} \right) + \nf \, \TR \, \frac{7}{48} 
    \simeq 0.94 \CA + 0.15 \TR \nf
    \, ,
  \end{align}
\end{subequations}
One notes that for small $R$ the result of
eq.~(\ref{eq:deltapt_pert_smallR_res}) is negative. 
The unspecified pure $\order{\as}$ term 
reflects the result's
dependence on the large-angle environment. It can be defined
unambiguously only in the threshold limit.
Neglecting it, one comes to the conclusion that with
$R=0.4$, a quark-induced jet has, on average, a $p_t$ that is about
$4-5\%$ smaller than the initiating parton, while a gluon jet's $p_t$
is $8-10\%$ smaller.

One can also evaluate the small-$R$ limit of the average difference
between the $p_t$ of a SISCone jet and (say) a $k_t$ jet (again
following~\cite{Dasgupta:2007wa})
\begin{equation}
  \label{eq:dkt-cone}
  \frac{\langle \delta p_t^{\SISCone} \rangle_{\rm pert} - 
    \langle \delta p_t^{k_t} \rangle_{\rm pert}}{p_t} = 
  \frac{\as}{\pi} \, K_i
\end{equation}
with
\begin{subequations}
  \begin{align}
    \label{eq:Kqg}
    K_q &=   \left(- \frac{15}{16} + \frac{9}{8} \ln 2 + \ln^2{2} \right) \CF 
    \simeq 0.323 \CF \, ,\\
    K_g &= 
    \left(- \frac{1321}{1152} + \frac{133}{96} \ln 2 + \ln^2 2 \right) \CA
    + \left(\frac{241}{576} - \frac{25}{48} \ln 2 \right) \nf \TR
    \simeq 0.294 \CA + 0.057 \nf \TR  \, .
  \end{align}
\end{subequations}
Numerically, $K_i \sim 0.3 L_i $, or equivalently the average
behaviour of SISCone and the $k_t$ (and related) algorithms are similar
perturbatively when $\ln R_{k_t} \simeq \, 0.3 + \ln R_\SISCone$, that
is $R_{k_t} \simeq 1.35 \, R_{\SISCone}$. This feature was originally
observed for a generic cone algorithm in~\cite{Kt-EllisSoper}.

\subsubsection{Jet mass}
\label{sec:jet-mass-Rdep}

Partons (except for heavy quarks) are essentially massless. Jets,
in particular those with significant substructure, are not.
Jet masses are interesting in part because hadronic decays of very
high-$p_t$ top quarks and electroweak bosons will be collimated by the
Lorentz boost factor and so form a single jet, whose invariant mass
might provide a means to identify the origin of the jet (cf.\
section~\ref{sec:substructure}).

The simplest quantity to examine is the mean squared invariant mass of
a jet. This was studied in a hadron-collider context
in~\cite{Ellis:2007ib} and it was pointed out that to first
non-trivial order,
\begin{equation}
  \label{eq:jet-pert-mass}
  \langle M^2\rangle \simeq C\cdot \frac{\as}{\pi} p_t^2 R^2\,,
\end{equation}
where $C$ is a coefficient that depends on the
relative fraction of quarks and gluons and on the type of jet
algorithm.\footnote{Results for jet masses at $\order{\as}$ are
  sometimes quoted as being NLO. It would be more accurate to state
  that they are LO results, since $\order{\as}$ is the first order at
  which the mass is non-zero. True NLO results would go up to
  $\order{\as^2}$. Jet masses can be calculated to NLO in dijet events
  using the 3-jet NLO component of a program like
  NLOJET++~\cite{NLOJet}.}
This is easily derived in the small-$R$ limit, e.g.\ for a
quark-induced jet:
\begin{equation}
  \label{eq:mass_pert_smallR}
  \langle M^2 \rangle_{\mathrm{pert}} \simeq  \int \frac{d \theta^2} 
  {\theta^2} \int dz \cdot  \underbrace{p_t^2 z(1-z) \theta^2}_{M^2} \cdot \,
  \frac{\as \big(\theta \,(1 - z) \, p_t \big)}{2 \pi} \, P_{qq} (z) \,
  \Theta \big( f_\alg(z) R - \theta \big) \, ,
\end{equation}
In a fixed-coupling approximation, the results can be summarised as
\begin{align}
  \label{eq:kt-mass-coeffs}
  C_q^{k_t} = \frac{3}{8}\CF\qquad C_g^{k_t} = \frac{7}{20}\CA + \frac{1}{20}\nf\TR\,,
\end{align}
for $k_t$-like algorithms and 
\begin{subequations}
  \label{eq:cone-mass-coeffs}
  \begin{align}
    C_q^{\mathrm{SIS}} &= \CF\left( \frac74 - \frac32 \ln 2
    \right) \simeq 0.71\CF\,, \\
    C_g^{\mathrm{SIS}} &= \CA\left( \frac{49}{24}-2\ln 2\right) +
    \nf\TR\left(\ln 2 - \frac7{12}\right)\simeq 0.66\CA +
    0.11\nf\TR\,,
  \end{align}
\end{subequations}
for SISCone type algorithms (consistent with the observation that
$k_t$ and SISCone type algorithms behave similarly for $R_{k_t} \simeq
1.35 \, R_{\SISCone}$).
These results coincide roughly with the rule of thumb given
in~\cite{Ellis:2007ib} that to within $25\%$, $\sqrt{\langle M^2
  \rangle} \simeq 0.2 R p_t$, with the exact value depending on the
mix of quarks and gluons, and subject also to finite-$R$ effects as
well as threshold modifications for high jet transverse momenta.
Ref.~\cite{Ellis:2007ib} also emphasises that
eqs.~(\ref{eq:cone-mass-coeffs}) will be subject to significant
higher-order corrections, associated with the fact that SISCone's
effective clustering reach is somewhat smaller for $z\simeq 1/2$ than
the two-parton reach $f(z=1/2)\simeq 2$, cf.\
fig.~\ref{fig:merging-contours}.
This is a point of some importance, and one should never lose sight of
the fact that \emph{all} the results given above are based on LO
perturbation theory and can be quite noticeably affected by
higher-order terms.

When using jet masses for tagging hadronically-decaying boosted heavy
objects, it is also of interest to know the distribution of the jet
mass. At leading order, $d\sigma/dM^2$ diverges with a logarithmic
enhancement $\sim \frac{\as}{M^2} \ln \frac{R^2 p_t^2}{M^2}$ for small
masses (cf.~the analytical result in \cite{Almeida:2008yp}). 
Higher order terms are enhanced by further powers of $\ln R p_t/M$ and
can in principle be resummed. 
Analytical results exist however only for certain specific cases in
\ee~\cite{Catani:1992ua,Burby:2001uz,Dasgupta:2001sh} and
DIS~\cite{Dasgupta:2002dc} and have not been extended to
hadron-collider jets, in part because of issues such as the
non-trivial process dependence~\cite{Oderda:1998en,Berger:2001ns} and
jet-algorithm dependence~\cite{Appleby:2002ke,Delenda:2006nf} of soft
logarithms associated with delimited (``non-global''
\cite{Dasgupta:2001sh,Dasgupta:2002bw,Banfi:2002hw}) regions of
phase-space.

\subsubsection{Other properties}
\label{sec:other-pt-jet-properties}

Many other properties of jets can be predicted perturbatively. Among
them one may mention the scale associated with subjets within a jet
\cite{Kt,JetsDisSchmell,NumSum}, multiplicities of subjets
\cite{Catani:1993yx,Forshaw:1999iv} and of particles (see e.g.\
\cite{Ochs:2008vg} and references therein) and jet
shapes~\cite{SeymourJetShapes} (radial moments and the fraction of
energy that is within a certain central core of the jet).
As well as providing important handles on our understanding of the QCD
dynamics within jets, these observables can also be useful for example
in discriminating quark and gluon jets. One such application is given
in~\cite{Chekanov:2004kz}.

\subsection{Hadronisation}
\label{sec:hadronisation}

The properties of jets are affected not just by perturbative
radiation, but also by low-$p_t$, non-perturbative effects.
It is useful to divide such effects into two classes: hadronisation
and the underlying event.
Hadronisation corresponds to the transition between partons and
hadrons, and occurs in all high-energy QCD processes ($\ee$, DIS and
$pp$).
The underlying event (UE) consists of the multiple low-$p_t$
interactions that occur between the two hadron remnants in a $pp$ or
a resolved $\gamma p$ collision.
Physically, in a $pp$ collision, hadronisation and the UE cannot be
unambiguously separated (the question of what hadronises depends on
what has interacted). Nevertheless it is useful to consider them
separately, because they affect jets in rather different ways.
Hadronisation is discussed here, and the UE in section~\ref{sec:jet-areas}.

With current techniques, the impact of hadronisation cannot be
calculated (or even easily defined) from first principles. However, in
the mid 1990's, methods were developed
\cite{Korchemsky:1994is,Dokshitzer:1995zt,Akhoury:1995sp,Ball:1995ni,Dokshitzer:1995qm,Gardi:2001di}
(reviewed in \cite{Beneke:1998ui}) that allowed one to predict the
main features of hadronisation, based on ambiguities that arise in
perturbative calculations related to the Landau pole.

A somewhat oversimplified statement of the idea is that if a perturbative
calculation involves an integral over $\as(\mu)$, then one can
estimate the size of the non-perturbative contribution by replacing
$\as(\mu)$ with a purely non-perturbative piece $\as^{\NP}(\mu) =
\Lambda \delta(\mu - \Lambda)$ where $\Lambda$ is commensurate with
the Landau scale. 
So, for example, to estimate the non-perturbative correction to a quark
jet's transverse momentum in the small-$R$ limit one takes
eq.~(\ref{eq:deltapt_pert_smallR}) and writes
\begin{subequations}
  \label{eq:deltapt_NP_smallR}
  \begin{align}
    \langle \delta p_t \rangle_{\NP} &\sim p_t \int \frac{d \theta^2}
    {\theta^2} \int dz \big( \max[z, 1 - z] - 1 \big) \,
    \frac{\as^{\NP} \big(\theta \, (1 - z) \, p_t \big)}{2 \pi} \,
    P_{qq} (z) \,
    \Theta \big( \theta - f_\alg(z) R \big) \, ,\\
    &\sim p_t \int_{R^2} \frac{d \theta^2} {\theta^2} \int dz \big( z
    - 1 \big) \, \frac{\as^{\NP} \big(\theta \, (1 - z) \, p_t
      \big)}{2 \pi} \, \frac{2\CF}{1-z} \,
    ,\smallskip\\
    &\sim 
    -\frac{2\CF \Lambda}{\pi R}\,,\label{eq:small-R-hadronisation}
  \end{align}
\end{subequations}
where in the second line one makes use of the knowledge that the
$\delta$-function will select $1-z = \Lambda/(\theta p_t) \ll R$.
For gluon jets the result is the same except for the replacement $\CF
\to \CA$.
A crucial idea in calculations such as
eqs.~(\ref{eq:deltapt_NP_smallR}) is that one can apply the same
procedure to a wide range of observables and the same value of
$\Lambda$ should hold for each.\footnote{As long as they all share the
  same $p_t$-dependence in the infrared --- a less oversimplified
  formulation of the idea in eq.~(\ref{eq:deltapt_NP_smallR}) is that
  observables with the same IR $p_t$-dependence are all sensitive to a
  common moment of the coupling in the infrared.}
This is known as ``universality''.
Universality has been investigated in some detail for event shapes in
$\ee$ and DIS collisions and there is some debate as to just how well
it works (e.g.\ \cite{Aktas:2005tz} as compared to
\cite{Chekanov:2006hv}). However for the purpose of understanding the
essentials  of the hadronisation of jets it is probably an adequate
assumption, and one can take $\Lambda \simeq 0.6\GeV$.

\begin{figure}
  \centering
  \includegraphics[angle=270,scale=0.6]{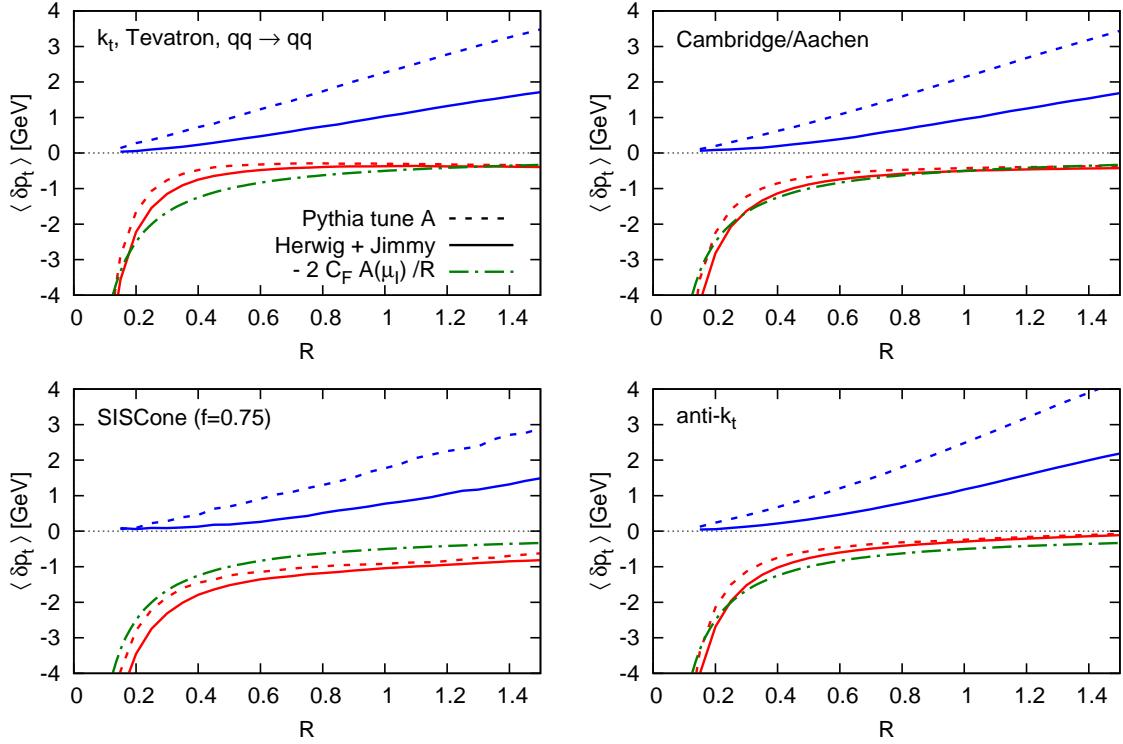}
  \caption{Modification of the $p_t$ of jets due to the underlying
    event (upper curves in each plot) and hadronisation (lower curves), for $qq \to
    qq$ scattering at the Tevatron Run II ($p \bar p$, $\sqrt{s} =
    1.96 \TeV$), comparing {\tt Pythia} 6.412~\cite{Sjostrand:2006za}
    (tune A, dashed lines) and {\tt Herwig} 6.510~\cite{Herwig} with {\tt
      Jimmy} 4.3~\cite{Butterworth:1996zw} (solid lines).  In the case of
    hadronisation, the Monte Carlo outputs are compared to the $1/R$
    part of the analytical result,
    eq.~(\ref{eq:qq2qq-hadronisation}) (dot-dashed lines). Dijet events are selected
    containing an underlying $qq \to qq$ scattering, and with the
    requirement that at parton-shower level the hardest jet has $55
    \GeV < p_{t} < 70 \GeV$. The non-perturbative corrections shown
    correspond to the average for the two hardest jets. Taken
    from~\cite{Dasgupta:2007wa}.}
  \label{fig:4stamp}
\end{figure}

The basic result that hadronisation removes transverse-momentum
$\order{\Lambda/R}$ from a jet was presented
in~\cite{Korchemsky:1994is} (and could be deduced from the results of
\cite{SeymourJetShapes}; hadronisation as a shift in $p_t$ was also
discussed in \cite{Mangano:1999sz}).
Aside from the
quark/gluon jet difference it is a process-independent result, as long
as $R$ is much smaller than the angle between jets.
It seems, however, that this result had largely been forgotten until the advent
of a more recent calculation~\cite{Dasgupta:2007wa}, which goes beyond
the small $R$ limit (in a threshold approximation for dijet
production).
As an example, the result for the case of the ${qq' \to qq'}$
subprocess of dijet production is
\begin{equation}
  \label{eq:qq2qq-hadronisation}
  \langle \delta p_t \rangle_{\NP}^{qq' \to qq'} = 
  \frac{\Lambda}{\pi}
  \left [- \frac{2}{R}
    C_F + \frac{1}{8} R \left( 5 C_F - \frac{9}{N_c} \right) 
    + {\cal O} \left( R^2 \right) \right]~.
\end{equation}
A feature of eq.~(\ref{eq:qq2qq-hadronisation}) is that the first
correction to the $1/R$ term is fairly small even for $R=1$ (less than
$20\%$). Consequently for most purposes it is adequate to take just
the $1/R$ piece. This is what is done in fig.~\ref{fig:4stamp}, whose
lower set of curves in each quadrant compares the hadronisation
correction as deduced from Herwig (solid lines) and from Pythia (dashed lines)
with the $1/R$ part of the analytical expectation given above
(dot-dashed lines).
Generally speaking the agreement is good, even in the large-$R$ region
where the $1/R$ approximation might be expected to break down.

To obtain a closer relation to studies of hadronisation for $\ee$ and
DIS event shapes (for a review see e.g.\ \cite{Dasgupta:2003iq}), one
may replace
\begin{equation}
  \label{eq:Lambda-to-Milan-A}
  \frac{\Lambda}{\pi} \to \frac{2 }{\pi} \cM \cA(\mu_I)\,.
\end{equation}
Here ${\cal A}(\mu_I)$ is defined as the integral over a
non-perturbative contribution to $\as$, $\delta \as$, up to some
infrared matching scale $\mu_I$ (usually $2\GeV$), ${\cal A}(\mu_I) =
\frac{1}{\pi} \int_0^{\mu_I} d \kappa_t \, \delta\alpha_s \left(
  \kappa_t \right)$. Following~\cite{Dokshitzer:1995zt}, it is often
written in terms of yet another parameter $\alpha_0(\mu_I)$, the
integral of the full coupling in the infrared $\alpha_0(\mu_I) =
\int_0^{\mu_I} d\kappa \as(\kappa)$, via
\begin{equation}
  \label{eq:ev-shp-alpha0}
  {\cal A}(\mu_I) = \frac{1}{\pi} \, \mu_I \left [ \alpha_0 \left( \mu_I 
    \right) - \alpha_s (p_t) - \frac{\beta_0}{2 \pi} \left(\ln \frac{p_t}{\mu_I} +  
      \frac{K}{\beta_0} + 1 \right) \alpha_s^2 (p_t) \right] \, ,
\end{equation}
where $K = C_A \left( \frac{67}{18} - \frac{\pi^2}{6} \right) -
\frac{5}{9} n_f$, and the subtracted terms in
eq.~(\ref{eq:ev-shp-alpha0}) remove double counting with
contributions already included in NLO calculations.
Fits to data in DIS and $\ee$ usually give $\alpha_0(2\GeV)\simeq
0.5$.  

The factor $\cM$ in eq.~(\ref{eq:Lambda-to-Milan-A}) is known as the Milan
factor~\cite{Dokshitzer:1997iz,Dokshitzer:1998pt,Dasgupta:1998xt,Dasgupta:1999mb,Smye:2001gq}.
It accounts for the corrections that arise when one considers two
non-perturbative ``gluons'' rather than a single one. 
For all known event shapes, $\cM$ has been calculated to be $\cM=1.49$
--- this ``universality'' of the Milan factor is due to the fact that
event-shape observables are effectively linear in the momenta of soft
gluons~\cite{Dokshitzer:1998pt}.
For jets it had been argued that only the anti-$k_t$ algorithm
satisfied this linearity property~\cite{Cacciari:2008gp}. This was
recently confirmed in an explicit calculation by Dasgupta and
Delenda~\cite{Dasgupta:2009tm}, who showed that the $k_t$ algorithm
instead has $\cM_{k_t} = 1.01$. 
This smaller value is consistent with the somewhat reduced
hadronisation corrections observed for the $k_t$ algorithm compared to
anti-$k_t$ in fig.~\ref{fig:4stamp}, though a detailed quantitative
comparison has not yet been performed.
Future calculations of the Milan factor for C/A and SISCone will
hopefully also fit in with the pattern of slight differences that are
observed in fig.~\ref{fig:4stamp} with respect to the
algorithm-independent behaviour that is given by
eqs.~(\ref{eq:small-R-hadronisation},\ref{eq:qq2qq-hadronisation}).

A point emphasised in~\cite{Mangano:1999sz} is that even if the
non-perturbative modification of a jet's $p_t$ is rather modest,
$\order{1\GeV}$, it can nevertheless have a significant impact on
steeply falling cross sections. Given a jet $p_t$ spectrum that falls
as $p_{t}^{-n}$, the full result for the jet spectrum can be expressed
in terms of the perturbative spectrum and the non-perturbative shift
as
\begin{equation}
  \label{eq:shift-v-effect-on-pt-spect}
  \frac{d\sigma_\mathrm{full}}{dp_t}(p_t) \to \frac{d\sigma_{\mathrm{PT}}}{dp_t}(p_t - \langle
  \delta p_t\rangle_{\NP}) \simeq \frac{d\sigma}{dp_t}(p_t)\cdot \left(1 - n\frac{\langle
      \delta p_t\rangle_{\NP}}{p_t}\right)\,.
\end{equation}
Thus for typical values of $n$ in an inclusive-jet spectrum, a $2\%$
change in $p_t$ can lead to a $10-15\%$ change in the cross section
(this observation holds also for $p_t$ shifts due to the underlying
event, which are discussed below).
These are the order of magnitudes often seen in experiments' Monte
Carlo studies of hadronisation, whose results also cast light on the
$R$-dependence of non-perturbative effects \cite{Abulencia:2007ez} and
on the differences between jet
algorithms~\cite{Aachen,ZEUS-akt-sis-prelim,Adloff:2003nr,D0kt}.

A final point is that the above methods can also be used to calculate
the non-perturbative corrections to the squared jet mass,
\begin{equation}
  \label{eq:squared-jet-mass-hadr}
  \langle \delta M^2 \rangle_{\NP} \simeq \frac{2\CF}{\pi} \Lambda p_t
  \left(R + \order{R^3}\right) 
  \equiv \frac{4\CF}{\pi} \cM {\cal A}(\mu_I)\, p_t \left(R + \order{R^3}\right)\,,
\end{equation}
where the $R^3$ terms have small coefficients~\cite{Dasgupta:2007wa}.
Note that for jet algorithms other than anti-$k_t$, the Milan factors
for $\langle \delta p_t \rangle_{\NP}$ and $\langle \delta M^2 \rangle_{\NP}$
will not be the same.

\subsection{UE, pileup, jet areas}
\label{sec:jet-areas}

While the process of hadronisation may well be reasonably universal
between $\ee$, DIS and $pp$ collisions, the latter have the additional
feature of the ``underlying event'' (UE), which can be thought of as
the semi- or non-perturbative interactions that occur between hadron
remnants in a $pp$ collision.
Our understanding of the UE is
somewhat less developed than that of hadronisation.
One way that one can model it is by saying that it induces an extra
amount of transverse momentum per unit rapidity,
$\Lambda_{UE}$.\footnote{Later we will talk of transverse momentum
  $\rho$ per unit area on the rapidity-azimuth plane; $\Lambda_{UE}
  =2\pi\rho_{UE}$.} %
In this case a jet should receive a position contribution to its $p_t$
from the UE that is proportional to the region of the rapidity-azimuth
region that it covers, \ie $\sim R^2$:
\begin{equation}
  \label{eq:delta-pt-UE}
  \langle \delta p_t \rangle_{\UE} \simeq \Lambda_{\UE} R J_1(R) = 
  \Lambda_{\UE} \left(\frac{R^2}{2} - \frac{R^4}{8} + \ldots\right)\,.
\end{equation}
where terms at $R^4$ and beyond~\cite{Dasgupta:2007wa} hold for the a
$4$-vector ($E$) recombination scheme.
The corresponding formula for the change to the squared jet mass is 
\begin{equation}
  \label{eq:delta-m2-UE}
  \langle \delta M^2 \rangle_{\UE} \simeq 
  \Lambda_{\UE}\, p_t \left(\frac{R^4}{4} + \frac{R^8}{4608} + \ldots\right)\,.
\end{equation}

In fig.~\ref{fig:4stamp}, the upper curves represent the UE
contributions to the $p_t$ of Tevatron jets, as determined from the UE models in
Pythia~\cite{Sjostrand:2006za} (tune A~\cite{Albrow:2006rt}) and
Jimmy~\cite{Butterworth:1996zw} (with the ATLAS
tune~\cite{Albrow:2006rt}).
Three features are worth commenting on: (a) the curves agree with a
rough $R^2$ dependence, (b) the two models disagree by a factor of two
even though they have both been tuned to Tevatron data and (c) the value
that one extracts for $\Lambda_\UE$, in the range $2-4\GeV$, is quite
a bit larger than the $p_t$ per unit rapidity that would be generated
by normal hadronisation for a quark or gluon dipole stretched between
the two beams (respectively $0.5\GeV$ and $1\GeV$).

\begin{figure}
  \centering
  \includegraphics[angle=-90,scale=0.6]{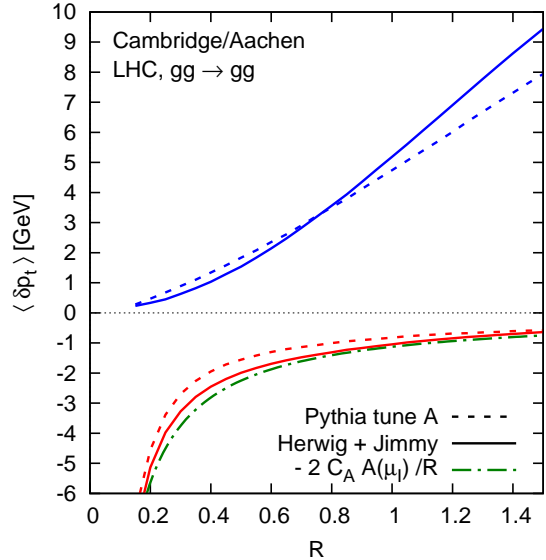}
  \caption{Similar to Fig.~\ref{fig:4stamp}, but for just one
    algorithm, at the LHC ($pp$, $\sqrt{s}=14\TeV$) rather than the
    Tevatron, and for $gg\to gg$ collisions (rather than $qq\to
    qq$). Taken from~\cite{Dasgupta:2007wa}.}
  \label{fig:lhc-gg}
\end{figure}

For the LHC, the models predict an even larger contribution from the
UE, cf.\ fig.~\ref{fig:lhc-gg} from which one deduces $\Lambda_{\UE}
\sim 10\GeV$, and for a large range of $R$ it dominates over
hadronisation. Furthermore pileup (multiple $pp$ collisions in a given
beam crossing) is expected to add up to an extra $100\GeV$ of soft
``junk'' per unit rapidity.

All of this implies that it is important to understand in more detail
how jets are affected by ``low-$p_t$'' noise that is roughly uniformly
distributed in rapidity. 
Two things can happen:
firstly, the soft junk can end up in the jet --- to study this it is
useful to refer to the ``jet area'' \cite{Cacciari:2008gn}, a measure
of the extent of the region in rapidity and azimuth over which a jet
captures UE or pileup;
secondly, the presence of the UE (and pileup) can modify the the way
non-UE particles get clustered into jets, a process named
back-reaction in~\cite{Cacciari:2008gn}.

\subsubsection{Jet areas}

A jet's ``area'' is a way of measuring its susceptibility to
contamination from soft radiation. Two definitions were proposed for
it in \cite{Cacciari:2008gn} and the results quoted here are all taken
from there.
The \emph{passive} area is a measure of the jet's susceptibility to
pointlike radiation. 
One introduces a ``ghost'' particle, $g(y,\phi)$ with infinitely low
transverse momentum and situated at some rapidity and azimuth
$y,\phi$, and then defines the area for jet $J$ in terms of the region
in $y,\phi$ over which the ghost is clustered with the jet:
\begin{equation}
  \label{eq:passive-area-def}
  a(J) \equiv \int dy\,d\phi\; f(g(y,\phi),J)\qquad\qquad  f(g,J) =
  \left\{\begin{array}{cc}
      1 & g \in J \\
      0 & g \notin J \\
    \end{array}
  \right. \; .
\end{equation}
For an infrared safe algorithm $J$ itself is of course unaffected by the
addition of the infinitely soft particle $g$.
If $J$ consists of a single (hard) particle then its passive area is
$\pi R^2$ for all algorithms.

An alternative definition of area is the \emph{active} area, which measures a
jet's susceptibility to diffuse radiation. Here one imagines a large
number of very soft ghost particles, uniformly distributed in rapidity
and azimuth (with some optional randomness). One can define a jet's
active area for a given ensemble ${\{g_i\}}$ of ghost particles,
\begin{equation}
  \label{eq:act_area_given_gi}
  A (J \,|\, {\{g_i\}})  =   \frac{\cNg(J)}{\nu_g} \,.
\end{equation}
where $\cNg(J)$ is the number that end up in the jet and ${\nu_g}$
their number density in $y,\phi$.
One then often considers the average active area, an average over many
ghost ensembles
\begin{equation}
  \label{eq:act_area}
  A(J) = \lim_{\nu_g \to \infty} \left \langle
  A (J \,|\, {\{g_i\}}) \right \rangle_g \,,
\end{equation}
taken in the limit of many ghost particles (with fixed infinitesimal total
$p_t$ per unit area).%
\footnote{
  These two areas are strictly speaking both ``scalar'' areas. Passive
  and active areas also come in ``4-vector'' variants, which take into
  account the ghosts' full impact on a jet's 4-vector. Though useful for
  subtracting pileup and other noise, most of their features are quite
  similar to those of scalar areas, so we shall not discuss them separately
  in this section.
}
A key difference between the passive and active areas, is that
in the latter, with many ghosts, the ghosts can cluster not only with
the real event particles but also with other ghosts and this modifies
the result for the area.

Given that one usually imagines the UE and pileup as being fairly diffuse (and UE/pileup
particles can cluster between themselves), it is
the active area that is probably the most natural measure of
sensitivity to the UE or pileup. However, there are two reasons why it is useful
to consider both passive and active areas. 
Firstly, the UE is actually somewhere in between diffuse and pointlike
and a full understanding of UE contamination benefits from considering
both limits.
Secondly, of the two, the passive area is often simpler to treat
analytically.

\begin{table}[One-particle active areas]
  \centering
  \begin{tabular}{lcccc}\toprule
    Algorithm      & $k_t$ & C/A & SISCone & anti-$k_t$\\\midrule
    {\small$\displaystyle\frac{A(\oPJ) \pm \Sigma(\oPJ)}{\pi R^2}$}  & $0.812\pm0.277$  & $0.814\pm0.261$  & $\frac14\pm
    0$ & $1\pm 0$ \\[10pt]
    {\small$\displaystyle\frac{A(\GJ) \pm \Sigma(\GJ)}{\pi R^2}$}  &
    $0.554\pm0.174$  & $0.551\pm0.176$  & --- & ---\\[5pt]
    \bottomrule
  \end{tabular}
  \caption{The average active area $A(\oPJ)$,
    eq.~(\ref{eq:act_area}), for an isolated one-particle
    jet in various jet algorithms and its standard deviation $\Sigma(\oPJ)$
    over ghost ensembles.  \label{tab:1particle-act-areas}
    Results are also given for the area of jets that are purely
    composed of ghosts (GJ), in the cases where this makes sense (in
    SISCone the result depends critically on the distribution of
    ghosts, while for anti-$k_t$ the distribution of ghost-jet areas
    has two peaks, one at $0$, the other at $\pi R^2$).
  }
\end{table}

Let's illustrate these points for the case of a jet that has just one,
hard particle (\oPJ).
As mentioned above, the passive area is $\pi R^2$. This statement
holds for all jet algorithms.
The average active area and its standard deviation over ghost
ensembles are given in table~\ref{tab:1particle-act-areas}. 
For one algorithm, anti-$k_t$, the active area is identical to the
passive area and $A (J \,|\, {\{g_i\}})$ is independent of the
particular ghost ensemble. This can be seen as advantageous, insofar
as it implies that the contamination of an anti-$k_t$ jet will be
independent of the detailed structure of the UE or pileup.

SISCone also has the property that its active area is independent of
the ghost ensemble, but the actual value is much smaller than for the
passive area (a consequence of the split--merge step which eats away
from the main jet). 
This is good insofar as it means less overall sensitivity to noise,
but bad because the exact amount of contamination will depend on the
details of just how pointlike the UE or pileup is (a feature that therefore
needs to be well tuned in Monte Carlo programs).

The $k_t$ and C/A algorithms are unlike the other two in that $A (J
\,|\, {\{g_i\}})$ depends significantly on the exact set of ghosts, as
indicated by the standard deviations in table~\ref{tab:1particle-act-areas},
which are about one third of the average area.
The non-zero standard deviation arises because ghosts tend to cluster
between themselves before clustering with the hard particle, and
slight shifts in the layout of the ghosts lead to significant
differences in the final clustering.
This implies an extra
source of fluctuations from UE and pileup contamination: one has not only the
intrinsic fluctuations in the amount of $p_t$ in a given event's UE/pileup,
but also a fluctuation in how much of the UE/pileup the jet actually
captures. 
The consequent (moderate) worsening of the kinematic resolution of the
jets seems to be an inevitable feature of jet algorithms with a
QCD-motivated hierarchical clustering sequence: the algorithm is trying
to assign meaningful QCD substructure to the jet, and the absence of
such substructure in the UE/pileup induces a degree of randomness in the
outcome of the clustering (this is related also to the irregularity of
the jet boundaries in fig.~\ref{fig:4algs}).

\begin{figure}
  \centering
  \includegraphics[width=0.9\tw]{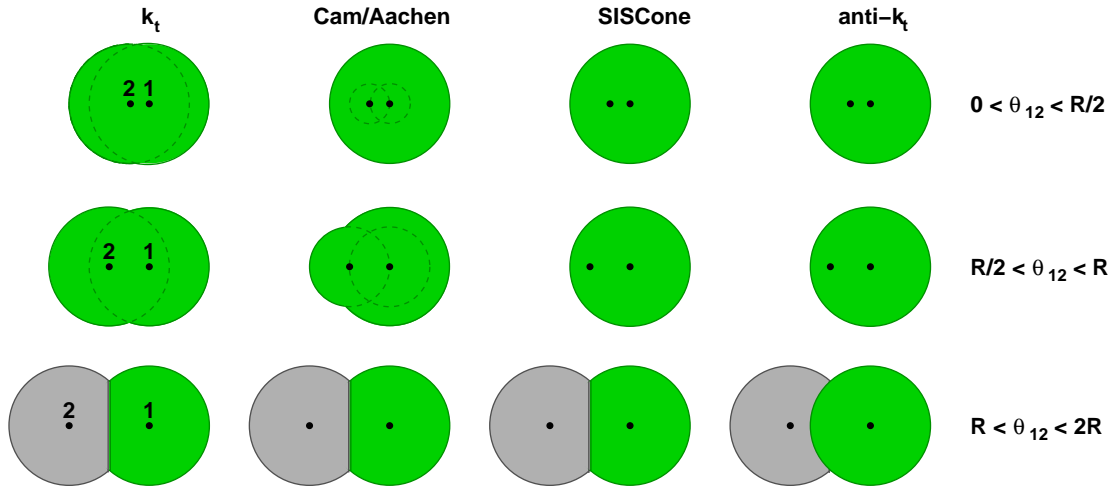}
  \caption{Schematic representation of the passive area of a jet
    containing one hard particle ``1'' and a softer one ``2'',
    $a_{\text{JA},R}(\theta_{12})$, for various separations between
    them and for the usual 4 jet algorithms. Different shadings represent
    distinct jets. Figure adapted from~\cite{Cacciari:2008gn}.}
  \label{fig:overlaps}
\end{figure}
The above results hold for 1-particle jets (1PJ). Real jets are more
complex because QCD branching gives them substructure. To a first
approximation, one can examine what happens if one adds a single soft
gluon at an angle $\theta$ with respect to the jet axis. This gives a
modified jet area, \eg $a_{\text{JA},R}(\theta)$ in the passive case,
illustrated in fig.~\ref{fig:overlaps}.
After integration over the (soft, collinear) QCD branching matrix
element one obtains
\begin{equation}
  \label{eq:anom-dim}
  \langle a_{\JA,R} \rangle = \pi R^2 +  d_{\text{JA},R}\,\frac{C_1}{\pi b_0} \ln
  \frac{\alpha_s(Q_0)}{\alpha_s(R p_{tJ})} \,,
\end{equation}
where the anomalous dimension (\ie the $p_t$-dependent rightmost term)
stems from the 
integration over the energy of the soft gluon, and its coefficient is
given by
\begin{equation}
  \label{eq:d}
  d_{\text{JA},R} = \int_0^{2R} \frac{d\theta}{\theta} (
  a_{\text{JA},R}(\theta) - \pi R^2)\,.
\end{equation}
In eq.~(\ref{eq:anom-dim}) $b_0 = \frac{11C_A -
  2n_f}{12\pi}$ and $C_1$ is the colour factor of the hard particle in
the jet ($\CF$ or $\CA$).
The scale $Q_0$ is a non-perturbative cutoff scale, introduced by
hand, and which is necessary because jet areas are not IR safe (except
in the case of anti-$k_t$). 
The physically natural value for $Q_0$ will depend on the
characteristics of the UE/pileup: given an amount of transverse
momentum $\rho$ per unit area, one expects $Q_0$ to be $\order{\rho
  R^3}$.\footnote{That is: a transverse momentum with respect to the
  beam $\order{\rho R^2}$, which translates to a transverse momentum
  with respect to the jet $\order{\rho R^3}$, it being the latter that
  corresponds to $Q_0$ in eqs.~(\ref{eq:anom-dim}) etc.}

\begin{table}\small
  \centering   
  \begin{tabular}{r|cc|cc|cc|cc}
                 & $a$(1PJ) & $A$(1PJ)  & $\sigma$(1PJ) & $\Sigma$(1PJ)  & $d$         &  $D$   & $s$     & $S$       \\\hline 
   $k_t$         & $1$      & $0.81$    & $0$           & $0.28$         & $\,\,0.56$  & $0.52$ & $0.45$  & $0.41$  \\ \hline  
   Cam/Aachen    & $1$      & $0.81$    & $0$           & $0.26$         & $\,\,0.08$  & $0.08$ & $0.24$  & $0.19$  \\ \hline  
   SISCone       & $1$      & $1/4$    & $0$           & $0$            & $\!\!-0.06$ & $0.12$ & $0.09$  &  $0.07$   \\ \hline 
   anti-$k_t$    & $1$      & $1$       & $0$           & $0$            & $0$         & $0$    & $0$     &  $0$      \\ \hline
  \end{tabular}
  \caption{A summary of main area results for our four jet algorithms:
    the passive ($a$) and active ($A$) areas for 
    1-particle jets (1PJ), the magnitude of the passive/active area
    fluctuations ($\sigma$, $\Sigma$), followed by the coefficients of
    the respective anomalous dimensions ($d$, $D$; $s$, $S$), in the
    presence of perturbative QCD radiation. All results are normalised
    to $\pi R^2$, and rounded to two decimal figures.  
    For algorithms other than anti-$k_t$, active-area
    results hold only in the small-$R$ limit, though finite-$R$
    corrections are small.}
  \label{tab:summary}
\end{table}

\begin{figure}
  \centering
  \qquad\includegraphics[angle=270,width=0.75\tw]{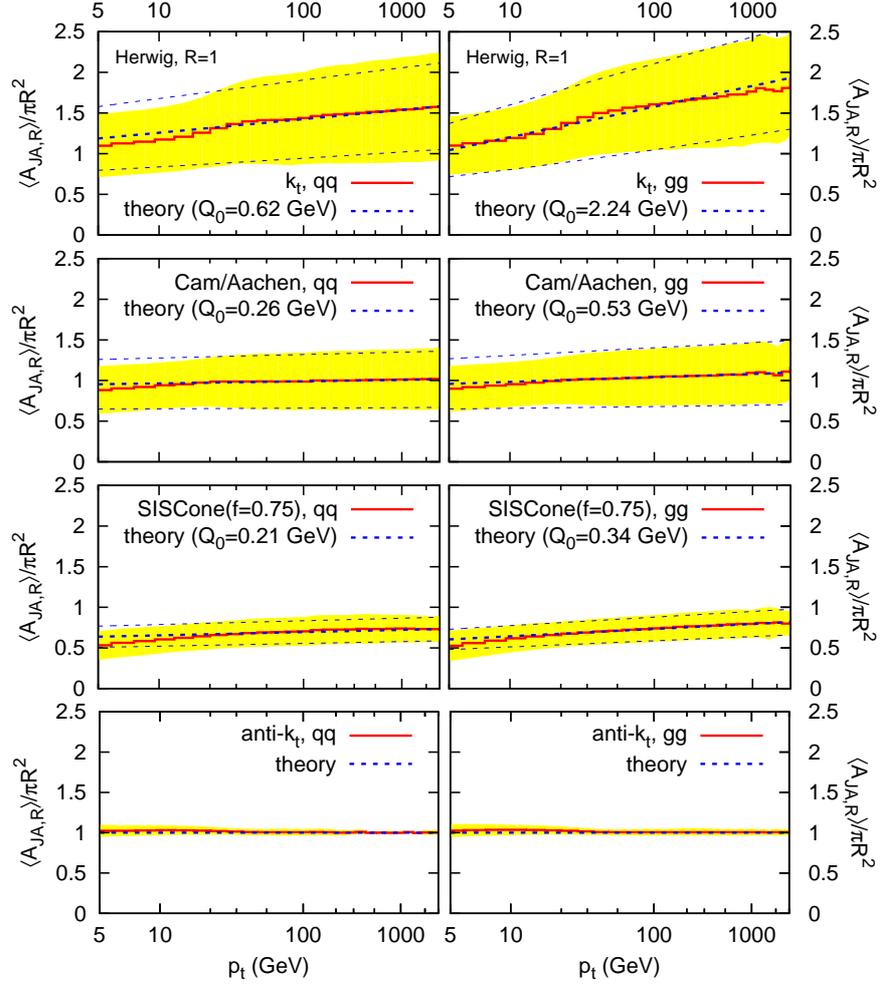}
  \caption{Average active area and standard deviation (solid lines and
    band) in
    simulated Herwig 6.5 events (default UE) compared to analytical
    expectations (dashed lines) with a fitted $Q_0$ value; shown separately for 4
    algorithms and for $qq\to qq$ and $gg\to gg$ events. Adapted
    from~\cite{Cacciari:2008gn,Cacciari:2008gp}.}
  \label{fig:anom-dim-areas}
\end{figure}

Formulae analogous to eqs.~(\ref{eq:anom-dim},\ref{eq:d}) hold for the
active area ($\pi R^2 \to A(\oPJ)$, $d\to D$), and also for the
standard deviations $\sigma_{\text{JA},R}$ of the passive area,
\begin{equation}
  \label{eq:delta-ajf-res2}
  \langle\sigma_{\text{JA},R}^2\rangle =
  s_{\text{JA},R}^2\frac{C_1}{\pi b_0}
  \ln \frac{\as({Q_0})}{\as(R p_{t1})}\;, \qquad
  s_{\text{JA},R}^2 = \int_0^{2R} \frac{d\theta}{\theta} ( a_{\text{JA},R}(\theta) -
  \pi R^2)^2\,
\end{equation}
and the active area 
\begin{equation}
  \label{eq:act-fluct-seq}
  \langle \Sigma^2_{\JA,R} \rangle =
  \Sigma^2_{\JA,R}(\oPJ) + S_{\text{JA},R}^2 \frac{C_1}{\pi b_0}
  \ln \frac{\as({Q_0})}{\as(R p_{t1})}\,.
\end{equation}
The various coefficients are all summarised in
table~\ref{tab:summary}, while in fig.~\ref{fig:anom-dim-areas} the
resulting predictions are compared to jet areas measured in Herwig
Monte Carlo simulations (with $Q_0$ fitted on a case-by-case
basis). One sees how $k_t$ has a fairly large area, large
fluctuations and strong $p_t$ dependence, C/A an area $\sim \pi R^2$
with moderate fluctuations and little $p_t$ dependence, SISCone an area
smaller than $\pi R^2$ (but still larger than $\pi R^2/4$), small
fluctuations and moderate $p_t$ dependence and, finally, anti-$k_t$ an
area very close to $\pi R^2$, with almost no fluctuations or $p_t$
dependence.
A corollary of the $k_t$ algorithm's strong $p_t$ dependence is that
is that if one increases the UE/pileup density $\rho$, the jet area will
\emph{shrink}, a consequence of the presence of $\as(Q_0)$ in
eq.~(\ref{eq:anom-dim}), with $Q_0 \sim \rho R^3$.

A final comment concerns the relation between passive and active
areas.
They differ in sparse events because there is an ambiguity in how one
assigns ``empty'' parts of the event to the jets --- the two kinds of
area simply consist of different prescriptions for doing this.
In very dense events, where the jet boundaries are well delimited by
the event's particles, the passive and active areas become identical
(as should any other sensible definition of area).

\subsubsection{Back reaction}

Suppose you have an event with particles numbered $1-100$, and in
which numbers $1-10$ end up in jet $a$. Now immerse that event in a
soft background with \emph{finite} $p_t$. The background will add its own particles to the jet,
but it can also alter the behaviour of the clustering with respect to the
original particles. So maybe only particles $1-9$ would end up in jet
$a$, or maybe jet $a$ would additionally contain particle $11$. This
is back reaction.

The study of back reaction bears similarities to that of jet areas: in
particular one can study it with pointlike noise, or with diffuse
noise. The former can be dealt with analytically, whereas the latter
is tractable only numerically.

Full details are given in \cite{Cacciari:2008gn}, but the basic
analytical result is that the average net change in a jet's $p_t$ due
to back reaction in the presence of diffuse noise has an asymptotic
behaviour of the form
\begin{equation}
  \label{eq:deltapt_back_active}
  \langle \Delta p_{t,\JA,R}^{BR} \rangle \simeq
  {\mathcal{B}}_{\JA,R} \, 
  \rho  \cdot \frac{C_1}{\pi b_0} \ln
  \frac{\as(\rho R^3)}{\as(p_{t1} R)}
  + \order{ \as \rho  } \,,
  \qquad\quad
\end{equation}
where the coefficients are ${\mathcal{B}}_{\kt,R} \simeq
{\mathcal{B}}_{\CamAachen,R} \simeq -0.10 \pi R^2$ and
${\mathcal{B}}_{\SISCone,R} = {\mathcal{B}}_{\akt,R} = 0$.
In practice, given the small size of the ${\mathcal{B}}$ coefficients
(and the fact that the $\frac{C_1}{\pi b_0}\ln \frac{\as(\rho
  R^3)}{\as(p_{t1} R)}$ factor is often $\order{1}$), the term of
order $\as \rho$ is usually as important as the formally leading
term. 
Both terms are generally small compared to the direct contamination of
the jet from UE/pileup noise, $\order{\rho \cdot \pi R^2}$.


\begin{figure}
  \centering
  \includegraphics[width=0.7\tw]{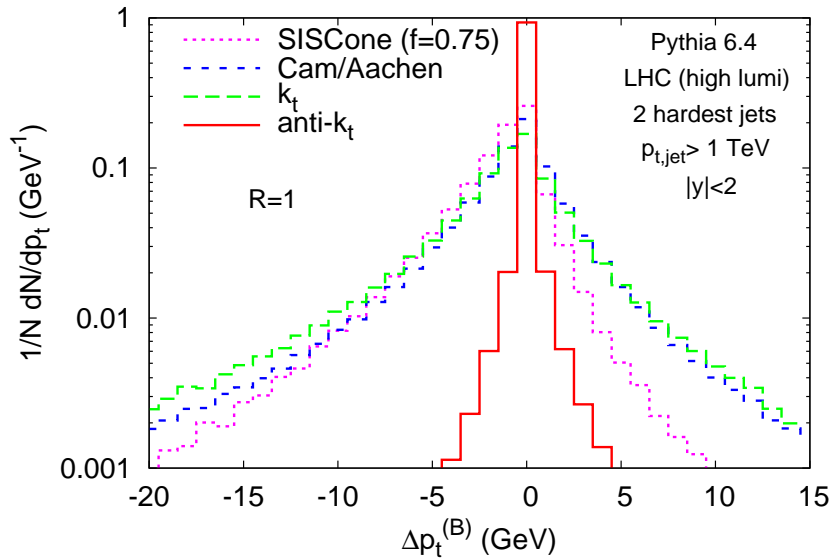}
  \caption{The distribution of back reaction for high-$p_t$ jets ($p_t
    > 1\TeV$) immersed in pileup corresponding to high-luminosity LHC
    running ($\rho \sim 15 \GeV$ per unit area). Simulated with
    Pythia 6.4 and shown for 4 algorithms. }
  \label{fig:back-reaction-plot}
\end{figure}

The concrete situation for the various algorithms is illustrated in
fig.~\ref{fig:back-reaction-plot}, which shows the distribution of
back reaction for a high-$p_t$ jet immersed in pileup ($\rho \sim 15
\GeV$). 
In about $1\%$ of events one has a back reaction of order of
$\rho$, except for anti-$k_t$, whose back reaction is far more
suppressed. 
Fig.~\ref{fig:back-reaction-plot} confirms that back reaction is a
modest effect compared to the direct contamination of a jet from
background noise. Essentially it is relevant only when trying to
determine a jet's energy to very high precision, or in the presence of
extreme noise (as in heavy-ion collisions).

\subsection{Summary}
\label{sec:jet-properties}

We have seen a number of results here. Let us summarise them:
\begin{itemize}
\item Most jet algorithms will cluster a pair of particles if they are
  within $R$ of each other; SISCone reaches out to $2R$ (somewhat less
  in real events) if the two particles are of similar hardness.
\item At small $R$, a jet's $p_t$ is reduced relative to a parton's by
  an amount $\sim \as p_t \ln 1/R$. With $R=0.4$, that's of order
  $5\%$ for a quark, $10\%$ for a gluon.
  The mean squared jet mass goes as $\as R^2 p_t^2$.
\item Hadronisation reduces a quark jet's $p_t$ by an amount $\sim
  0.5\GeV/R$ at small $R$ (roughly double this for a gluon jet), with modest
  differences between algorithms.
\item The underlying event and pileup induce contaminations
  proportional to $\sim R^2$.
  Because the intrinsic energy scale associated with the underlying
  event ($p_t$ per unit rapidity of $10-15\GeV$ at the LHC) is an
  order of magnitude larger than that from hadronisation (and pileup
  is yet another order of magnitude larger), one should devote special
  effort to understanding different jet algorithms' susceptibility to
  them. This can be done via the concept of jet areas.
  The $k_t$ algorithm has the largest jet area (with noticeable $p_t$
  dependence and fluctuations), SISCone the smallest, and anti-$k_t$
  has the most stable jet area, nearly always $\pi R^2$.
\item The UE (and pileup) modifies how an event's original particles get
  clustered into jets --- this is back reaction. Its impact is an
  order of magnitude smaller than the direct contamination, but can be
  relevant for precision studies. It is essentially zero for the \akt\ 
  algorithm.
\end{itemize}
The differing impact of various physical effects across algorithms and
as a function of $R$ might seem like a source of considerable
complication in jet finding, and in some ways it is.
However, it can also be used to our advantage. 
One example is in studies like top-mass measurements that are in part
limited by physics-modelling systematics. 
If one uses multiple jet definitions with different sensitivities to
UE/pileup, gluon radiation and hadronisation, and then finds that the final
Monte Carlo-corrected top mass is independent of the choice of jet
definition, then this provides a powerful cross check of the physics
modelling within the Monte Carlo generator.
%
%
And, in the next \toplevel, we shall see how our understanding of jets
can help guide the choice of ``optimal'' jet definitions for various
reconstruction tasks.




\section{Using jets}
\label{sec:using-jets}

So far at hadron colliders, jets have mostly been used as fixed
objects --- universal, if imperfect proxies for partons.
Generally, experiments have settled on one or two main jet definitions
for nearly all their analyses: for example at CDF the Midpoint
algorithms with $R=0.7$ for most inclusive-jet
studies,\footnote{Though in recent years they also studied the $k_t$
  algorithm with three $R$ values~\cite{CDFkt,Abulencia:2007ez}, and
  this probably played a key role in convincing the LHC experiments
  that the $k_t$ algorithm is viable.} and JetClu with $R=0.4$ for
top-quark physics and for searches.

Such a strategy was probably not too far from optimal at the Tevatron,
where most of the physics being looked at is in a modest range of
scales, from a few tens of GeV to a few hundred, and pileup is present,
but not overwhelming.

The LHC, in contrast, will cover a broader range of scales from a few
tens of GeV to a few TeV, events with multiple simultaneous scales
will be common (\eg EW bosons and top quarks with $p_t \gg m$) and
pileup will range from almost none to $20-30$ simultaneous $pp$
interactions in each bunch crossing.
This begs the question: could analyses benefit from more flexible
jet finding?
%

The work examined below tries to examine this question by
concentrating on two characteristic types of analysis --- standard
mass reconstructions, with attention also to the issue of pileup; and
the task of identifying highly boosted massive particles.
Our discussion will be restricted to studies at ``particle level''
(also referred to as hadron level) and won't go into
detector-specific effects. For a discussion of the latter, see for
example~\cite{Buttar:2008jx:MarioEtAl,CMS-Jet-PAS,Aad:2009wy}.

\subsection{Choosing an algorithm and a radius}
\label{sec:choosing-radius}

Which jet algorithm is ``best''? This is a widespread question, and a
natural follow-on question is ``which $R$ should one use''?
This question cannot be answered in isolation. It inevitably goes with
the issue of what one wants to use the jet algorithm for.

The most reliable way of answering the question is to carry out a
detailed study of the process one is interested in, with many jet
algorithms, and many $R$ values for each. Then one may devise some
``quality measure'' and establish which algorithm optimises it.
This can be a big job (4 algorithms, maybe $10$ $R$ values) and is
seldom done in a systematic way.
In what follows we'll see how even crude analytical estimates can give
guidance on the question, and then examine some Monte Carlo studies.

\subsubsection{Analytical study}
\label{sec:analytical-radius}

In a QCD measurement like that of the inclusive jet spectrum, one
will compare data to a perturbative QCD prediction. 
At moderate $p_t$, one of the largest ambiguities in the comparison comes from
non-perturbative effects (since perturbative effects are calculable to
some accuracy), often comparable with PDF and experimental
uncertainties.
Therefore one might want to choose an algorithm and $R$ value
that minimise hadronisation and UE contributions.
One can choose to ignore the relatively modest differences between
algorithms, and just take the analytic formulae of
sections~\ref{sec:hadronisation},~\ref{sec:jet-areas}. From these, one
deduces a value of $R$ that minimises the squared sum of hadronisation
and UE pieces~\cite{Dasgupta:2007wa},
\begin{equation}
  \label{eq:optimal-R-hadr-UE}
  R = \sqrt{2} \left( \frac{2C_i \cM {\cal A}(\mu_I)}{\pi \Lambda_\UE} 
  \right)^{1/3}\,,
\end{equation}
where only the leading $R$ terms have been used, $C_i$ is the
appropriate colour factor ($\CF,\CA$) for quark/gluon jets,
and we have assumed a jet area of $\pi R^2$ for simplicity
The resulting numerical $R$ values are given in
table~\ref{tab:best-R-non-pert}.\footnote{In practice, an additional
  issue is that perturbative uncertainties from missing higher-order
  contribution may also depend on $R$. The interplay between this and
  non-perturbative uncertainties has not been studied.}

\begin{table}
  \centering
  \begin{tabular}{lcc}\toprule
                        & quark jets & gluon jets \\ \midrule
    Tevatron            &  0.56 & 0.73 \\
    LHC ($14\TeV$)      &  0.41 & 0.54\\ \bottomrule
  \end{tabular}
  \caption{$R$ values that minimise the two non-perturbative
    contributions in various circumstances for Tevatron and LHC
    running, based on eq.~(\ref{eq:optimal-R-hadr-UE}), with
    $2\cM {\cal A}(\mu_I)/\pi = 0.19\GeV$ and $\Lambda_{UE} = 4\GeV$ ($10\GeV$) at
    the Tevatron (LHC).}
  \label{tab:best-R-non-pert}
\end{table}

\begin{figure}
  \centering
  \includegraphics[width=0.45\tw]{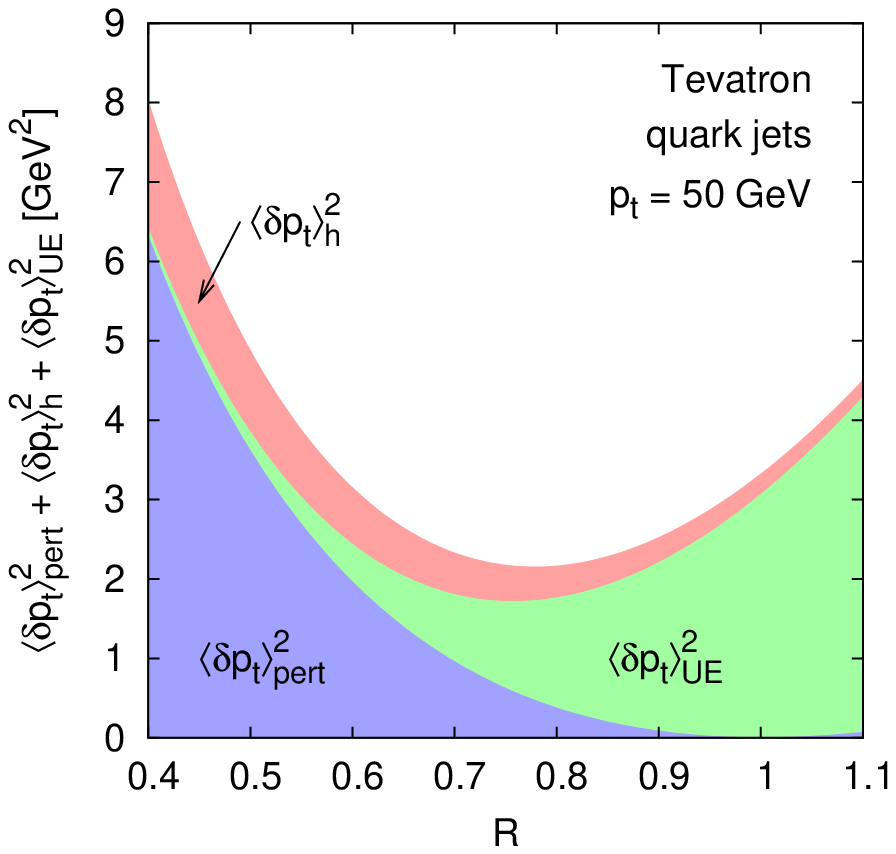}\hfill
  \includegraphics[width=0.53\tw]{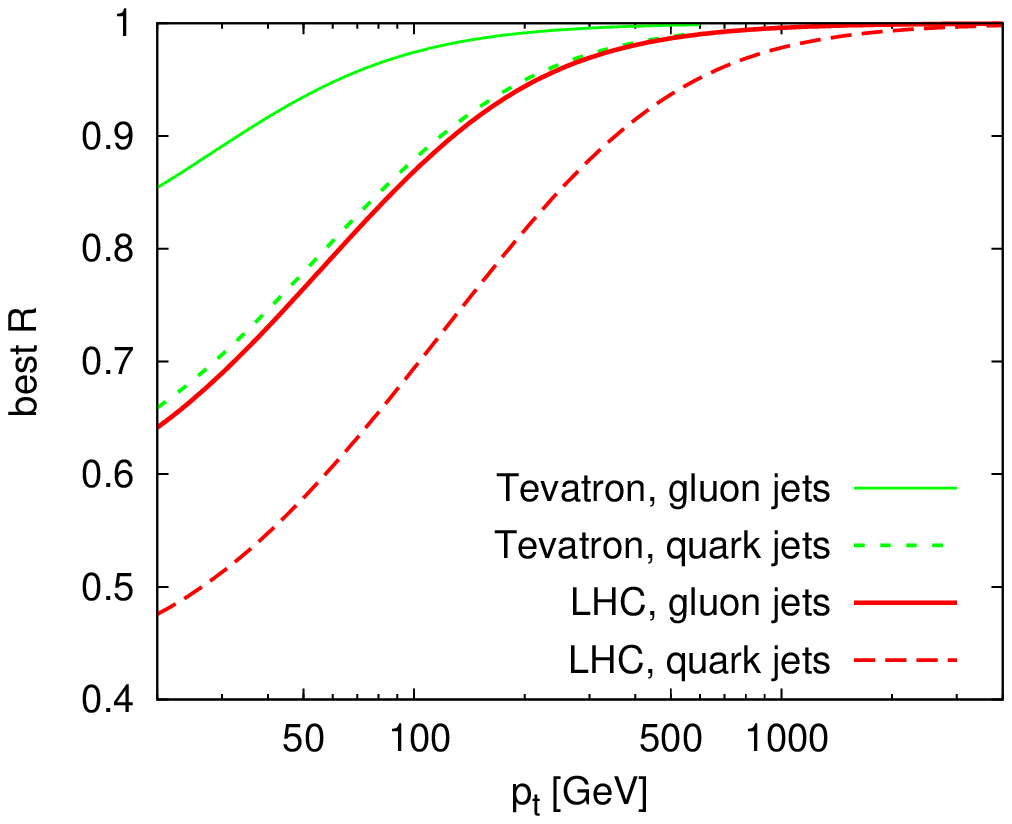}
  \caption{Left: sum of the squares of the mean shifts of a jet's
    momentum due to perturbative gluon radiation, hadronisation and
    the UE, as a function of $R$ for $p_t\sim 50 \GeV$ quark jets at
    the Tevatron; right: the resulting crude estimate for the ``best''
    $R$ as a function of jet $p_t$, for quark and gluon jets at the
    Tevatron and LHC ($14\TeV$). 
    These values are to be taken as indicative of general trends
    rather than reliable estimates of the best $R$.
    The plots use the same parameters as
    table~\ref{tab:best-R-non-pert} and the perturbative contribution
    is taken in the small-$R$ limit. 
    Taken from \cite{Dasgupta:2007wa}.}
  \label{fig:hard-best-R}
\end{figure}

If one uses jets for kinematic reconstruction, the considerations are
different: when trying to identify a mass peak, for example, it is of
little consolation that one can calculate the perturbative degradation
of the peak if that degradation in any case causes the peak to
disappear under the background.
A \emph{very} crude estimate of what goes on can be had by assuming
that fluctuations in a jet's momentum due to perturbative radiation,
hadronisation and UE are each proportional to their average
effect. Adding the squared averages in quadrature gives
fig.~\ref{fig:hard-best-R} (left) and the minimum provides an idea of
the optimal $R$ (as before, ignore differences between algorithms),
and illustrates how the main relevant interplay is between
perturbative radiation and the UE.
The right-hand plot shows how the resulting optimal $R$ varies with
$p_t$: gluon jets and high $p_t$ jets prefer larger $R$ values
(because of the greater relative importance of perturbative
radiation), while one needs smaller $R$ values at the LHC than at the
Tevatron (the former has more UE).

While fig.~\ref{fig:hard-best-R} is useful for understanding general
trends (notably the need for large $R$ at high $p_t$), it is not
quantitatively reliable: the fluctuations in a jet's kinematics and
the mean energy-loss due to gluon radiation are for example \emph{not}
proportional to each other; also, for $R\sim 1$, the $\as p_t \ln R$
approximation for perturbative energy loss is in itself poor
since it neglects terms of $\order{\as p_t}$; finally, for simplicity,
it has been obtained neglecting differences between algorithms and
this is not entirely legitimate.

\subsubsection{Numerical studies}
\label{sec:numerical-radius}

Given that a complete analytic treatment is not yet available, one
can use Monte Carlo event simulation to examine the
optimisation of the choice of jet definition.

Historically the approach taken to studying the quality of jet definitions
has been to take hard partons in a Monte Carlo, let them shower and
hadronise, and then see how closely the reconstructed hadron-level
jets match the original partons (see for
example~\cite{Buttar:2006zd}). 
This procedure has the drawback that it is conceptually impossible to
extend it to advanced Monte Carlo tools like
MC@NLO~\cite{Frixione:2002ik}, because the ``original parton'' is no
longer identifiable.
Furthermore it leads to several distinct quality measures: whether the
number of jets is equal to the number of ``partons'', the angular
distance between the jets and the partons, and the $p_t$ difference
between the jets and the partons.
It is then often not clear which of these measures is most
representative of the algorithm's usefulness in a real physics
analysis.

The most robust way of proceeding would be, for each possible
experimental study, to carry out a full signal and background analysis
with a wide variety of jet definitions and then see which provides the
best signal to background (or root-background) ratio.
As well as being a major undertaking, this can have subtleties: for
example, optimal cuts in an analysis may depend on the jet definition
and so may need to be reoptimised for each new jet definition.

An approach taken in
\cite{Buttar:2008jx:Us,Buttar:2008jx:KlausEtAl,Cacciari:2008gd,VariableR}
attempts to carry out a simplified version of this procedure. 
It takes a physical process, for example the production of a narrow
$Z'$ that decays to $q\bar q$, or a $t\bar t$ event ($t\to$~hadrons)
and attempts to reconstruct the massive object.
The ``better'' the reconstructed mass peak, the better the jet
algorithm.
This ignores issues like how the jet algorithm performs with respect
to background events (which was however additionally studied
in~\cite{VariableR}), but is well-defined (no reference to partons)
and avoids the appearance of multiple quality measures (angular
dispersion, energy dispersion, etc.).
Note: the physical process itself need not necessarily be realistic
--- e.g.\ it is highly unlikely that there exists an as-yet
undiscovered, hadronically decaying $Z'$ with mass $100\GeV$. But it
still serves as a useful stand-in for a generic $q\bar q$ resonance, whose
mass scale can easily be varied.

\begin{floatingfigure}{0.45\textwidth}
  \includegraphics[width=0.43\textwidth]{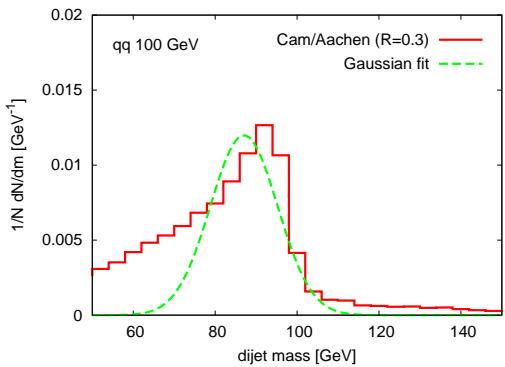}
  \caption{Distribution of the reconstructed invariant mass of a
    $100\GeV$ $q\bar q$ system for the C/A algorithm with $R=0.3$,
    simulated with Pythia 6.4, together with a Gaussian fit. 
    \label{fig:fit-histo-cam}
  }
\end{floatingfigure}
Quantifying the quality of the mass peak is, as it turns out, a
complex issue.
Two properties that matter are the position of the mass peak (how
close it is to the mass scale of the particle being reconstructed) and
its width.
The position matters, for example, when trying to accurately
reconstruct the mass of a hadronically decaying resonance. 
The width matters in terms of being able to identify clear signal
peaks over a smooth background.
In this section we will concentrate of the width.
That the width is not trivial to determine is evident from
fig.~\ref{fig:fit-histo-cam}. It is tempting to measure the peak
quality by fitting a Gaussian. However, the fit is poor; the results
of the fit depend on the choice of fit-window; and then it's not clear
which parameters of the Gaussian would serve as the quality measure:
the normalisation?  The width? Some combination of the two?
The approach
of~\cite{Buttar:2008jx:Us,Buttar:2008jx:KlausEtAl,Cacciari:2008gd} is
to avoid the fit-function and instead to find the width of the
smallest window that contains a specified fraction $z$ of the events,
$Q^w_{f=z}$. A sharper (\ie better) peak corresponds to a lower
$Q^w_{f=z}$ value. %
There is still some arbitrariness in this, for example the choice of
the fraction of events $z$ (defined with respect to all
events before cuts). However $z$ is easily
varied to check the robustness of the procedure and one can also
examine other quality measures.\footnote{One alternative is to fix the
  window width to be $x\GeV$, place the window so as to maximise the number of events that
  it contains, and then use the inverse fraction of contained events
  $Q^{1/f}_{w=x}$ as the quality measure. A better peak concentrates
  more events in a given window, giving a lower result for
  $Q^{1/f}_{w=x}$.\label{foot:Qb}}

\begin{figure}
  \centering
  \includegraphics[width=0.8\textwidth]{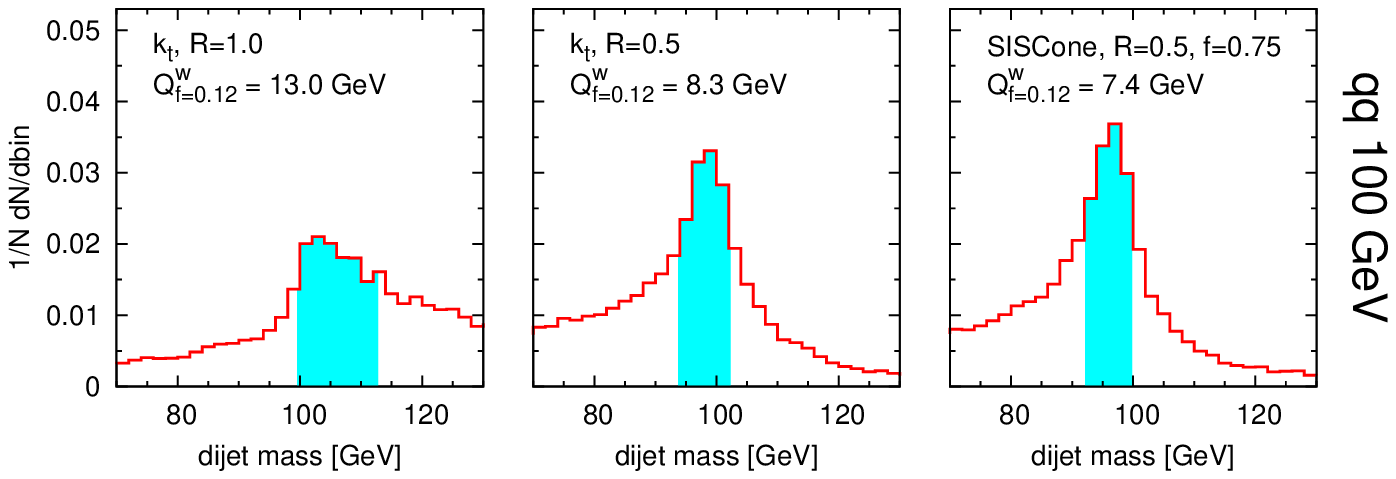}
  \includegraphics[width=0.8\textwidth]{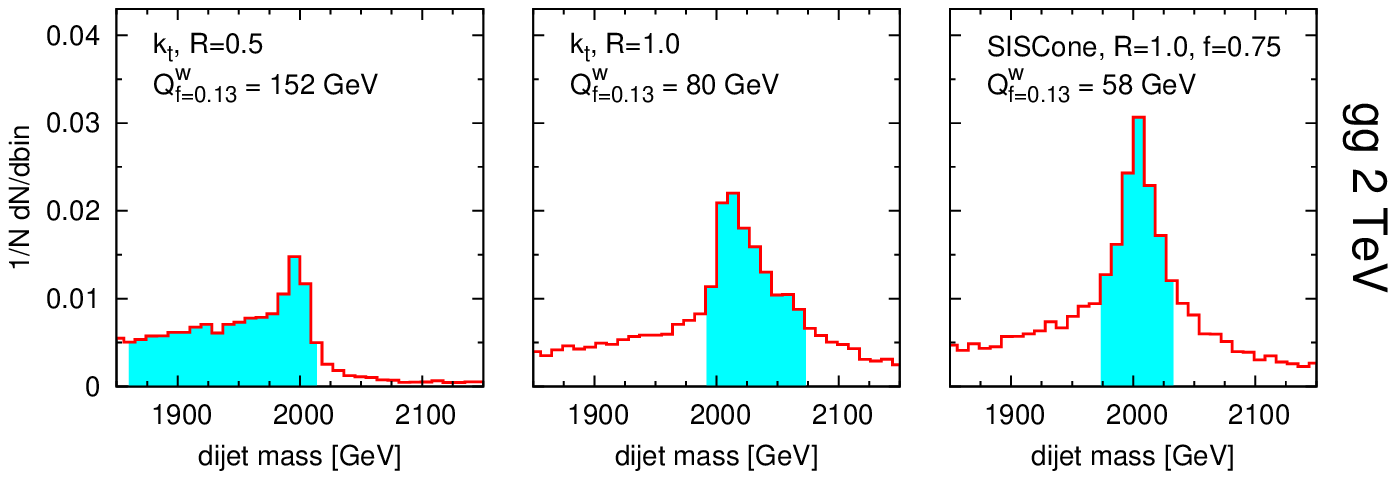}
  \caption{Illustrative dijet invariant mass distributions for two
    processes (above: \qq case at $M=100\GeV$; below: \gluglu case
    at $M=2\TeV$), comparing three jet definitions for each
    process. The shaded bands indicate the regions used when obtaining
    the $Q^w_z$ quality measure. Note that different values of
    $R$ have been used for the \qq and \gluglu cases.}
  \label{fig:quality-distributions}
\end{figure}

Fig.~\ref{fig:quality-distributions} illustrates the procedure for a
$100\GeV$ \qq resonance and a $2\TeV$ \gluglu resonance, examining
three jet definitions (the use of $z=0.12$ corresponds to taking about
$25\%$ of events after cuts).\footnote{This and other figures in this
  subsection are all taken from~\cite{Cacciari:2008gd}, to which the
  reader is referred for full details. In the dijet case, a mass is
  reconstructed from the two hardest jets, with a cut $|\Delta y| < 1$
  on the rapidity interval between the two jets (because in studies
  with background, such a cut greatly reduces the background).
} %
One sees how better peaks have lower values for \Qa{z}, together with
the extent of the differences between algorithms, and the relevance of
the choice of $R$.

\begin{figure}[p]
  \centering
  \includegraphics[width=0.48\textwidth]{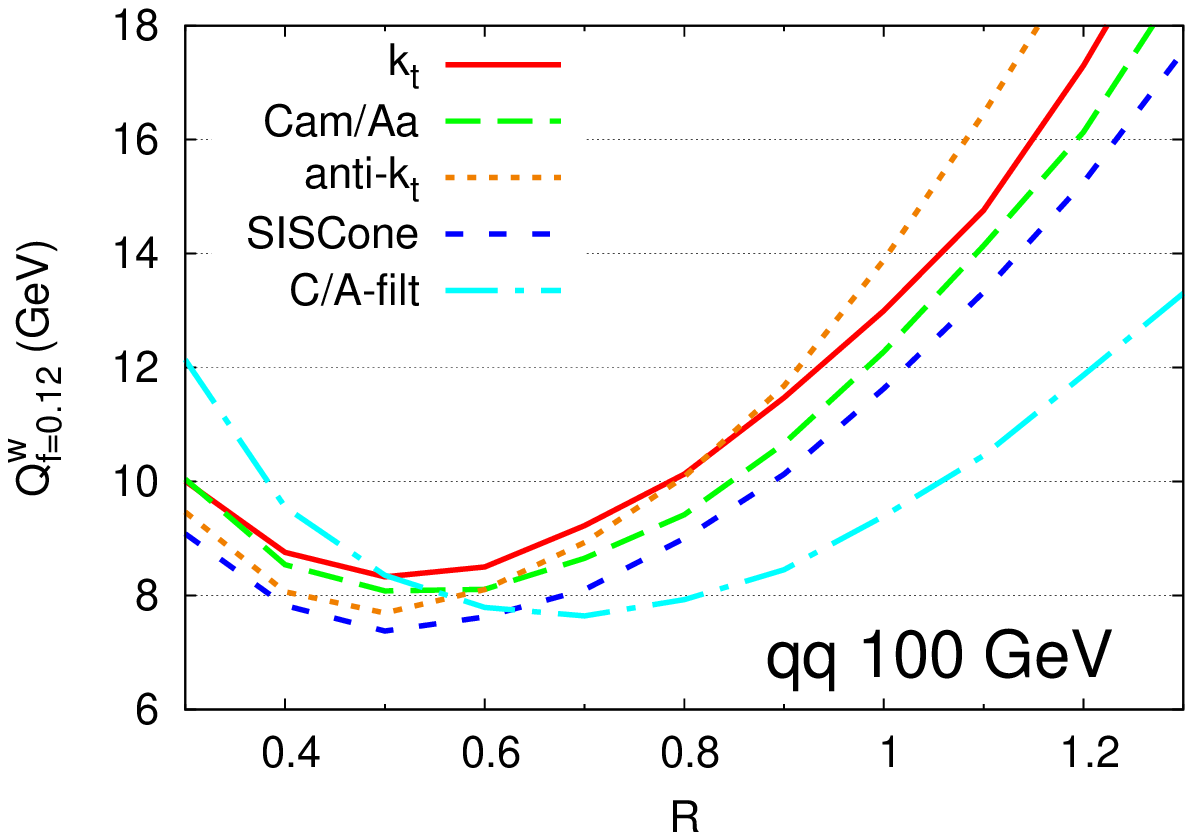}
  \includegraphics[width=0.48\textwidth]{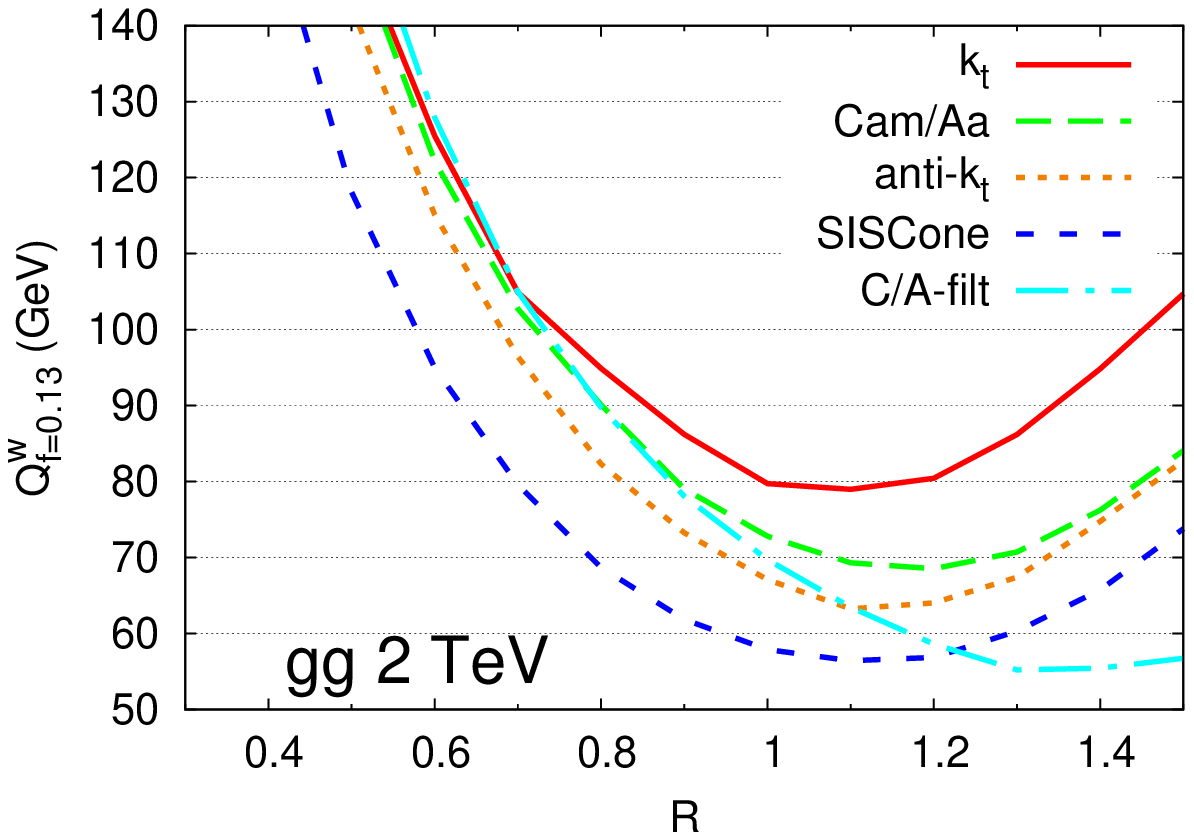}
  \caption{\small The quality measure \Qa{z}, for different jet
    algorithms as a function of $R$, for the $100\GeV$ \qq case (left)
    and $2\TeV$ \gluglu (right).}
  \label{fig:all_merit} 
\end{figure}

\begin{figure}[t]
  \centering
  \includegraphics[width=0.48\textwidth]{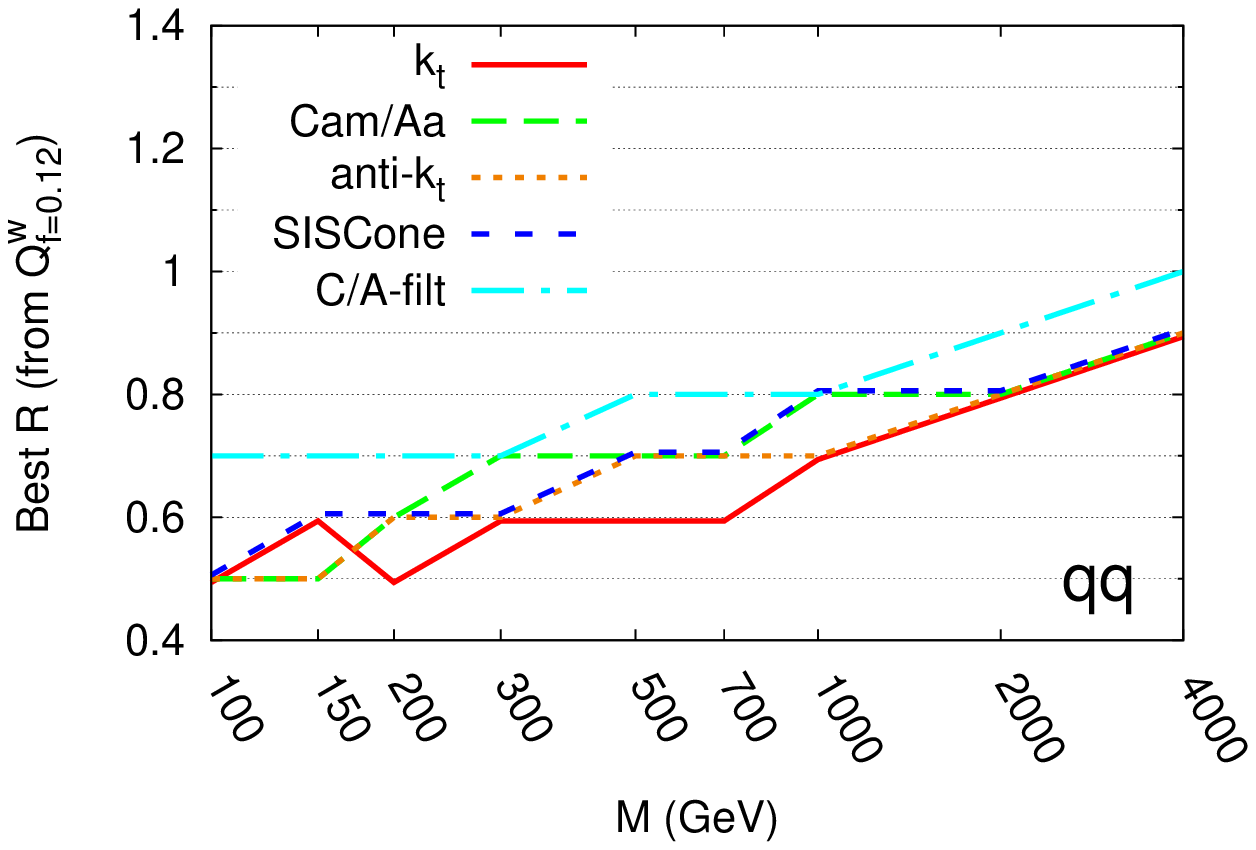}
  \includegraphics[width=0.48\textwidth]{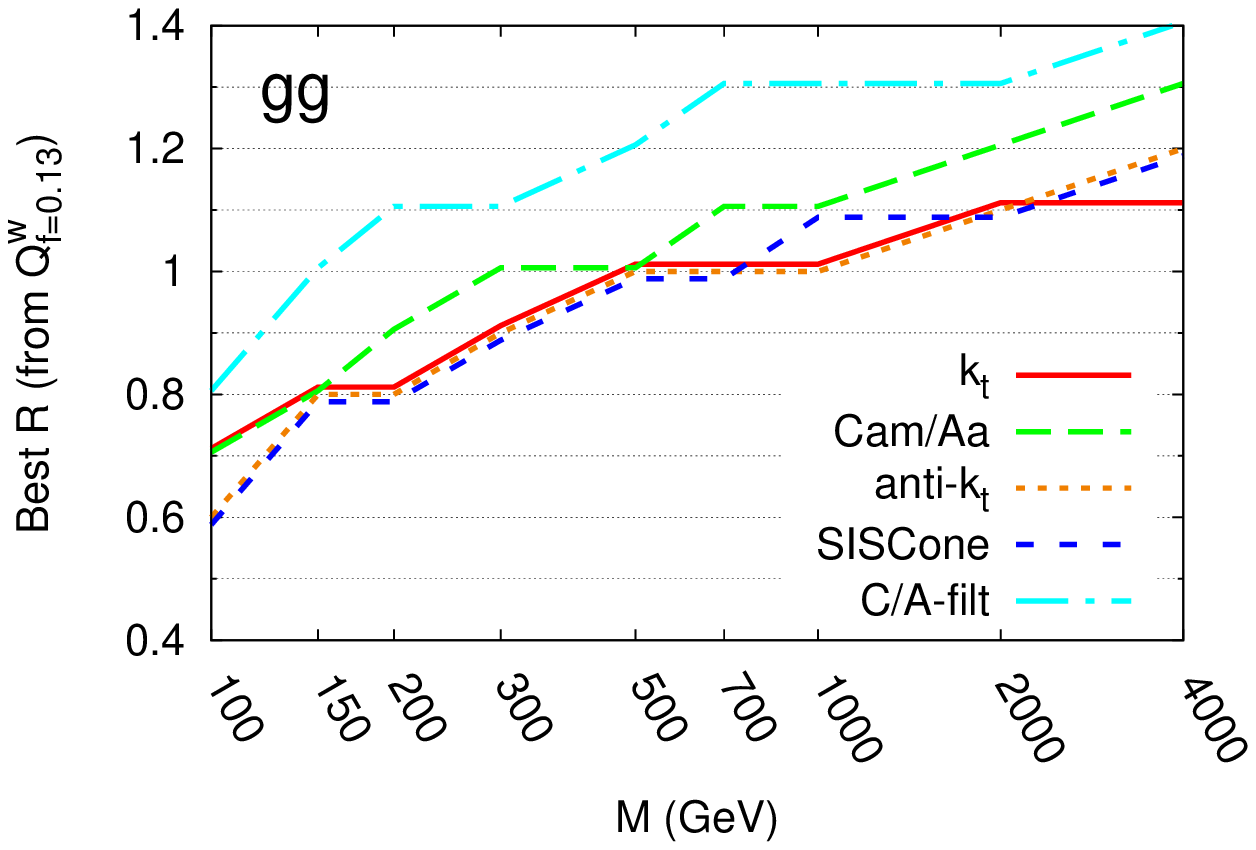}
  \caption{\small The optimal value for $R$ as a function of the mass
    of the \qq/\gluglu system (left/right), as determined from the
    $\Qa{z}$ quality measure for various jet algorithms. 
    Note that the exact results for the optimal $R$ depend a little on
    the choice of quality measure, however the observed trends do not.
  }
  \label{fig:Rbest-no-PU} 
\end{figure}

The full $R$-dependence of the quality measure is shown for $5$
algorithms in fig.~\ref{fig:all_merit}, for the same two physics
cases. 
The minimum of each curve indicates the best $R$ for that particular
algorithm, while the differences between the curves are illustrative
of the different behaviour of the various algorithms.
In particular, one sees the preference for larger $R$ in the
$2\TeV$ $gg$ case. This is in accord with the expectations discussed
in section~\ref{sec:analytical-radius}: the larger the importance of
perturbative gluon radiation, the larger the preferred $R$ value.
The optimal $R$ as a function of mass scale, for the different
algorithms and for the \qq and \gluglu cases, is illustrated in
fig.~\ref{fig:Rbest-no-PU}. 
The overall trend is not unlike the rough analytical estimate,
fig.~\ref{fig:hard-best-R} (right), but the details differ: for
example the full result doesn't show as rapid an $R$ dependence, and
it is not clear to which extent the optimal $R$ saturates at the
highest scales in fig.~\ref{fig:Rbest-no-PU}.

Figure~\ref{fig:all_merit} shows that even at the optimal $R$ there
are differences between algorithms. The origin of these differences
has not been analysed in full detail, but can almost certainly be
traced back to the very different area properties of the various
algorithms: $k_t$ fares worst because its larger area allows more UE
into the jet, causing enhanced fluctuations of the kinematics from
event to event; meanwhile SISCone, with its small area, fares well, as
does the filtered version of the C/A algorithm, which resolves each
jet on an angular scale $R/2$ and takes just the two hardest subjets
(\cf section~\ref{sec:other-seq-rec}). Ref.~\cite{Krohn:2009th} has
shown that additional modifications of the filtering procedure
(``trimming'') and tuning of its parameters can yield yet further
improvements, as does noise ``subtraction'', discussed in
section~\ref{sec:noise-subtraction}.

A key question in examining results such as those in
fig.~\ref{fig:quality-distributions} and \ref{fig:all_merit} is how
much the differences between algorithms (and $R$-values) matter:
from this information an experiment can then decide whether it is
worthwhile investing in the calibration of multiple jet definitions
(or in the inherent flexibility that allows easy use of any jet
definition).
One way of making such an estimate is to assume that the peak is being
reconstructed in the context of a search with significant background. 
If one can assume that the amount of background basically scales as
the width of the window that comes out of the $\Qa{z}$ quality
measure, then the significance $S/\sqrt{B}$ of any signal will just be
inversely proportional to  $\sqrt{\Qa{z}}$. 
One can then define a measure $\rho_L$, which is the extra factor in
luminosity that is needed to see the signal with given significance
when using one jet definition (JD) relative to another.

\begin{figure}[p]
  \centering
  \includegraphics[width=0.8\textwidth]{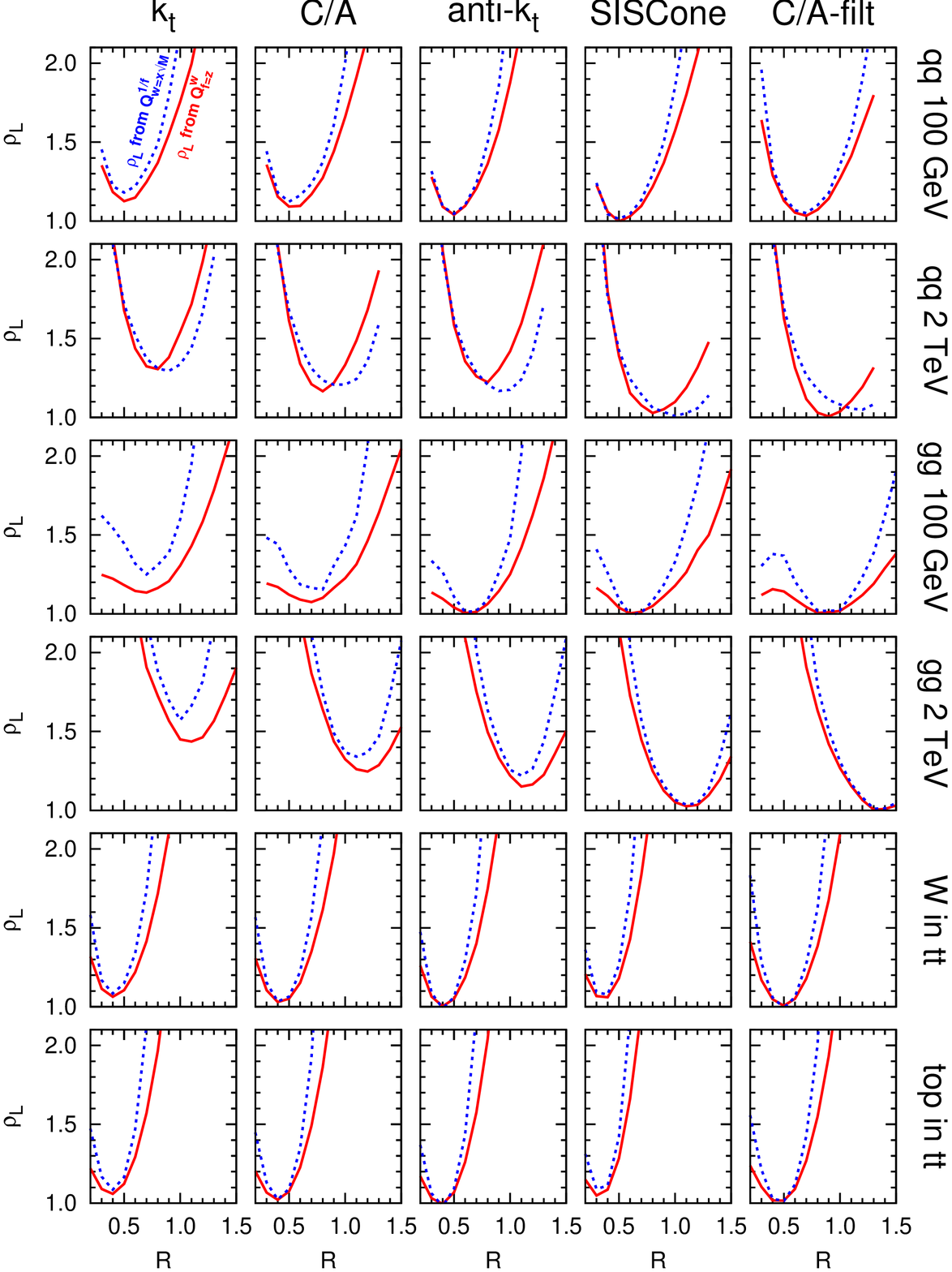}
  \caption{
    For each process (one per row) this plot shows the extra factor in
    luminosity, $\rhoL$, required in order to obtain the same
    significance as with the best jet definition, as a function of
    $R$. The (red) solid line corresponds to the estimate of $\rhoL$
    based on the minimal width \Qa{z}, while the (blue) dotted line
    corresponds to that based on the maximal fraction \Qb{1.25} (cf.\
    footnote~\ref{foot:Qb}). }
  \label{fig:summary-no-PU}
\end{figure}

Results for $\rho_L$ are given in fig.~\ref{fig:summary-no-PU}. 
One sees that the impact of the jet algorithm choice is greatest at
large mass scales (not surprising perhaps, given that large scales
prefer large $R$, where the area-sensitivity matters particularly).
That figure also illustrates how, at large mass scales, especially for
gluon jets (as dicussed first in \cite{Buttar:2008jx:KlausEtAl}),
standard choices of $R\sim 0.5$ are extremely poor --- requiring up to
twice as much luminosity to see a mass peak above a background.
This conclusion is relatively robust: fig.~\ref{fig:summary-no-PU}
actually has results for two different quality measures, which agree
remarkably well (the solid line derives from $\Qa{z}$, discussed
above, while the dashed line stems from an alternative measure, cf.\
footnote~\ref{foot:Qb}, or \cite{Cacciari:2008gd} for full details).

A reader who wishes to examine these quality measures further is
encouraged to consult a web-tool \cite{quality.fastjet.fr}, which
provides access to over 100,000 plots, two different quality measures
and $z$ values, with histograms for a wide range of jet definitions
and mass scales as well as summary plots of the quality measures and
resulting $\rhoL$ values.

Three  comments are due concerning the above
discussion. Firstly, it applies directly only to simple dijet events.
There have also been studies with multijet events from top-quark
decays, cf.\ the two bottom rows of fig.~\ref{fig:summary-no-PU}, as
well as an extensive analysis in \cite{Nojiri:2008ir}. 
What emerges from these studies is that the best choices in dijet
events (where SISCone works very well) may not be optimal in multijet
cases. For example, in \cite{Nojiri:2008ir} SISCone had more
difficulty resolving all relevant jets, while in
fig.~\ref{fig:summary-no-PU} the acceptable range of $R$ is somewhat
narrower for SISCone than for other algorithms.
A related point is that in multijet events the conclusion about the
need for larger $R$ at high scales is likely to conflict with the need
to resolve the multiple jets.
These are important issues and call for further study.

A second comment is that
refs.~\cite{Buttar:2008jx:Us,Buttar:2008jx:KlausEtAl,Cacciari:2008gd}
did not carry out any detailed tests of how the presence of realistic
background events affects the relation between mass scale and optimal
$R$.
However, the analysis to be described below, ref.~\cite{VariableR},
did include a study with background events, and confirms that at high
masses large-$R$ values are preferred (though the cuts and other
details differ somewhat from those in \cite{Cacciari:2008gd}).

The final comment concerns the relation between the results here and
those in table~\ref{tab:best-R-non-pert}. Here we have seen that for
high masses, large $R$ values are preferred, whereas
table~\ref{tab:best-R-non-pert} showed that in order to minimise
non-perturbative corrections one should use smaller $R$ values. 
These results are not in contradiction. 
In the case of table~\ref{tab:best-R-non-pert}, we had in mind
observables such as the inclusive jet spectrum, where the effects of
perturbative radiation just shift the distribution slightly, in a way
that is accounted for within the NLO QCD predictions to which one
compares experimental data (as long as $R$ is not so small that the
perturbative expansion breaks down). Therefore our aim was simply to
minimise the poorly controlled non-perturbative contributions, leading
to a preference for small $R$ values.
In the discussion in this section we considered the reconstruction of
a hadronically decaying resonance.
Perturbative radiation loss deforms the peak making it less visible
above a smooth background. 
Though we could calculate that deformation pertubatively, our ultimate
aim is to make the peak more visible and so a larger $R$ value, which
retains more perturbative radiation, becomes preferable.

\paragraph{Variable $R$.} One should be aware that there may also be
benefits to be had by moving away from the use of a single $R$ value
even when studying a single mass scale. For example, one might choose
to adapt $R$ according to the amount of noise (UE, pileup) in each
given event. Alternatively, other kinematic variables, like the total
transverse energy in the event~\cite{Kaplan:2008ie} or the rapidity
separation between the leading jets, can also be correlated with the
optimal $R$ choice on an event-by-event basis.

This last point was studied recently by Krohn, Thaler and
Wang (KRT)~\cite{VariableR} with a variable-$R$ jet algorithm (cf.\
section~\ref{sec:other-seq-rec}). 
Their $R$ value is actually not directly a function of the rapidity
difference, $\Delta y$, between the two hardest jets, but rather
scales as $1/p_{t,jet}$, which for two hard jets stemming from a
resonance of given fixed mass translates to a rapidity dependence $R
\sim \cosh \frac{\Delta y}{2}$.
This was motivated on the grounds that jets from a resonance decay
emit gluons on an angular scale that is independent of whether the
jets are transverse or along the beam direction; in the centre-of-mass
frame of the resonance, for two partons at rapidity $y/2$, separated
by an angle $\theta_{ij} \ll 1$, the boost-invariant angular distance
$\Delta R_{ij}$ is given by $\theta_{ij} \cosh \frac{y}{2}$. Hence the
scaling used for the jet radius.%
\footnote{The optimal $R$ choice should of course depend also
  on initial state radiation and the underlying event, and further
  studies might benefit by taking into account this information too.
}

\begin{table}
  \centering
  \begin{tabular}{lcccc}\toprule
    Algorithm & $500\GeV$ & $1\TeV$ & $2\TeV$ & $3\TeV$ \\ \midrule
    anti-$k_t$ $\to$ anti-$k_t$ VR & $18\%$ (0.9, 200) & $14\%$ (1.0,
    450) &$10\%$ (1.2, 1000) &$8\%$ (1.3, 1500) \\
    C/A $\to$ C/A VR & $17\%$ (0.9, 175) & $14\%$ (1.0,
    400) &$7\%$ (1.2, 1000) &$9\%$ (1.3, 1500) \\
    \bottomrule
  \end{tabular}
  \caption{Percentage improvement in the number of events from a resonance
    $X$ that have been reconstructed in the mass window $m_X\pm 25
    \GeV$, comparing a fixed-$R$ algorithm at its best $R$ (first
    number in brackets) with the variable-$R$ algorithm (the second
    number in brackets, $\rho/\!\GeV$, sets the jet radius as $R(p_t)
    = \rho/p_t$). Results taken from \cite{VariableR}.}
  \label{tab:KRT-var-R}
\end{table}

Table~\ref{tab:KRT-var-R} illustrates the improvements in signal
reconstruction that are obtained with this approach as compared to
fixed-$R$ algorithms. The benefit is at the level of tens of
percent. This is similar in magnitude to the improvement seen above by
optimising the choice of algorithm, or optimising a fixed $R$ as
compared to a standard $R=0.5$ or $R=0.7$ choice.

The KRT analysis was performed using the two leading jets
reconstructed from all particles with $|\eta| < 3$. This means that
significant numbers of events involve two leading jets with large
rapidity separations.
In this respect, the KRT analysis differs from that discussed above
\cite{Cacciari:2008gd}, which only studied events in which the leading
jet pair was separated by $|\Delta y| < 1$. With the latter
requirement, since $\cosh\frac{\Delta y}{2}$ would be close to $1$,
one might expect the $p_t$-dependent variable-$R$ choice to have a
more modest impact.

KRT also studied jet performance for resonance reconstruction that
includes a dijet background. Here too they found improvements with a
variable $R$ choice (and again at the level of about $10-15\%$), but
only if they supplemented their analysis with a ``jet quality'' cut
which requires that the energy be deposited centrally within the jet.
The corresponding fixed-$R$ analysis confirmed the need for large $R$
values at high mass scales.

\subsection{Pileup subtraction}
\label{sec:noise-subtraction}

The LHC will collide protons with an unprecedented instantaneous
luminosity of up to $10^{34}\cm^{-2}\s^{-1}$ and a bunch spacing of
$25\ns$, corresponding to $0.25\mb^{-1}$ per bunch crossing.
While this high luminosity is essential for many searches of rare new
physics processes at high energy scales, it also complicates analyses,
because at each bunch crossing there will be of the order of 20
minimum bias $pp$ interactions, which pollute any interesting hard
events with many soft particles.
The beams at LHC will have a longitudinal spread, and it may be
possible experimentally to associate each charged particle with a
distinct primary vertex that corresponds to a single $pp$ interaction
and so eliminate some fraction of the soft contamination. 
However, for neutral particles this is not possible, and many jet
measurements are in any case expected to be carried out with
calorimeter-cell or cluster information, for which there is not
sufficient angular resolution to reconstruct the original primary
vertex.
Therefore kinematic measurements for jets will be adversely affected
by pileup (PU), with resolution and absolute energy measurements
suffering significantly.

\begin{figure}[t]
  \centering
  \includegraphics[width=0.85\textwidth]{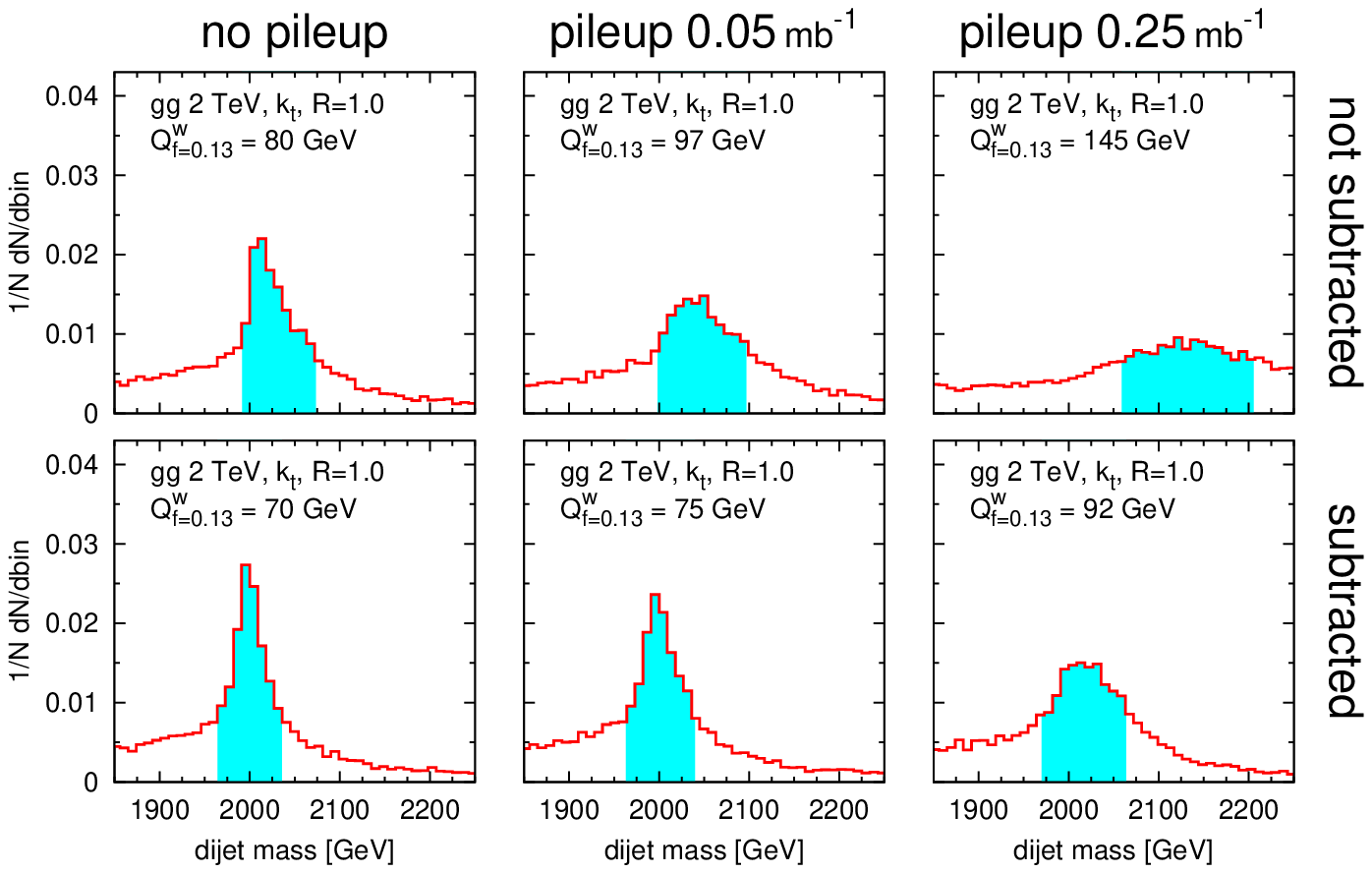}
  \caption{Invariant mass distributions for the $2\TeV$ \gluglu process of
    section~\ref{sec:numerical-radius}, for the $k_t$ algorithm with
    $R=1$, shown with no pileup (left), low pileup (middle) and high
    pileup (right), without subtraction (upper row) and with pileup
    subtraction as outlined in \cite{Cacciari:2007fd} (lower row). The
    shaded bands indicate the region used to calculate the \Qa{z}
    quality measure in each case. Figure taken
    from~\cite{Cacciari:2008gd}.}
  \label{fig:histos-with-PU}
\end{figure}

The impact of PU is illustrated in the upper row of
fig.~\ref{fig:histos-with-PU}, which shows histograms for the same
$2\TeV$ \gluglu resonance used above, but now with varying degrees of
pileup: none, low-luminosity LHC running ($0.05\mb^{-1}$ per
bunch crossing) and high-luminosity running ($0.25\mb^{-1}$ per
bunch crossing). The degradation of the
peak and its shift to higher masses are clearly evident here. While
the shift is perhaps not overly consequential (it could be corrected
for by comparing to MC simulation of the pileup), the loss of
resolution is a serious issue.

Both the Tevatron and LHC experiments have examined the question of
pileup. Some approaches to limiting its impact are based on average
correction procedures, for example the requirement that final measured
distributions should be independent of luminosity~\cite{CDFkt}, or a
correction to each jet given by some constant times the number of
primary interaction vertices (minus one)~\cite{CDFCone}.
These approaches have the advantage of being simple, but their
averaged nature limits the extent to which they can restore resolution
lost through pileup.
Other approaches involve event-dependent corrections that are applied
to calorimeter towers either before or during the
clustering~\cite{CMS-sub,AliceHija}. These can give better
restoration of resolution than average-based methods. 
One drawback that they have is that they are tightly linked to the
specific experimental setup (for example calorimeter cell-size), and
require ad-hoc transverse-momentum thresholds to distinguish pileup
from hard jets.
Additionally they are sometimes tied to specific (legacy) jet
algorithms, and so may not always be readily applied to more modern
jet algorithms.

The above issues triggered the development of an
experiment-independent pileup subtraction approach in
\cite{Cacciari:2007fd}.
The essential observation is that pileup roughly modifies a jet's
$p_t$ as follows:
\begin{equation}
  \label{eq:deltapt-for-sub}
  \Delta p_t =  A \rho \pm \sigma \sqrt{A} + \Delta p_t^{B}\,,
\end{equation}
where $A$ is the jet's area (after addition of the pileup), $\rho$ is
the mean amount of transverse momentum per unit area that has been
added to the event by pileup; $\sigma$ measures the fluctuations of
the pileup from point-to-point within the event (defined as the
standard deviation of the distribution of pileup across many squares
of area $1$); and $\Delta p_t^{B}$ is the net change in transverse
momentum due to back reaction.
Fluctuations in the jet $p_t$ arise because $A$ varies from jet-to-jet,
$\rho$ from event-to-event, because the pileup density fluctuates
from place to place within an event
(the $\sigma$ term) and because of back reaction.
The approach of \cite{Cacciari:2007fd} involves first running the jet
algorithm, estimating $\rho$ and then subtracting the $A \rho$ term
from each jet. 
This leaves just the $\sigma \sqrt{A}$ piece and the back-reaction
term and should significantly reduce both the fluctuations and the mean
offset in jet energy.\footnote{One might eliminate the back reaction
  too if one could subtract the PU before running the jet-finder ---
  subtracting PU directly at particle level leads, however, to
  negative momenta, with consequent non-trivial issues within many
  jet-algorithms.}

\begin{figure}[tp]
  \includegraphics[width=0.48\textwidth]{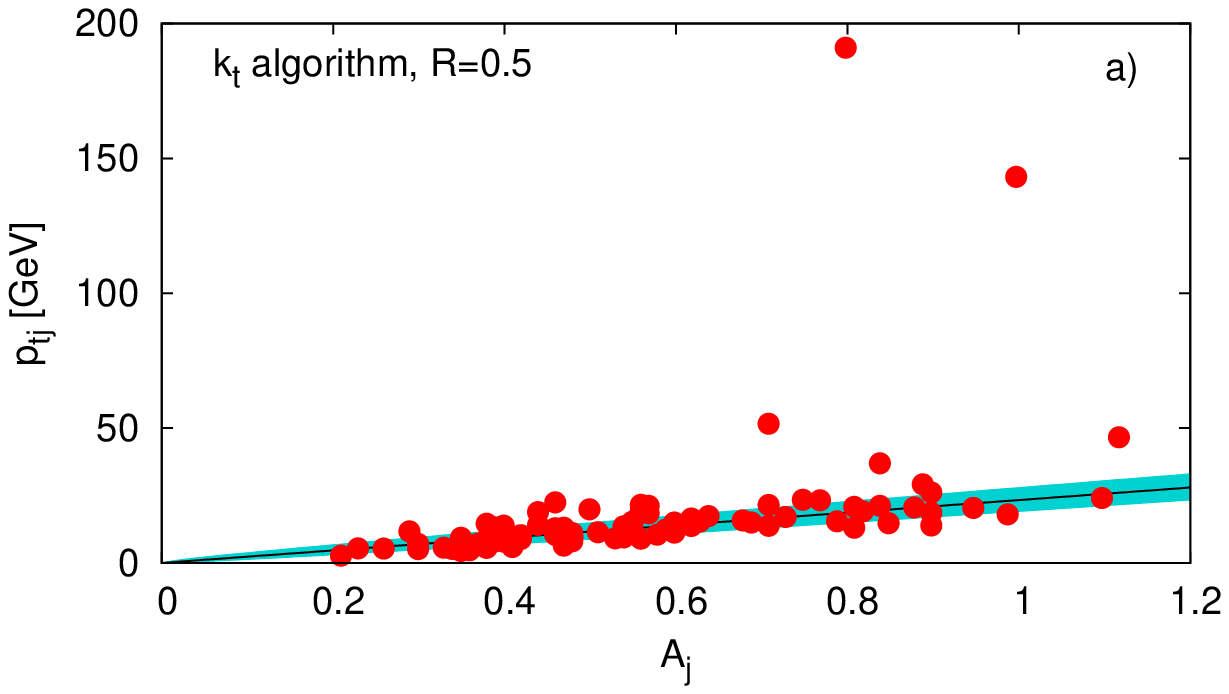}\hfill
  \includegraphics[width=0.48\textwidth]{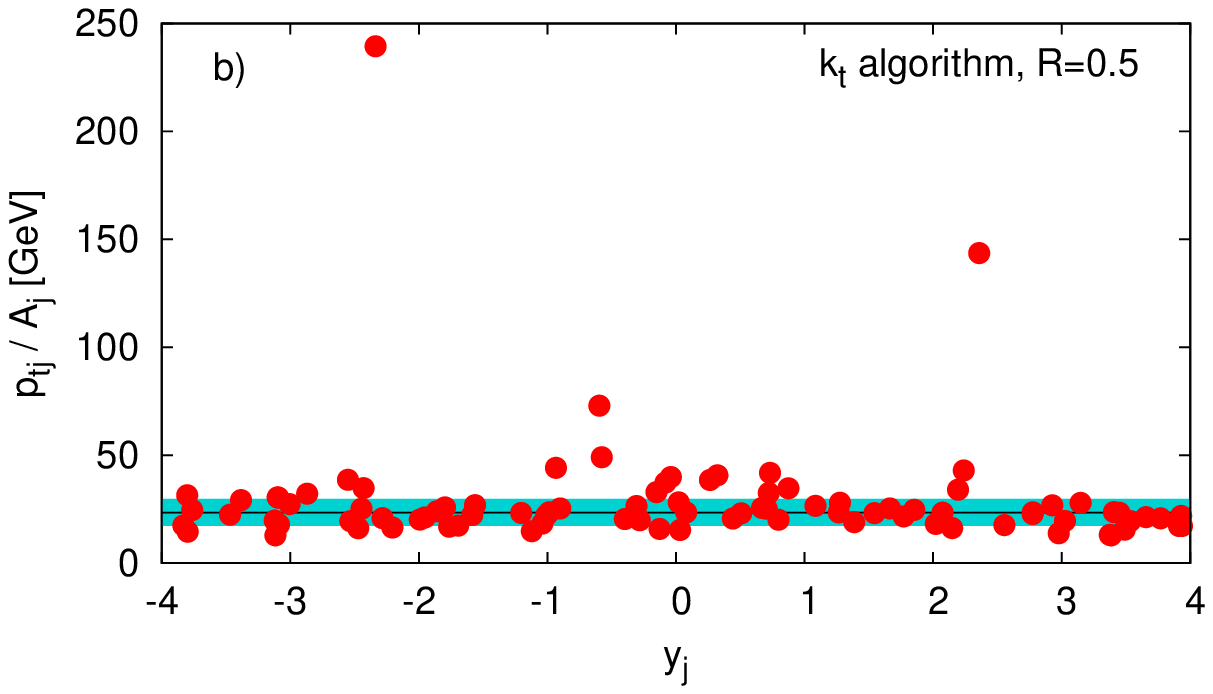}
  \caption{\small a) Scatter plot of the jet transverse momentum
    $p_{tj}$ versus its area $A_j$, for an LHC dijet event with a
    pileup of 22 minimum bias interactions (simulated with the default
    tune of Pythia~6.325~\cite{Pythia63}). The line and band are given
    by $\rho A_j \pm \sigma \sqrt{A_j}$.  b) The ratio $p_{tj}/A_{j}$
    as a function of the rapidity, $y_j$, for the same event; the line
    and band are given by $\rho \pm \sigma / \sqrt{\langle
      A\rangle}$. Taken from~\cite{Cacciari:2007fd}.}
  \label{fig:pt-v-area}
  \label{fig:pt_over_area-v-rap}
\end{figure}

The estimation of $\rho$ (without using detector-specific information,
such as the origin of tracks) is non-trivial because one must decide
what part of the event belongs to the hard event and which part comes
from pileup. Ideally one wants to do this without introducing any
explicit threshold to distinguish the two, given that the natural
threshold would vary significantly from event to event.
One observation of \cite{Cacciari:2007fd} is that this can be done
using the jets themselves. Fig.~\ref{fig:pt-v-area}a shows a scatter
plot of jet $p_t$ versus jet area for a single Monte Carlo event and
one sees a clear correlation from nearly all the
jets. Fig.~\ref{fig:pt-v-area}b shows $p_t/A$ for each jet as a
function of rapidity for the same event and one sees that the $p_t/A$
results cluster around a value that is fairly independent of rapidity
(it is still unclear just how true this will actually be at the LHC).
These two features led to the proposal to take the distribution of
$p_t/A$ for all jets in an event, up to some maximum rapidity, and to
then use its median (robust with respect to outliers, \ie hard jets)
as an estimate
of $\rho$.\footnote{One technical detail is that only certain jet
  algorithms, essentially $k_t$ and C/A, are suitable for estimating
  $\rho$, while the subtraction itself can be performed for any
  algorithm. When trying to optimise the subtraction there are also
  issues with the choice of the correct $R$ value in the JD used to
  estimate $\rho$, with $R\sim 0.5-0.6$ being a generally reasonably
  choice.} 
That estimate gives the black line in
fig.~\ref{fig:pt-v-area}, while the  band's width is controlled
by the value of $\sigma$ obtained from examining the width of the
$p_t/A$ distribution.

With $\rho$ estimated in this fashion, one can correct each jet by an
amount:
\begin{equation}
  \label{eq:area-correction}
  p_{tj}^\mu \to p_{tj}^\mu - \rho A_j^\mu\,,
\end{equation}
where $A_j^\mu$ is the jet's 4-vector area.
This approach was used to obtain the lower row of
fig.~\ref{fig:histos-with-PU}, illustrating the substantial gain in
peak quality that is to be had with the method (as well as nearly
correct reconstruction of peak position). 
In this specific case there is even an improvement in the peak quality
in the case without pileup, a consequence of the fact that the above
method also subtracts UE.

One comment is that pileup subtraction does not completely
eliminate the effect of pileup. In the area-based approach just
described, this is because of the last two terms on the RHS of
eq.~(\ref{eq:deltapt-for-sub}) are still present after subtraction.
Nevertheless it reduces the impact of pileup sufficiently that
conclusions about optimal jet definitions drawn from
fig.~\ref{fig:summary-no-PU} in the absence of pileup also hold with
pileup. This is important, because it means that analyses that use
data taken at different luminosities can successfully use a common jet
algorithm, independently of the pileup.

\paragraph{Subtraction in heavy-ion collisions.} 
The techniques described above have the potential to be useful also in
heavy-ion (HI) jet finding, where the problem is to identify jets given the
large soft background of particles that results from the hot dense
matter that is formed in a heavy-ion collision. 
This is of interest in the heavy-ion community (see for example the
review \cite{d'Enterria:2009vg}) because the modification of jets as
they traverse the hot dense medium may provide insight into the nature
of the medium.

The proposal for area-based pileup subtraction \cite{Cacciari:2007fd} also included
an application to the HI case and it has been investigated by the STAR
collaboration at RHIC in a first preliminary measurement of jet-cross
sections in Au\,Au gold collisions at $\sqrt{S_{NN}} = 200\GeV$
\cite{Salur:2008hs}.

The value of $\rho$ in the heavy-ion case is up to an order of
magnitude larger than in high-luminosity $pp$ running. This places
particularly stringent constraints on the accuracy that is required in
the subtraction and has spurred various ongoing investigations. 
Given the similarities between HI and high-pileup $pp$ jet finding, it
is to be expected that these investigations will be beneficial in both
environments.

Among the issues that are being considered is that of how to estimate
$\rho$ without recourse to the whole event, given that there is
significant rapidity dependence (and even azimuth dependence in some
cases) in the production of soft particles, both in HI and $pp$
collisions (the event in fig.~\ref{fig:pt_over_area-v-rap}b is a
little unusual in its degree of rapidity-independence).
Other issues are those of minimising systematic residual shifts from
back reaction (for which the anti-$k_t$ algorithm is beneficial) and
reducing the impact of point-to-point fluctuations in the noise (in
this respect C/A filtering seems to offer a promising avenue).
%


\subsection{Substructure}
\label{sec:substructure}

A key feature of the LHC is that it will be the first collider
to probe scales that are significantly above the EW scale.
This is what will allow the LHC to investigate the nature of
electroweak symmetry breaking and explore new territory in the search
for particles and phenomena beyond those of the standard model.

The importance of physics at transverse momenta $p_t \gg m_Z$ has
implications for the structure of the final state because at high
transverse momenta, ``signature'' particles, W's, Z's, Higgs bosons
and top quarks, have very collimated decays (due to their relativistic
boost).
Standard approaches for identifying these particles (i.e.  recombining
different jets) fail because all the decay products end up in a
single jet.

The work so far on identifying hadronic decays of boosted heavy
particles has fallen into two broad classes: particles with
two-pronged decays (the EW bosons), and those with three-pronged
decays (top quarks).
In each case, the mass of the jet is one indicator of its
origin (as discussed recently for example
in~\cite{Fitzpatrick:2007qr,Skiba:2007fw,Holdom:2007nw,Holdom:2007ap,Agashe:2007ki}).  
However, even for massless partons, QCD branching generates a
significant fraction of jets with large masses (or equivalently
with 2 or 3-pronged substructure): assuming a given jet $p_t$, the
leading-order (fixed-coupling) differential QCD jet-mass distribution
goes as
\begin{equation}
  \label{eq:jet-mass-dist}
  \frac{1}{n}\frac{dn}{dM^2} \sim \frac1{M^2} \frac{\as C_i}{\pi} \left(\ln \frac{R^2
    p_t^2}{M^2} + \order{1}\right)
\end{equation}
(see \cite{Almeida:2008yp} for more detailed analytic expressions, or
\cite{Catani:1992ua,Dasgupta:2001sh} for corresponding resummed
results in \ee collisions) and the logarithm can in part compensate
the smallness of $\as$, especially at larger $p_t$.
Two main questions that need to be answered are then: how can one
reduce the background of QCD jets of a given mass, and how can one get
the best resolution on jet mass so as to be able to use a small
jet-mass window in selecting candidate heavy particles?

\subsubsection{Two-pronged decays}
\label{sec:two-pronged-decays}

The first detailed discussion of advanced jet techniques for
two-pronged decays, over 15 years ago, was given by Seymour
in~\cite{Seymour:1993mx} in the context of a search for a heavy Higgs
boson decaying to $WW$ with one $W$ decaying leptonically, the other
hadronically.
He mainly considered the issue of mass resolution and investigated two
approaches.
One method involved the (inclusive) $k_t$ algorithm, with $R=1$, in
which the clustering sequence for the hardest jet was essentially
undone by one step, so as to resolve the jet into the two subjets from
the $W$ decay. 
The resulting separation of the subjets could then be used to set a
smaller $R$ for a second run of the $k_t$ algorithm, which helped
improve the mass resolution.
Another method involved the use of a cone algorithm with quite small
$R$, $\sim 0.25$ in order to directly identify the two subjets. 
This small $R$ was needed in order to robustly resolve the two
subjets, but that then caused it to lose significant gluon radiation
from the $W\to q\bar q$ system, giving worse mass resolution than the
$k_t$ algorithm.
The basic observation was therefore that the $k_t$ algorithm's
intrinsic internal information on substructure allowed one to be more
flexible in the compromise between identifying substructure and
capturing the bulk of the relevant radiation.

%

The next development on the subject was made by Butterworth, Cox and
Forshaw \cite{Butterworth:2002tt} who examined $WW$ scattering, again
with one leptonically and one hadronically decaying $W$.
They observed that the distribution of $k_t$ distance, $d_{ij}$
(eq.~(\ref{eq:ppkt-dist})), between the two $W$ subjets was close to
the $W$ mass in $W$ decays, but tended to have lower values in generic
massive jets.
This allowed them to obtain a substantial reduction in the background.
The same idea was used later for electroweak-boson reconstruction in
the context of a SUSY search~\cite{Butterworth:2007ke}.
The tool associated with this technique is often referred to as
``Y-splitter''.

It is worthwhile looking at some simple analytic results that relate
to the techniques of~\cite{Butterworth:2002tt} and
\cite{Seymour:1993mx}.
For a quasi-collinear splitting into two objects $i$ and $j$, the
total mass is $m^2 \simeq p_{ti} p_{tj} \Delta R_{ij}^2$. Labelling
$i$ and $j$ such that $p_{tj} < p_{ti}$ and defining $z = p_{tj}/p_t$
($p_t = p_{ti}+ p_{tj}$), then
\begin{align}
  \label{eq:m2-boosted}
  m^2 &\simeq z(1-z) p_t^2 \Delta R_{ij}^2\,,\\
  d_{ij} &= z^2 p_t^2 \Delta R_{ij}^2 \simeq \frac{z}{(1-z)} m^2\,.
\end{align}
Electroweak bosons decay with a fairly uniform
distribution in $z$ (exactly uniform for a Higgs boson), whereas a QCD
splitting has a soft divergence, e.g.
\begin{equation}
  \label{eq:Pgq-for-dij}
  P_{gq} \propto \frac{1+(1-z)^2}{z}\,.
\end{equation}
This means that for a fixed mass window, the background will have lower
$d_{ij}$ values than the signal.
Indeed, the logarithm in eq.~(\ref{eq:jet-mass-dist}) comes from the
integral over the $1/z$ divergence in eq.~(\ref{eq:Pgq-for-dij}), with
lower limit $z\gtrsim m^2/p_t^2 R^2$.
If one places a cut on $d_{ij}$, or analogously on $z$,
then one eliminates that logarithm, thus reducing the
QCD background (one can even calculate, analytically, what the optimal
cut is for given signals and backgrounds).

A second set of observations concerns mass resolution. Firstly, with a
small cone of size $R \ll \Delta R_{ij}$ used to reconstruct the two
prongs of a colour-singlet $q\bar q$ state, then there will be an
average loss of (squared) mass, and correspondingly of mass
resolution, dominated by a contribution from perturbative gluon
radiation,
\begin{equation}
  \label{eq:pt-mass-loss}
  \langle \delta m^2\rangle \simeq 2 m^2 \cdot \frac{\as L_q}{\pi} \left(\ln
    \frac{R}{\Delta R_{ij}} 
    + \order{1}\right)\,,\qquad R\ll \Delta R_{ij}\,,
\end{equation}
with $L_q\simeq \CF$ as given in eq.~(\ref{eq:Lqg-res}).
If instead a single jet is used to reconstruct the whole $q\bar q$
system, then one can show that most of the perturbative radiation from
the $q \bar q$ system will be contained in the jet. However there may 
then be significant contamination from the UE and pileup,
\begin{equation}
  \label{eq:UE-mass-gain}
  \langle \delta m^2\rangle \simeq  \rho \;p_t \frac{\pi R^4}{2}\,,
\end{equation}
for a circular jet (cf.\ eq.~(\ref{eq:delta-m2-UE}), with $\rho \equiv
\Lambda_{UE}/2\pi$), with an additional contribution coming also from
perturbative radiation from the beam.
Even though the above two equations represent major
oversimplifications of the full dynamics, one can
understand the task of optimising mass resolution as one of minimising
both types of contribution (in analogy with
section~\ref{sec:analytical-radius}).

This understanding provided the backdrop to a two-pronged subjet
technique given in \cite{Butterworth:2008iy}, used there for a
high-$p_t$ Higgs boson search in association with a back-to-back
high-$p_t$ vector boson.
The approach involved the Cambridge/Aachen algorithm, because its
sequential recombination in increasing angular distance is ideally
suited to dealing with problems that involve multiple or unknown
angular scales.
The basic procedure that was used to identify a $H\to b\bar b$ decay
went as follows:
\begin{enumerate}
\item Break a C/A jet $j$ into two subjets by undoing its last stage
  of clustering. Label the two subjets $j_1, j_2$ such that $m_{j_1} >
  m_{j_2}$.
\item If there was a significant mass drop (MD), $m_{j_1} < \mu
  m_{j}$, and the splitting is not too asymmetric, $y=
  \frac{\min(p_{tj_1}^2, p_{tj_2}^2)}{m_j^2} \Delta R_{j_1,j_2}^2 >
  y_\cut$, then deem $j$ to be the heavy-particle neighbourhood and
  exit the loop ($\mu$ was taken to be $0.67$ and $y_\cut =
  0.09$). Note that $y \simeq
  \min(p_{tj_1},p_{tj_2})/\max(p_{tj_1},p_{tj_2})$.\footnote{This
    $y_\cut$ is related to, but not the same as, that used to
    calculate the splitting scale
    in~\cite{Butterworth:2002tt,Butterworth:2007ke}, which use a
    dimensionful $d_\cut$.}
\item Otherwise redefine $j$ to be equal to $j_1$ and go back to step 1.
\end{enumerate}
The search for a mass-drop, step~2, served to identify the point in
the decomposition that involved significant hard substructure and, in
the context of a Higgs-boson search, one can verify that the two
subjets at that stage both have a $b$-tag.
The cut on $y \simeq z/(1-z)$ allows one to kill the logarithm for
(fake $b$-tag) QCD backgrounds in eq.~(\ref{eq:jet-mass-dist}).
By virtue of angular ordering~\cite{Coherence}, the two C/A subjets
produced at that stage, each with opening angle equal to $\Delta
R_{j_1,j_2}$, should contain nearly all the perturbative radiation
from the $b\bar b$ system (\ie eq.~(\ref{eq:pt-mass-loss}) is close to
zero).
They still tend to include too much contamination from the UE however,
so one can then apply a filtering technique in which the two
subjets are reexamined on a smaller angular scale $R_{\filt}$ and only
the three hardest components (\ie $b\bar b g$) were retained. This
essentially reduces the coefficient of the UE contamination in
eq.~(\ref{eq:UE-mass-gain}). The value used for $R_{\filt}$ was
specific to the jet, $R_{\filt} = \min(0.3, R_{b\bar b}/2)$, though
this could perhaps be further optimised.

\begin{table}
  \centering
  \begin{tabular}{lrrr}\toprule
    Jet definition & $\sigma_{S}/$fb & $\sigma_{B}$/fb &
    $\sigma_S/\sqrt{\sigma_B\!\cdot\mathrm{fb}}$ \\ \midrule
    C/A, $R=1.2$, MD-F         & 0.57 & 0.51 & 0.80 \\
    $k_t$, $R=1.0$, $y_{cut}$    & 0.19 & 0.74 & 0.22 \\
    SISCone, $R=0.8$           & 0.49 & 1.33 & 0.42 \\
    anti-$k_t$, $R=0.8$          & 0.22 & 1.06 & 0.21 \\ 
    \bottomrule
  \end{tabular}
  \caption{Cross section for signal $(\sigma_S)$ and the $Z+$jets
    background ($\sigma_B$) in the
    leptonic $Z$ 
    channel of $HZ$ production at a $14\TeV$ LHC, for $200 <
    p_{TZ}/\!\GeV < 600$ and $110 < m_J/\!\GeV <
    125$, with perfect $b$-tagging; the C/A algorithm uses the
    procedure outlined in the text; the $k_t$ algorithm uses the first
    step of decomposition to identify two subjets with a cut on
    $y_{ij}$ as for C/A; SISCone and anti-$k_t$ do not use any subjet
    analysis, but each require two $b$-tags within the jet.
    In each case $R$ has been chosen to give near optimal 
    significance with that algorithm.}
  \label{tab:jet-perf}
\end{table}

A comparison of different jet algorithms for the $ZH$ search channel
for $m_H = 115\GeV$, with $Z\to \ee, \mu^+\mu^-$ for $p_{tH}, p_{tZ} >
200\GeV$ (and other cuts detailed in \cite{Butterworth:2008iy}) is
shown in table~\ref{tab:jet-perf}. 
The C/A algorithm with the mass-drop and filtering (MD-F) is clearly the
best both at extracting the signal and limiting the background.
The $k_t$ algorithm fares poorly mainly because of its poor mass
resolution (its larger area and fluctuations,
cf.~section~\ref{sec:jet-areas}, make it intrinsically worse than the
C/A algorithm, and it is shown without any filtering).
SISCone does quite well on the reconstruction of the signal, mainly
because of its particularly low sensitivity to UE contamination, but
does poorly on the background rejection because it fails to correlate
the $b$ tagging with the subjet momentum structure, as does
anti-$k_t$.
It is probably fair to say that the defects of the algorithms could to
some extent be resolved with refinements such as the use of
jet finding with multiple $R$ values. However it is only in the C/A
algorithm that the use of multiple $R$ values fits in naturally within
the context of a single run of the jet finder, and the C/A algorithm
provides an internal representation of the jet structure that makes it
particularly easy to establish the right $R$ values.

\begin{figure}[t]
  \begin{center}
  \includegraphics[width=0.6\linewidth]{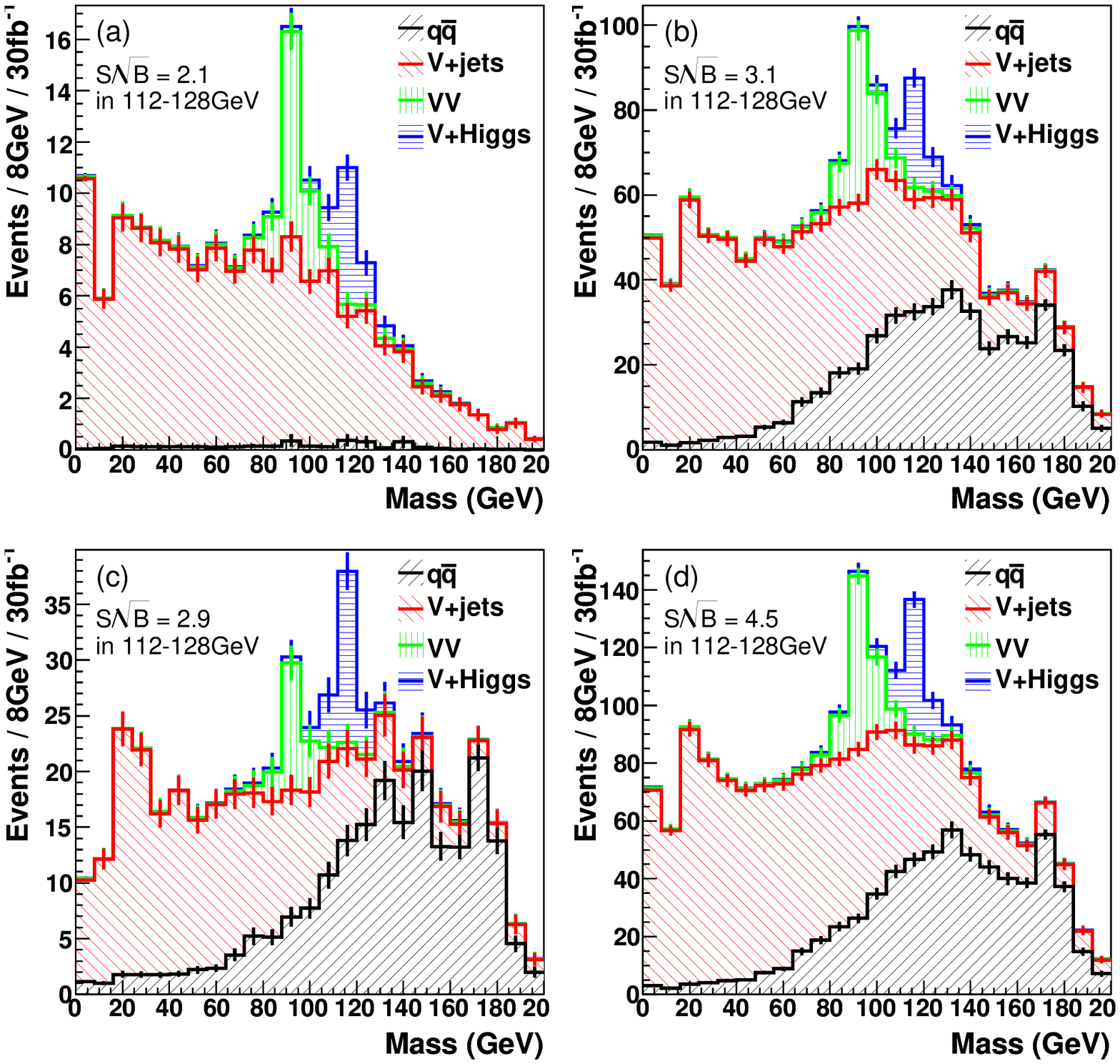}
  \caption{Signal and background for a 115~GeV SM Higgs in the $pp\to
    VH$ channels, with $H\to b\bar b$, simulated
    using Herwig 6.5 and Jimmy 4.31 (an ATLAS tune), C/A~MD-F with $R=
    1.2$ and $p_t > 200$~GeV, for 30 fb$^{-1}$. The $b$ tag efficiency
    is assumed to be 60\% and a fake-tag probability of 2\% is used. The
    $q\bar{q}$ sample includes dijets and $t\bar{t}$. The vector boson
    selections are (a) two leptons, (b) missing energy and (c) lepton
    plus missing energy, while (d) 
    shows the sum of all three channels (see \cite{Butterworth:2008iy}
    for details). 
    The errors reflect the statistical uncertainty on the simulated
    samples, and correspond to integrated luminosities $>
    30$~fb$^{-1}$.\vspace{-1.5em}}
  \label{fig:optlep}
  \end{center}
\end{figure}

For completeness, fig.~\ref{fig:optlep} shows the results of the Monte
Carlo simulation (particle-level) of the boosted $pp\to HV$ Higgs-boson search,
illustrating how this becomes a relevant search channel at the LHC
(and one that provides clean access to the product of $Hbb$ and $ZZH$,
$WWH$ couplings).
Subsequent to the writing of the original version of this review,
ATLAS confirmed that results similar to those of
\cite{Butterworth:2008iy} are obtained when accounting for detector
effects~\cite{AtlasHWHZ}.
Related methods were also examined for a $t\bar tH$
search~\cite{Plehn:2009rk} and for Higgs-boson production in new
physics events~\cite{Kribs:2009yh}.

\subsubsection{Three-pronged decays}
\label{sec:three-pronged-decays}

Three-pronged decays have been studied mainly in the context of top
decays.\footnote{Though there has also recently been work on the
  three-pronged hadronic decay of a neutralino in an $R$-parity
  violating SUSY scenario~\cite{Butterworth:2009qa}.}
This is motivated in part because high-mass $t\bar t$
resonances are a feature of many new physics scenarios (see for
example
\cite{Agashe:2006hk,Fitzpatrick:2007qr,Lillie:2007yh,Frederix:2007gi,Baur:2008uv}
and references therein).

The use of subjet structure in identifying hadronically decaying tops
is a much more recent topics than for EW bosons, having developed
mainly in the last two years. However many of the ideas are directly
inspired from the two-pronged case.
Aside from examining the jet-mass (whose distribution is calculated in
detail at leading order in \cite{Almeida:2008yp}),
techniques that have been investigated include subjet-decomposition
with the $k_t$ algorithm \cite{Thaler:2008ju,Brooijmans:ATL08},
C/A subjet techniques \cite{Kaplan:2008ie} and pruning~\cite{Ellis:2009su,Ellis:2009me}.
Among the discriminating variables that are used, there are $d_{ij}$
type variables \cite{Brooijmans:ATL08,AtlasTop09}, $z$-type variables
\cite{Thaler:2008ju,Kaplan:2008ie,Ellis:2009su,Ellis:2009me,Plehn:2009rk} and
event-shape variables \cite{Thaler:2008ju,Almeida:2008yp} (in both
cases, a spherocity-like variable in the plane transverse to the jet),
constraints on a $W$-subjet mass \cite{Thaler:2008ju,Kaplan:2008ie}
as well as other interjet correlation variables (a helicity angle
$\theta_h$ in \cite{Kaplan:2008ie}).
Most of the work has been geared to quite high-$p_t$ boosted tops in
the simple environment of a resonance decaying to $t\bar bt$, though one
study has also been carried out of moderately boosted top-tagging in
the busy environment of $pp\to t\bar t H$~\cite{Plehn:2009rk}.

\begin{table}
  \centering
    \begin{tabular}[t]{llcc}\toprule
      & Method                                & efficiency & fake fraction\\\midrule
      %
      %
       (from \cite{Thaler:2008ju})     & just jet mass           &  50\%  & 10\% \\
       %
       %
       ATLAS \cite{Brooijmans:ATL08,AtlasTop09}      & 3,4 $k_t$
       subjets, $d_{cut}$            & 45\% (85\%)      & 5\% (15\%)\\
       %
       %
       Thaler \& Wang \cite{Thaler:2008ju} & 2,3 $k_t$ subjets, $z_{cut}$ + shape    & 40\%       & 5\%\\
       %
       %
       Kaplan et al. \cite{Kaplan:2008ie}  & 3,4 C/A subjets, $z_{cut}$ + $\theta_h$ & 40\%       & $1\%$\\
       Ellis et al. \cite{Ellis:2009su,Ellis:2009me}  & Pruning & 15\%       & $0.05\%$\\
       CMS \cite{CMS-boosted-top} & variant of \cite{Kaplan:2008ie} &
       $45\%$ & $3\%$\\
       \bottomrule
    \end{tabular}
    \caption{
      Efficiencies for reconstructing top quarks with $p_t\sim
      1\TeV$ and the fraction of normal QCD jets that get a fake ``top-tag''.
      Shown for the various tagging methods that quoted numbers easily
      amenable to interpretation in this manner (numbers in brackets
      are for alternative sets of cuts). 
      Insofar as results involve different detector and resolution
      assumptions, Monte Carlo
      generator choices, as well as slightly different $p_t$ cuts,
      the comparison should be considered indicative   
      rather than precise.
      Furthermore the results of the various methods have
      different dependences on the transverse momenta being studied.
    }
  \label{tab:top-eff}
\end{table}

\begin{figure}
  \centering
  \includegraphics[width=0.5\textwidth]{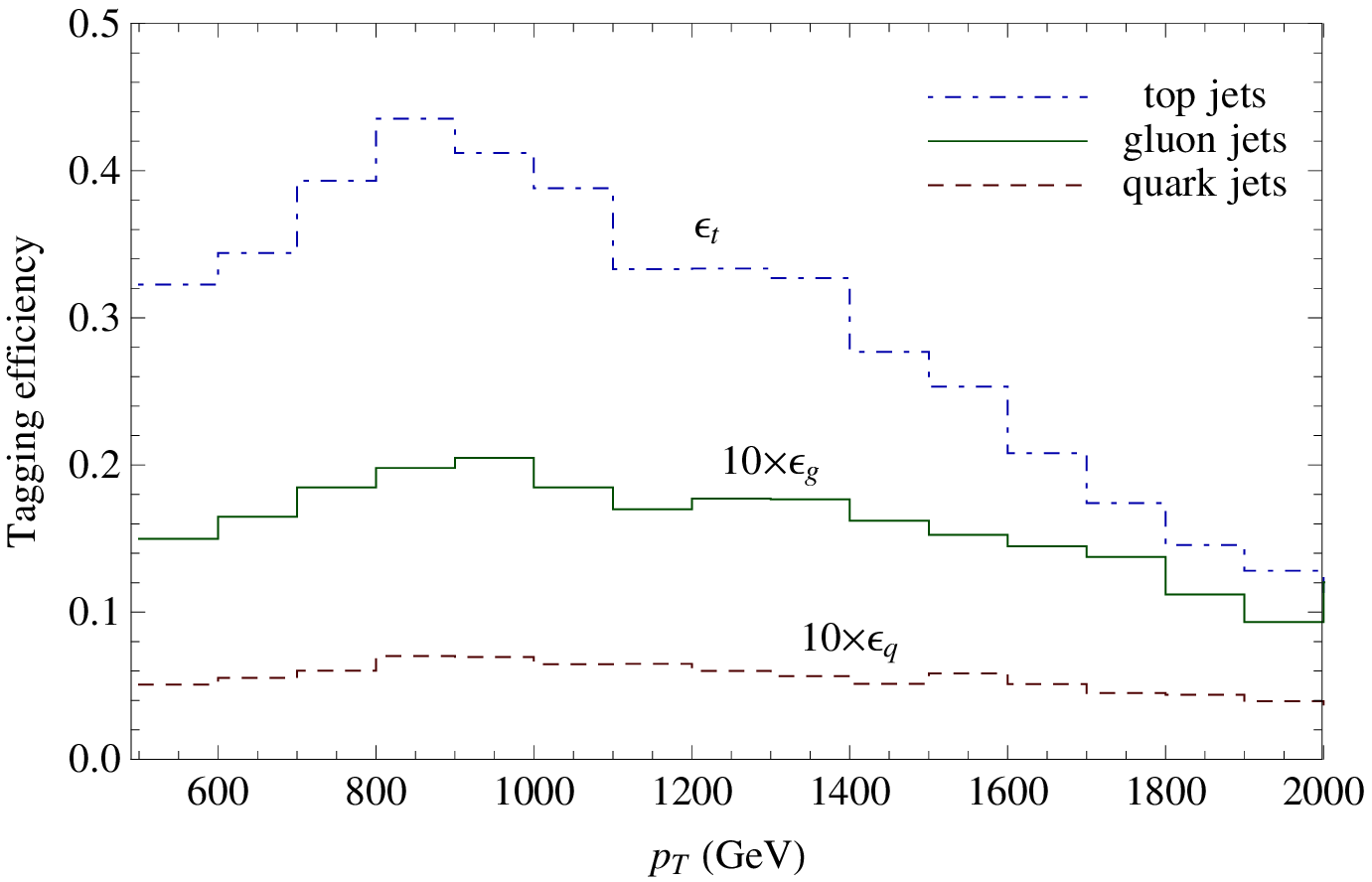}\qquad\qquad
  \includegraphics[width=0.3\textwidth]{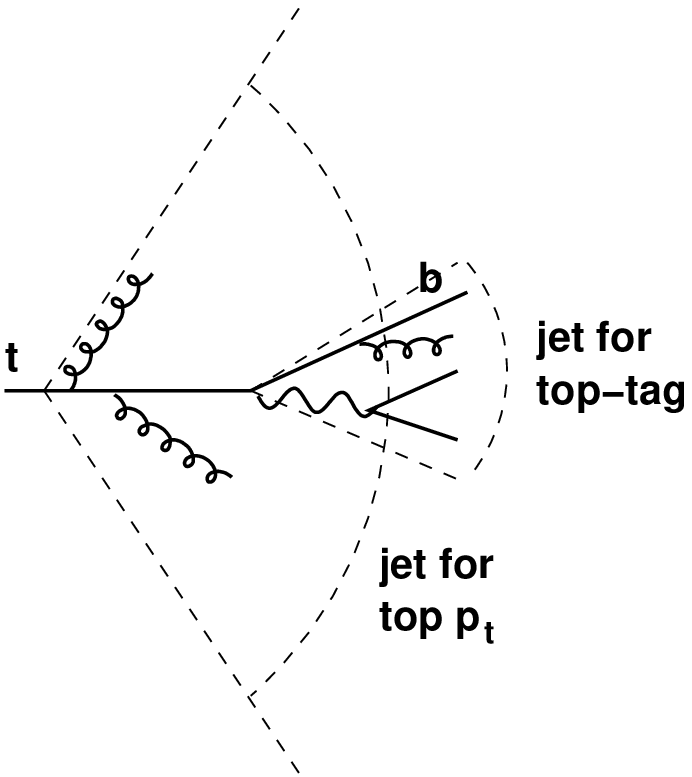}
  \caption{Left: signal efficiency for boosted top ID, $\epsilon_t$, and fake-tag rates
    for quark and gluon jets ($\epsilon_q,\epsilon_g$, both multiplied
    by 10 for visibility) for the Kaplan et al.\ C/A-based top-tagger,
    as a function of jet $p_t$ (reproduced from~\cite{Kaplan:2008ie}).
    Right: the use of two jet sizes for top reconstruction: the inner
    cone, of order a few times $m/p_t$, includes the top decay
    products, but excludes radiation from the top quark itself
    (dead-cone). To capture that radiation and reconstruct the correct
    top $p_t$, one should use the outer cone.}
  \label{fig:boosted-top-illust}
\end{figure}

A summary of results for top-tagging efficiencies and fake rates in
the various high-$p_t$ methods is given in table~\ref{tab:top-eff}. What emerges
from this is that the C/A-based approach of~\cite{Kaplan:2008ie} seems to
offer good efficiencies and very good QCD-jet rejection, with the best
signal significance and signal-to-background ratio (``better than
$b$-tagging at high $p_t$'').
Compared to the C/A-based method of \cite{Butterworth:2008iy} for
Higgs decays, particularly relevant differences are that it avoids any
reference to mass-drops (thus simplifying the method), and introduces
a minimal distance between subjects, which needs to be adjusted with
the jet $p_t$.
Its signal efficiency and fake-tag rates are shown as a function of
transverse momentum in fig.~\ref{fig:boosted-top-illust} (left).  The
decrease in efficiencies at high transverse momenta is simply a
consequence of the inclusion of finite calorimeter tower sizes
($0.1\times0.1$), and could perhaps be alleviated experimentally with
the use of tracking and electromagnetic calorimetry, both of which
have finer angular resolution.
In tests by CMS~\cite{CMS-boosted-top} of a variant of the method, the
efficiency instead saturates near $45\%$ at high-$p_t$, while it is
the fake rate that progressively degrades.

%

One important point in top-tagging is that to obtain the best tagging
at high transverse momenta, one should use an $R$ value that scales as $1/p_t$,
because the top-quark decay is mostly contained in a cone of width of
order $2-3$ times $m/p_t$.
Using a jet-opening angle that is much larger than this will lead to
considerable degradation in mass resolution, not only because of UE
contamination (as in the colour-singlet two-body decay case), but also
because the top quark, a coloured object, itself radiates gluons, which
will tend to increase the jet mass.
The C/A-based approach of~\cite{Kaplan:2008ie} is to some extent able
to find the right $R$ automatically for a given top-decay, and this is
part of its strength.

Finding the top quark is only half the task however: one must also
establish its momentum.
Barely any of the gluons emitted from a fast-moving top quark are
contained in the small jet used to identify the top --- this is the
dead-cone phenomenon for radiation from a massive quark.
To capture them one should instead use a jet with large opening angle,
as one would for a high-$p_t$ light quark~\cite{Cacciari:2008gd}, cf.\
section~\ref{sec:choosing-radius}.
This is essential if one is to obtain good mass resolution on a $t\bar
t$ resonance and is summarised in fig.~\ref{fig:boosted-top-illust}
(right).

Thus, when studying highly boosted tops, one needs to examine the
event on two angular scales: quite small $R\sim 3m_t/p_t$ to tag
the top-decay structure, and large $R\sim1$ to reconstruct the
top-quark momentum before the top started emitting gluons.

\subsection{Summary}
\label{sec:use-summary}

It is probably fair to say that the question of how best to use jets
is still in its infancy. Nevertheless, some clear results have emerged
from the above discussion.
\begin{itemize}
\item There will not be a single ``best'' jet definition (\ie $R$ and
  jet-algorithm) at the LHC.
  What's optimal will depend on what one wants to measure. 
  The trade-off will be between resolving separate jets (not really
  discussed here), capturing their perturbative radiation and limiting
  UE contamination, and depends on the momentum scale of an event, the
  number of jets, and so forth.
  In particular, kinematic reconstructions prefer larger $R$ values at
  high $p_t$ and for gluon jets (even $R\gtrsim 1$), because of the
  increased importance of capturing perturbative emission from the
  jets.

\item Monte Carlo studies of dijet resonances confirm this
  picture. They also indicate that among the different algorithms,
  $k_t$ is worst and SISCone and Cambridge/Aachen with filtering are
  best (cf.\ fig.~\ref{fig:summary-no-PU}). This is in accord with
  expectations based on their areas, \ie 
  their sensitivity to the UE, and is most relevant at high scales.
  Differences between algorithms, expressed as the extra luminosity
  needed to obtain a given (toy) signal significance, are at the level
  of a few tens of percent.

  Furthermore, it seems that event-dependent choices for the $R$ value
  can lead to additional improvements of a similar order of magnitude.

\item Pileup is a major issue, and significantly degrades kinematic
  reconstructions, even at high momentum scales.
  One can devise tools to measure the amount of pileup event-by-event
  and to subtract it jet-by-jet. 
  This leads to a noticeable improvement in kinematic reconstruction
  quality, though it does not quite restore it to the level of the
  no-pileup case.

\item At the LHC's highest momentum scales, electroweak-scale
  particles appear to be light. Their decays are collimated into
  single jets.
  Sequential-recombination jet algorithms provide a clean way of
  resolving the consequent substructure, and the most flexible seems
  to be Cambridge/Aachen, again with filtering to reduce UE
  contamination.
\end{itemize}
There are many remaining open questions. Among them: how to reconcile
the need for large $R$ at high $p_t$ with the task of resolving
complex multi-jet events; how this connects with the use of
substructure is resolving highly boosted decays; how to calculate the
optimal $R$ analytically, perhaps using the resulting information
event-by-event; how to choose the parameters of filtering and how this
ties in with possible improvements in pileup subtraction; and how all
of this works in full physics studies, including realistic backgrounds
and detector effects.
It is to be hoped that future work will cast light on these questions.


\section{Conclusions}

This review has covered a range of developments in the practical and
theoretical aspects of jet finding over the past few years. These are
steps on the way to a fully developed science of the use of jets,
``jetography''.

One important development is that LHC now has access to a range of
fast, infrared- and collinear-safe algorithms, together with methods
that allow any of the algorithms to be used in a high-luminosity LHC
environment. 
IRC safety is essential if the LHC is to benefit maximally from the
huge predictive effort that is ongoing within the QCD theory
community. 
Practicality is a necessary condition for the algorithms to be used in
an experimental context.

A number of these advances have been taken up by the LHC
experiments. For example both ATLAS and CMS incorporate FastJet within
their software frameworks.
At the time of writing (v2 of the arXiv version of this report), all 4
LHC experiments have seen first collisions at a centre-of-mass energy
of $\sqrt{s} = 900\GeV$ and it appears that both ATLAS and CMS have
used the anti-$k_t$ algorithm for finding jets in this initial data. 
%
These are welcome developments given the importance of IRC safety for
straightforward comparisons with perturbative QCD predictions and for
the use of perturbative methods in generally thinking about jets.

The second main development is that theoretical work has started on
the question of how best to use jets in an LHC-type environment. 
This is an important question because the LHC spans two orders of
magnitude in jet energy and has substantial (and variable) pileup, and
no single jet definition will work optimally for the whole range of
LHC phenomena.

Progress has been outlined here (\toplevel~\ref{sec:understanding}) on
our analytical understanding of how jets behave, and in
\toplevel~\ref{sec:using-jets} we have seen a handful of examples that
benefited significantly from the use of the ``right'' jet-finding
approach.
Currently these two aspects of work on jets are connected
qualitatively: the understanding of \toplevel~\ref{sec:understanding}
helped to interpret the results and inspire some of the methods of
\toplevel~\ref{sec:using-jets}.
However a rigorous, quantitative link is still missing, and
\toplevel~\ref{sec:using-jets} in any case covered only a small fraction
of the possible use-cases for jets.
This highlights a clear path for future work: that of bringing our
analytical tools to bear on the full range of uses of jets at the LHC,
so as to identify optimal jet-finding solutions across the board.



\section*{Acknowledgements}
\addcontentsline{toc}{section}{Acknowledgements}

My initial interest in the subject of jet finding owes much to talks,
exchanges, arguments, comments, encouragement by/with/from Jon
Butterworth, G\"unther Dissertori, Joey Huston, Mike Seymour, Markus
Wobisch, as well as many others.
The bulk of my direct involvement in the subject has been in
collaboration with Matteo Cacciari and Gregory Soyez and they have
done a lot to transform the subject --- without their ideas,
determination and enthusiasm, the advances that I've been involved
with would not have happened.
Many others have contributed, both as collaborators, Andrea Banfi, Jon
Butterworth, Adam Davison, John Ellis, Mrinal Dasgupta, Lorenzo
Magnea, Are Raklev, Juan Rojo, Mathieu Rubin, Sebastian Sapeta, Giulia
Zanderighi; and in terms of more informal exchanges: Timothy Chan,
Olivier Devillers (both from the field of computational geometry) as
well as Patrick Aurenche, Siegmund Brandt, Tancredi Carli,
Pierre-Antoine Delsart, Yuri Dokshitzer, Steve Ellis, David
d'Enterria, Michel Fontannaz, Peter Jacobs, Thomas Kluge, David
Kosower, Bruce Knuteson, Peter Loch, Michelangelo Mangano, Arthur
Moraes, Andreas Oehler, Klaus Rabbertz, Christof Roland, Philipp
Schieferdecker, Torbj\"orn Sj\"ostrand, Peter Skands, George Sterman,
Brock Tweedie, Monica Vazquez Acosta.
I am also grateful to David Kosower and an anonymous reviewer for
their detailed comments on this manuscript, and to Florent Fayette and
Adam Falkowski for pointing out ambiguities and typographical errors.
%
%
%

\end{document}